%% file: main.tex
\pdfoutput=1
\documentclass[aps,prx,twocolumn,groupedaddress]{revtex4-2}
\usepackage[utf8]{inputenc}
\usepackage[T1]{fontenc}
\usepackage{graphicx}
\usepackage{amsfonts,amsmath,amssymb,amsthm}
\usepackage{mathtools}
\usepackage{xcolor}
\usepackage{hyperref}

\DeclareMathOperator{\sign}{sign}
\DeclareMathOperator*{\extr}{extr}

\DeclareMathOperator*{\argmax}{argmax}
\DeclareMathOperator{\erf}{erf}
\DeclareMathOperator{\tr}{Tr}

\DeclareMathOperator{\cov}{Cov}

\begin{document}
\title{Statistical physics analysis of graph neural networks: \\ Approaching optimality in the contextual stochastic block model}

\author{\includegraphics[height=0.79em]{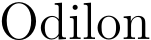} Duranthon}
\email{firstname.secondname@epfl.ch}
\author{Lenka Zdeborová}
\email{firstname.secondname@epfl.ch}
\affiliation{%
Statistical physics of computation laboratory,\\
École polytechnique fédérale de Lausanne, Switzerland
}%

\date{\today}

\begin{abstract}
Graph neural networks (GNNs) are designed to process data associated with graphs. They are finding an increasing range of applications; however, as with other modern machine learning techniques, their theoretical understanding is limited. GNNs can encounter difficulties in gathering information from nodes that are far apart by iterated aggregation steps. This situation is partly caused by so-called oversmoothing; and overcoming it is one of the practically motivated challenges. We consider the situation where information is aggregated by multiple steps of convolution, leading to graph convolutional networks (GCNs). We analyze the generalization performance of a basic GCN, trained for node classification on data generated by the contextual stochastic block model. We predict its asymptotic performance by deriving the free energy of the problem, using the replica method, in the high-dimensional limit. Calling {\it depth} the number of convolutional steps, we show the importance of going to large depth to approach the Bayes-optimality. We detail how the architecture of the GCN has to scale with the depth to avoid oversmoothing. The resulting large depth limit can be close to the Bayes-optimality and leads to a continuous GCN. Technically, we tackle this continuous limit via an approach that resembles dynamical mean-field theory (DMFT) with constraints at the initial and final times. An expansion around large regularization allows us to solve the corresponding equations for the performance of the deep GCN. This promising tool may contribute to the analysis of further deep neural networks.
\end{abstract}

\maketitle

\section{Introduction}

\subsection{Summary of the narrative}

Graph neural networks (GNNs) emerged as the leading paradigm when learning from data that are associated with a graph or a network. Given the ubiquity of such data in sciences and technology, GNNs are gaining importance in their range of applications, including chemistry \cite{wang23gnnMolecules}, biomedicine \cite{li22gnnBiomedecine}, neuroscience \cite{bessadok21gnnNeuroscience}, simulating physical systems \cite{sanchez2020gnnPhysique}, particle physics \cite{shlomi20gnnPhysiquePart} and solving combinatorial problems \cite{peng20gnnOpt,cappart23gnnOpt}. As common in modern machine learning, the theoretical understanding of learning with GNNs is lagging behind their empirical success. In the context of GNNs, one pressing question concerns their ability to aggregate information from far away parts of the graph: the performance of GNNs often deteriorates as depth increases \cite{morris24positionGNN}. This issue is often attributed to oversmoothing \cite{li18oversmoothing,oono20oversmoothing}, a situation where a multi-layer GNN averages out the relevant information. Consequently, mostly relatively shallow GNNs are used in practice or other strategies are designed to avoid oversmoothing \cite{li19GNNskipConn,chen20GNNskipConn}.

Understanding the generalization properties of GNNs on unseen examples is a path towards yet more powerful models. Existing theoretical works addressed the generalization ability of GNNs mainly by deriving generalization bounds, with a minimal set of assumptions on the architecture and on the data, relying on VC dimension, Rademacher complexity or a PAC-Bayesian analysis; see for instance \cite{ju23pacBayes} and the references therein. Works along these lines that considered settings related to one of this work include \cite{tang23generalizationGNN}, \cite{cong21depthGCN} or \cite{esser21generalizationCSBM}. However, they only derive loose bounds for the test performance of the GNN and they do not provide insights on the effect of the structure of data. \cite{tang23generalizationGNN} provides sharper bounds; yet they do not take into account the data structure and depend on continuity constants that cannot be determined a priori. In order to provide more actionable outcomes, the interplay between the architecture of the GNN, the training algorithm and the data needs to be understood better, ideally including constant factors characterizing their dependencies on the variety of parameters. 

Statistical physics traditionally plays a key role in understanding the behaviour of complex dynamical systems in the presence of disorder. In the context of neural networks, the dynamics refers to the training, and the disorder refers to the data used for learning. In the case of GNNs, the data is related to a graph. The statistical physics research strategy defines models that are simplified and allow analytical treatment. One models both the data generative process, and the learning procedure. A key ingredient is a properly defined thermodynamic limit in which quantities of interest self-average. One then aims to derive a closed set of equations for the quantities of interest, akin to obtaining exact expressions for free energies from which physical quantities can be derived. While numerous other research strategies are followed in other theoretical works on GNNs, see above, the statistical physics strategy is the main one accounting for constant factors in the generalization performance and as such provides invaluable insight about the properties of the studied systems. This line of research has been very fruitful in the context of fully connected feed-forward neural networks, see e.g. \cite{seung1992statistical,loureiro2021learning,mei2022generalization}. It is reasonable to expect that also in the context of GNNs this strategy will provide new actionable insights.    

The analysis of generalization of GNNs in the framework of the statistical physics strategy was initiated recently in \cite{shi2022statistical} where the authors studied the performance of a single-layer graph convolutional neural network (GCN) applied to data coming from the so-called contextual stochastic block model (CSBM). The CSBM, introduced in \cite{yan2021covariate,cSBM18}, is particularly suited as a prototypical generative model for graph-structured data where each node belongs to one of several groups and is associated with a vector of attributes. The task is then the classification of the nodes into groups. Such data are used by practitioners as a benchmark for performance of GNNs \cite{cong21depthGCN,chien20pageRank,fu21pLaplacianGNN,lei22evenNet}. On the theoretical side, the follow-up work \cite{dz24gcn} generalized the analysis of \cite{shi2022statistical} to a broader class of loss functions but also alerted to the relatively large gap between the performance of a single-layer GCN and the Bayes-optimal performance. 

In this paper, we show that the close-formed analysis of training a GCN on data coming from the CSBM can be extended to networks performing multiple layers of convolutions. With a properly tuned regularization and strength of the residual connection this allows us to approach the Bayes-optimal performance very closely. Our analysis sheds light on the interplay between the different parameters --mainly the depth, the strength of the residual connection and the regularization-- and on how to select the values of the parameters to mitigate oversmoothing. On a technical level the analysis relies on the replica method, with the limit of large depth leading to a continuous formulation similar to neural ordinary differential equations \cite{chen18NODE} that can be treated analytically via an approach that resembles dynamical mean-field theory with the position in the network playing the role of time. We anticipate that this type of infinite depth analysis can be generalized to studies of other deep networks with residual connections such a residual networks or multi-layer attention networks.

\subsection{Further motivations and related work}

\subsubsection{Graph neural networks:}
\label{sec:introGNN}
In this work we focus on graph neural networks (GNNs). GNNs are neural networks designed to work on data that can be represented as graphs, such as molecules, knowledge graphs extracted from encyclopedias, interactions among proteins or social networks. GNNs can predict properties at the level of nodes, edges or the whole graph. Given a graph $\mathcal G$ over $N$ nodes, its adjacency matrix $A\in\mathbb R^{N\times N}$ and initial features $h_i^{(0)}\in\mathbb R^M$ on each node $i$, a GNN can be expressed as the mapping
\begin{align}
h_i^{(k+1)} = f_{\theta^{(k)}}\left(h_i^{(k)}, \mathrm{aggreg}(\{h_j^{(k)}, j\sim i\})\right) \label{eq:gnn}
\end{align}
for $k=0,\dots,K$ with $K$ being the depth of the network. 
where $f_{\theta^{(k)}}$ is a learnable function of parameters $\theta^{(k)}$ and $\mathrm{aggreg}()$ is a function that aggregates the features of the neighboring nodes in a permutation-invariant way. A common choice is the sum function, akin to a convolution on the graph
\begin{align}
\mathrm{aggreg}(\{h_j, j\sim i\}) = \sum_{j\sim i}h_j = (Ah)_i\ .
\end{align}
Given this choice of aggregation the GNN is called graph convolutional network (GCN) \cite{kipf17GCN}. For a GNN of depth $K$ the transformed features $h^{(K)}\in\mathbb R^{M'}$ can be used to predict the properties of the nodes, the edges or the graph by a learnt projection.

In this work we will consider a GCN with the following architecture, that we will define more precisely in the detailed setting part \ref{sec:setup}. We consider one trainable layer $w\in\mathbb R^M$, since dealing with multiple layers of learnt weights is still a major issue \cite{hugo23reseauProfondProp}, and since we want to focus on modeling the impact of numerous convolution steps on the generalization ability of the GCN.
\begin{align}
h^{(k+1)} &= \left(\frac{1}{\sqrt N}\tilde A+c_kI_N\right)h^{(k)} \\
\hat y &= \sign\left(\frac{1}{\sqrt N}w^Th^{(K)}\right)
\end{align}
where $\tilde A$ is a rescaling of the adjacency matrix, $I_N$ is the identity, $c_k\in\mathbb R$ for all $k$ are the residual connection strengths and $\hat y\in\mathbb R^N$ are the predicted labels of each node. We will call the number of layers $K$ the depth, but we reiterate that only the layer $w$ is learned.

\subsubsection{Analyzable model of synthetic data:}
Modeling the training data is a starting point to derive sharp predictions. A popular model of attributed graph, that we will consider in the present work and define in detail in sec.~\ref{sec:cSBM}, is the contextual stochastic block model (CSBM), introduced in \cite{yan2021covariate,cSBM18}. It consists in $N$ nodes with labels $y\in\{-1,+1\}^N$, in a binary stochastic block model (SBM) to model the adjacency matrix $A\in\mathbb R^{N\times N}$ and in features (or attributes) $X\in\mathbb R^{N\times M}$ defined on the nodes and drawn according to a Gaussian mixture. $y$ has to be recovered given $A$ and $X$. The inference is done in a semi-supervised way, in the sense that one also has access to a train subset of $y$.

A key aspect in statistical physics is the thermodynamic limit, how should $N$ and $M$ scale together. In statistical physics we always aim at a scaling in which quantities of interest concentrate around deterministic values, and the performance of the system ranges between as bad as random guessing to as good as perfect learning. As we will see, these two requirements are satisfied in the high-dimensional limit $N\to\infty$ and $M\to\infty$ with $\alpha=N/M$ of order one. This scaling limit also aligns well with the common graph datasets that are of interest in practice, for instance Cora \cite{mccallum00cora} ($N=3.10^3$ and $M=3.10^3$), Coauthor CS \cite{shchur18coauthor} ($N=2.10^4$ and $M=7.10^3$), CiteSeer \cite{giles98citeseer} ($N=4.10^3$ and $M=3.10^3$) and PubMed \cite{sen08pubmed} ($N=2.10^4$ and $M=5.10^2$).

A series of works that builds on the CSBM with lower dimensionality of features that is $M=o(N)$ exists. Authors of \cite{fountoulakis21GC} consider a one-layer GNN trained on the CSBM by logistic regression and derive bounds for the test loss; however, they analyze its generalization ability on new graphs that are independent of the train graph and do not give exact predictions. In \cite{baranwal23clipGNN} they propose an architecture of GNN that is optimal on the CSBM with low-dimensional features, among classifiers that process local tree-like neighborhoods, and they derive its generalization error. In \cite{wang24GCNmultiConv} the authors analyze the structure and the separability of the convolved data $\tilde A^KX$, for different rescalings $\tilde A$ of the adjacency matrix, and provide a bound on the classification error. Compared to our work these articles consider a low-dimensional setting (\cite{baranwal23clipGNN}) where the dimension of the features $M$ is constant, or a setting where $M$ is negligible compared to $N$ (\cite{fountoulakis21GC} and \cite{wang24GCNmultiConv}). 

\subsubsection{Tight prediction on GNNs in the high-dimensional limit:}
Little has been done as to tightly predicting the performance of GNNs in the high-dimensional limit where both the size of the graph and the dimensionality of the features diverge proportionally. The only pioneering references in this direction we are aware of are \cite{shi2022statistical} and \cite{dz24gcn}, where the authors consider a simple single-layer GCN that performs only one step of convolution, $K=1$, trained on the CSBM in a semi-supervised setting. In these works the authors express the performance of the trained network as a function of a finite set of order parameters following a system of self-consistent equations.

There are two important motivations to extend these works and to consider GCNs with a higher depth $K$. First, the GNNs that are used in practice almost always perform several steps of aggregation, and a more realistic model should take this in account. Second, \cite{dz24gcn} shows that the GCN it considers is far from the Bayes-optimal (BO) performance and the Bayes-optimal rate for all common losses. The BO performance is the best that any algorithm can achieve knowing the distribution of the data, and the BO rate is the rate of convergence toward perfect inference when the signal strength of the graph grows to infinity. Such a gap is intriguing in the sense that previous works \cite{mignacco20gaussMixt,aubin20glmReg} show that a simple one-layer fully-connected neural network can reach or be very close to the Bayes-optimality on simple synthetic datasets, including Gaussian mixtures. A plausible explanation is that on the CSBM considering only one step of aggregation $K=1$ is not enough to retrieve all information, and one has to aggregate information from further nodes. Consequently, even on this simple dataset, introducing depth and considering a GCN with several convolution layers, $K>1$, is crucial.

\begin{figure}[t]
 \centering
 \includegraphics[width=0.9\linewidth]{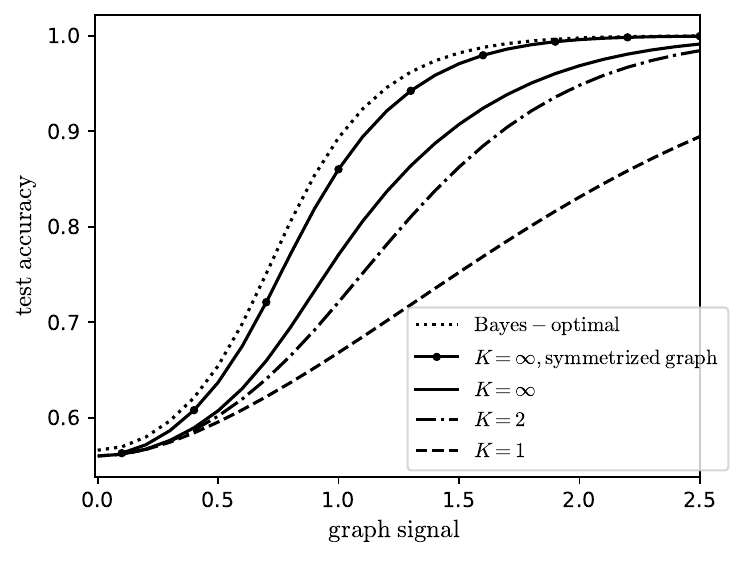}
 \caption{\label{fig:introduction} Test accuracy of the graph neural network on data generated by the contextual stochastic block model vs the signal strength. We define the model and the network in section \ref{sec:setup}. The test accuracy is maximized over all the hyperparameters of the network. The Bayes-optimal performance is from \cite{dz23csbm}. The line $K=1$ has been studied by \cite{shi2022statistical,dz24gcn}; we improve it to $K>1$, $K=\infty$ and symmetrized graphs. All the curves are theoretical predictions we derive in this work.}
\end{figure}

In the present work we study the effect of the depth $K$ of the convolution for the generalization ability of a simple GCN. A first part of our contribution consists in deriving the exact performance of a GCN performing several steps of convolution, trained on the CSBM, in the high-dimensional limit. We show that $K=2$ is the minimal number of steps to reach the BO learning rate. As to the performance at moderate signal strength, it appears that, if the architecture is well tuned, going to larger and larger $K$ increases the performance until it reaches a limit. This limit, if the adjacency matrix is symmetrized, can be close to the Bayes optimality. This is illustrated on fig.~\ref{fig:introduction}, which highlights the importance of numerous convolution layers.

\subsubsection{Oversmoothing and residual connections:}
Going to larger depth $K$ is essential to obtain better performance. Yet, GNNs used in practice can be quite shallow, because of the several difficulties encountered at increasing depth, such that vanishing gradient, which is not specific to graph neural networks, or oversmoothing \cite{li18oversmoothing,oono20oversmoothing}. Oversmoothing refers to the fact that the GNN tends to act like a low-pass filter on the graph and to smooth the features $h_i$, which after too many steps may converge to the same vector for every node. A few steps of aggregation are beneficial but too many degrade the performance, as \cite{keriven22oversm} shows for a simple GNN, close to the one we study, on a particular model. In the present work we show that the model we consider can suffer from oversmoothing at increasing $K$ if its architecture is not well-tuned and we precisely quantify it.

A way to mitigate vanishing gradient and oversmoothing is to allow the nodes to remember their initial features $h_i^{(0)}$. This is done by adding residual (or skip) connections to the neural network, so the update function becomes
\begin{align}
h_i^{(k+1)} = c_kh_i^{(k)}+f_{\theta^{(k)}}\left(h_i^{(k)}, \mathrm{aggreg}(\{h_j^{(k)}, j\sim i\})\right) \label{eq:res-gnn}
\end{align}
where the $c_k$ modulate the strength of the residual connections. The resulting architecture is known as residual network or resnet \cite{he15resnet} in the context of fully-connected and convolutional neural networks. As to GNNs, architectures with residual connections have been introduced in \cite{pham17columnNet} and used in \cite{li19GNNskipConn,chen20GNNskipConn} to reach large numbers of layers with competitive accuracy. \cite{xu21GNNskipConnGradient} additionally shows that residual connections help gradient descent. In the setting we consider we prove that residual connections are necessary to circumvent oversmoothing, to go to larger $K$ and to improve the performance.

\subsubsection{Continuous neural networks:}
Continuous neural networks can be seen as the natural limit of residual networks, when the depth $K$ and the residual connection strengths $c_k$ go to infinity proportionally, if $f_{\theta^{(k)}}$ is smooth enough with respect to $k$. In this limit, rescaling $h^{(k+1)}$ with $c_k$ and setting $x=k/K$ and $c_k = K/t$, the rescaled $h$ satisfies the differential equation
\begin{align}
\frac{\mathrm dh_i}{\mathrm dx}(x) = tf_{\theta(x)}\left(h_i(x), \mathrm{aggreg}(\{h_j(x), j\sim i\})\right)\ . \label{eq:gnn-cont}
\end{align}
This equation is called a neural ordinary differential equation \cite{chen18NODE}. The convergence of a residual network to a continuous limit has been studied for instance in \cite{sander22convergenceNODE}. Continuous neural networks are commonly used to model and learn the dynamics of time-evolving systems, by usually taking the update function $f_\theta$ independent of the time~$t$. For an example \cite{ling16turbulenceNN} uses a continuous fully-connected neural network to model turbulence in a fluid. As such, they are a building block of scientific machine learning; see for instance \cite{rackauckas20nodeScience} for several applications. As to the generalization ability of continuous neural networks, the only theoretical work we are aware of is \cite{marion23genNODE}, that derives loose bounds based on continuity arguments.

Continuous neural networks have been extended to continuous GNNs in \cite{poli19cGNN,xhonneux20cGNN}. For the GCN that we consider the residual connections are implemented by adding self-loops $c_kI_N$ to the graph. The continuous dynamic of $h$ is then
\begin{align}
\frac{\mathrm dh}{\mathrm dx}(x) = t\tilde Ah(x)\ ,
\end{align}
with $t\in\mathbb R$; which is a diffusion on the graph. Other types of dynamics have been considered, such as anisotropic diffusion, where the diffusion factors are learnt, or oscillatory dynamics, that should avoid oversmoothing too; see for instance the review \cite{han23revueDynGNN} for more details. No prior works predict their generalization ability. In this work we fill this gap by deriving the performance of the continuous limit of the simple GCN we consider.

\subsection{Summary of the main results:}
We first generalize the work of \cite{shi2022statistical,dz24gcn} to predict the performance of a simple GCN with arbitrary number $K$ of convolution steps. The network is trained in a semi-supervised way on data generated by the CSBM for node classification. In the high-dimensional limit and in the limit of dense graphs, the main properties of the trained network concentrate onto deterministic values, that do not depend on the particular realization of the data. The network is described by a few order parameters (or summary statistics), that satisfy a set of self-consistent equations, that we solve analytically or numerically. We thus have access to the expected train and test errors and accuracies of the trained network.

From these predictions we draw several consequences. Our main guiding line is to search for the architecture and the hyperparameters of the GCN that maximize its performance, and check whether the optimal GCN can reach the Bayes-optimal performance on the CSBM. The main parameters we considers are the depth $K$, the residual connection strengths $c_k$, the regularization $r$ and the loss function.

We consider the convergence rates towards perfect inference at large graph signal. We show that $K=2$ is the minimal depth to reach the Bayes-optimal rate, after which increasing $K$ or fine-tuning $c_k$ only leads to sub-leading improvements. In case of asymmetric graphs the GCN is not able to deal the asymmetry, for all $K$ or $c_k$, and one has to pre-process the graph by symmetrizing it.

At finite graph signal the behaviour of the GCN is more complex. We find that large regularization $r$ maximizes the test accuracy in the case we consider, while the loss has little effect. The residual connection strengths $c_k$ have to be tuned to a same optimal value $c$ that depends on the properties of the graph.

An important point is that going to larger $K$ seems to improve the test accuracy. Yet the residual connection $c$ has to vary accordingly. If $c$ stays constant with respect to $K$ then the GCN will perform PCA on the graph $A$, oversmooth and discard the information from the features $X$. Instead, if $c$ grows with $K$, the residual connections alleviate oversmoothing and the performance of the GCN keeps increasing with $K$, if the diffusion time $t=K/c$ is well tuned.

The limit $K\to\infty, c\propto K$ is thus of particular interest. It corresponds to a continuous GCN performing diffusion on the graph. Our analysis can be extended to this case by directly taking the limit in the self-consistent equations. One has to further jointly expand them around $r\to+\infty$ and we keep the first order. At the end we predict the performance of the continuous GCN in an explicit and closed form. To our knowledge this is the first tight prediction of the generalization ability of a continuous neural network, and in particular of a continuous graph neural network. The large regularization limit $r\to+\infty$ is important: on one hand it appears to lead to the optimal performance of the neural network; on another hand, it is instrumental to analyze the continuous limit $K\to\infty$ and it allows to analytically solve the self-consistent equations describing the neural network.

We show that the continuous GCN at optimal time $t$ performs better than any finite-$K$ GCN. The optimal $t$ depends on the properties of the graph, and can be negative for heterophilic graphs. This result is a step toward solving one of the major challenges identified by \cite{morris24positionGNN}; that is, creating benchmarks where depth is necessary and building efficient deep networks.

The continuous GCN as large $r$ is optimal. Moreover, if run on the symmetrized graph, it approaches the Bayes-optimality on a broad range of configurations of the CSBM, as exemplified on fig.~\ref{fig:introduction}. We identify when the GCN fails to approach the Bayes-optimality: this happens when most of the information is contained in the features and not in the graph, and has to be processed in an unsupervised manner.

We provide the code that allows to evaluate our predictions in the supplementary material.

\section{Detailed setting}
\label{sec:setup}

\subsection{Contextual Stochastic Block Model for attributed graphs}
\label{sec:cSBM}
We consider the problem of semi-supervised node classification on an attributed graph, where the nodes have labels and carry additional attributes, or features, and where the structure of the graph correlates with the labels. We consider a graph $\mathcal G$ made of $N$ nodes; each node $i$ has a binary label $y_i=\pm 1$ that is a Rademacher random variable.

The structure of the graph should be correlated with $y$. We model the graph with a binary stochastic block model (SBM): the adjacency matrix $A\in\mathbb R^{N\times N}$ is drawn according to
\begin{equation}
A_{ij} \sim \mathcal B\left(\frac{d}{N}+\frac{\lambda}{\sqrt N}\sqrt{\frac{d}{N}\left(1-\frac{d}{N}\right)}y_iy_j\right)
\end{equation}
where $\lambda$ is the signal-to-noise ratio (snr) of the graph, $d$ is the average degree of the graph, $\mathcal B$ is a Bernoulli law and the elements $A_{ij}$ are independent for all $i$ and $j$. It can be interpreted in the following manner: an edge between $i$ and $j$ appears with a higher probability if $\lambda y_iy_j>1$ i.e. for $\lambda>0$ if the two nodes are in the same group. The scaling with $d$ and $N$ is chosen so that this model does not have a trivial limit at $N\to\infty$ both for $d=\Theta(1)$ and $d=\Theta(N)$. Notice that we take $A$ asymmetric.

Additionally to the graph, each node $i$ carries attributes $X_i\in\mathbb R^M$, that we collect in the matrix $X\in\mathbb R^{N\times M}$. We set $\alpha=N/M$ the aspect ratio between the number of nodes and the dimension of the features. We model them by a Gaussian mixture: we draw $M$ hidden Gaussian variables $u_\nu\sim\mathcal N(0,1)$, the centroid $u\in \mathbb R^M$, and we set
\begin{equation}
X = \sqrt\frac{\mu}{N}yu^T+W \label{eq:csbm}
\end{equation}
where $\mu$ is the snr of the features and $W$ is noise whose components $W_{i\nu}$ are independent standard Gaussians. We use the notation $\mathcal N(m, V)$ for a Gaussian distribution or density of mean $m$ and variance $V$. The whole model for $(y, A, X)$ is called the contextual stochastic block model (CSBM) and was introduced in \cite{yan2021covariate,cSBM18}.

We consider the task of inferring the labels $y$ given a subset of them. We define the training set $R$ as the set of nodes whose labels are revealed; $\rho=|R|/N$ is the training ratio. The test set $R'$ is selected from the complement of~$R$; we define the testing ratio $\rho'=|R'|/N$. We assume that $R$ and $R'$ are independent from the other quantities. The inference problem is to find back $y$ and $u$ given $A$, $X$, $R$ and the parameters of the model.

\cite{cSBM18,cSBM20} prove that the effective snr of the CSBM is
\begin{equation}
\mathrm{snr}_\mathrm{CSBM}=\lambda^2+\mu^2/\alpha\ ,
\end{equation}
in the sense that in the unsupervised regime $\rho=0$ for $\mathrm{snr}_\mathrm{CSBM}<1$ no information on the labels can be recovered while for $\mathrm{snr}_\mathrm{CSBM}>1$ partial information can be recovered. The information given by the graph is $\lambda^2$ while the information given by the features is $\mu^2/\alpha$. As soon as a finite fraction of nodes $\rho>0$ is revealed the phase transition between no recovery and weak recovery disappears.

We work in the high-dimensional limit $N\to\infty$ and $M\to\infty$ while the aspect ratio $\alpha=N/M$ is of order one. The average degree $d$ should be of order $N$, but taking $d$ growing with $N$ should be sufficient for our results to hold, as shown by our experiments. The other parameters $\lambda$, $\mu$, $\rho$ and $\rho'$ are of order one.

\subsection{Analyzed architecture}
In this work, we focus on the role of applying several data aggregation steps. With the current theoretical tools, the tight analysis of the generic GNN described in eq.~\eqref{eq:gnn} is not possible: dealing with multiple layers of learnt weights is hard; and even for a fully-connected two-layer perceptron this is a current and major topic \cite{hugo23reseauProfondProp}. Instead, we consider a one-layer GNN with a learnt projection $w$. We focus on graph convolutional networks (GCNs) \cite{kipf17GCN}, where the aggregation is a convolution done by applying powers of a rescaling $\tilde A$ of the adjacency matrix. Last we remove the non-linearities. As we will see, the fact that the GCN is linear does not prevent it to approach the optimality in some regimes. The resulting GCN is referred to as simple graph convolutional network; it has been shown to have good performance while being much easier to train \cite{wu19simpleGCN,zhu21simpleGCN}. The network we consider transforms the graph and the features in the following manner:
\begin{equation}
h(w) = \prod_{k=1}^K\left(\frac{1}{\sqrt N}\tilde A+c_kI_N\right)\frac{1}{\sqrt N}Xw \label{eq:GCNdiscret}
\end{equation}
where $w\in\mathbb R^M$ is the layer of trainable weights, $I_N$ is the identity, $c_k\in\mathbb R$ is the strength of the residual connections and $\tilde A\in\mathbb R^{N\times N}$ is a rescaling of the adjacency matrix defined by
\begin{equation}
\tilde A_{ij}=\left(\frac{d}{N}\left(1-\frac{d}{N}\right)\right)^{-1/2}\left(A_{ij}-\frac{d}{N}\right),\;\mathrm{for\;all\;}i,\,j.
\end{equation}
The prediction $\hat y_i$ of the label of $i$ by the GNN is then $\hat y_i=\sign h(w)_i$.

$\tilde A$ is a rescaling of $A$ that is centered and normalized. In the limit of dense graphs, where $d$ is large, this will allow us to rely on a Gaussian equivalence property to analyze this GCN. The equivalence \cite{lesieur2017constrained,cSBM18,shi2022statistical} states that in the high-dimensional limit, for $d$ growing with $N$, $\tilde A$ can be approximated by the following spiked matrix $A^\mathrm{g}$ without changing the macroscopic properties of the GCN:
\begin{equation}
A^\mathrm{g}=\frac{\lambda}{\sqrt N}yy^T+\Xi\ ,
\end{equation}
where the components of the $N\times N$ matrix $\Xi$ are independent standard Gaussian random variables. The main reason for considering dense graphs instead of sparse graphs $d=\Theta(1)$ is to ease the theoretical analysis. The dense model can be described by a few order parameters; while a sparse SBM would be harder to analyze because many quantities, such as the degrees of the nodes, do not self-average, and one would need to take in account all the nodes, by predicting the performance on one realization of the graph or by running population dynamics. We believe that it would lead to qualitatively similar results, as for instance \cite{dz23glmSbm} shows, for a related model.

The above architecture corresponds to applying $K$ times a graph convolution on the projected features $Xw$. At each convolution step $k$ a node $i$ updates its features by summing those of its neighbors and adding $c_k$ times its own features. In \cite{shi2022statistical,dz24gcn} the same architecture was considered for $K=1$; we generalize these works by deriving the performance of the GCN for arbitrary numbers $K$ of convolution steps. As we will show this is crucial to approach the Bayes-optimal performance.

Compared to \cite{shi2022statistical,dz24gcn}, another important improvement towards the Bayes-optimality is obtained by symmetrizing the graph, and we will also study the performance of the GCN when it acts by applying the symmetrized rescaled adjacency matrix $\tilde A^\mathrm{s}$ defined by:
\begin{equation}
\tilde A^\mathrm{s} = \frac{1}{\sqrt 2}(\tilde A+\tilde A^T)\ ,\quad A^\mathrm{g,s}=\frac{\lambda^\mathrm{s}}{\sqrt N}yy^T+\Xi^\mathrm{s}\ . \label{eq:symétrisation}
\end{equation}
$A^\mathrm{g,s}$ is its Gaussian equivalent, with $\lambda^\mathrm{s}=\sqrt 2\lambda$, $\Xi^\mathrm{s}$ is symmetric and $\Xi^\mathrm{s}_{i\le j}$ are independent standard Gaussian random variables.
In this article we derive and show the performance of the GNN both acting with $\tilde A$ and $\tilde A^\mathrm{s}$ but in a first part we will mainly consider and state the expressions for $\tilde A$ because they are simpler. We will consider $\tilde A^\mathrm{s}$ in a second part while taking the continuous limit. To deal with both cases, asymmetric or symmetrized, we define $\tilde A^\mathrm{e}\in\{\tilde A, \tilde A^\mathrm{s}\}$ and $\lambda^\mathrm{e}\in\{\lambda, \lambda^\mathrm{s}\}$.

The continuous limit of the above network \eqref{eq:GCNdiscret} is defined by
\begin{equation}
h(w) = e^{\frac{t}{\sqrt N}\tilde A^\mathrm{e}}\frac{1}{\sqrt N}Xw \label{eq:GCNcontinu_vsTetR}
\end{equation}
where $t$ is the diffusion time. It is obtained at large $K$ when the update between two convolutions becomes small, as follows:
\begin{equation}
\left(\frac{t}{K\sqrt N}\tilde A^\mathrm{e}+I_N\right)^K\underset{K\to\infty}{\longrightarrow}e^{\frac{t}{\sqrt N}\tilde A^\mathrm{e}}\ . \label{eq:GCNcontinuBis}
\end{equation}
$h$ is the solution at time $t$ of the time-continuous diffusion of the features on the graph $\mathcal G$ with Laplacian $\tilde A^\mathrm{e}$, defined by $\partial_xX(x)=\frac{1}{\sqrt N}\tilde A^\mathrm{e}X(x)$ and $X(0)=X$. The discrete GCN can be seen as the discretization of the differential equation in the forward Euler scheme. The mapping with eq.~\eqref{eq:GCNdiscret} is done by taking $c_k=K/t$ for all $k$ and by rescaling the features of the discrete GCN $h(w)$ as $h(w)\prod_kc_k^{-1}$ so they remain of order one when $K$ is large. For the discrete GCN we do not directly consider the update $h_{k+1}=(I_N+c_k^{-1}\tilde A/\sqrt N)h_k$ because we want to study the effect of having no residual connections, i.e. $c_k=0$. The case where the diffusion coefficient depends on the position in the network is equivalent to a constant diffusion coefficient. Indeed, because of commutativity, the solution at time $t$ of $\partial_xX(x)=\frac{1}{\sqrt N}\mathrm a(x)\tilde A^\mathrm{e}X(x)$ for $\mathrm a:\mathbb R\to\mathbb R$ is $\exp\left(\int_0^t\mathrm dx\,\mathrm a(x)\frac{1}{\sqrt N}\tilde A^\mathrm{e}\right)X(0)$.

The discrete and the continuous GCNs are trained by empirical risk minimization. We define the regularized loss
\begin{equation}
L_{A,X}(w) = \frac{1}{\rho N}\sum_{i\in R}\ell(y_ih_i(w))+\frac{r}{\rho N}\sum_\nu \gamma(w_\nu)
\label{eq:loss}
\end{equation}
where $\gamma$ is a strictly convex regularization function, $r$ is the regularization strength and $\ell$ is a convex loss function. The regularization ensures that the GCN does not overfit the train data and has good generalization properties on the test set. We will focus on $l_2$-regularization $\gamma(x)=x^2/2$ and on the square loss $\ell(x)=(1-x)^2/2$ (ridge regression) or the logistic loss $\ell(x)=\log(1+e^{-x})$ (logistic regression). Since $L$ is strictly convex it admits a unique minimizer $w^*$. The key quantities we want to estimate are the average train and test errors and accuracies of this model, which are
\begin{align}
E_\mathrm{train/test} = \mathbb E\ \frac{1}{|\hat R|}\sum_{i\in\hat R}\ell(y_ih(w^*)_i) \label{eq:erreur} \\
\mathrm{Acc}_\mathrm{train/test} = \mathbb E\ \frac{1}{|\hat R|}\sum_{i\in\hat R\ ,}\delta_{y_i=\sign{h(w^*)_i}} \label{eq:precision}
\end{align}
where $\hat R$ stands either for the train set $R$ or the test set $R'$ and the expectation is taken over $y$, $u$, $A$, $X$, $R$ and $R'$. $\mathrm{Acc}_\mathrm{train/test}$ is the proportion of train/test nodes that are correctly classified. A main part of the present work is dedicated to the derivation of exact expressions for the errors and the accuracies. We will then search for the architecture of the GCN that maximizes the test accuracy $\mathrm{Acc}_\mathrm{test}$.

Notice that one could treat the residual connection strengths $c_k$ as supplementary parameters, jointly trained with $w$ to minimize the train loss. Our analysis can straightforwardly be extended to deal with this case. Yet, as we will show, to take trainable $c_k$ degrades the test performances; and it is better to consider them as hyperparameters optimizing the test accuracy.

\begin{table}[h!]
\caption{\label{tab:paramètres} Summary of the parameters of the model.}
\begin{ruledtabular}
\begin{tabular}{cl}
$N$ & number of nodes \\
$M$ & dimension of the attributes \\
$\alpha=N/M$ & aspect ratio \\
$d$ & average degree of the graph \\
$\lambda$ & signal strength of the graph \\
$\mu$ & signal strength of the features \\
$\rho = |R|/N$ & fraction of training nodes \\
$\ell$, $\gamma$ & loss and regularization functions \\
$r$ & regularization strength \\
$K$ & number of aggregation steps \\
$c_k$, $c$, $t$ & residual connection strengths, diffusion time
\end{tabular}
\end{ruledtabular}
\end{table}

\subsection{Bayes-optimal performance:}
An interesting consequence of modeling the data as we propose is that one has access to the Bayes-optimal (BO) performance on this task. The BO performance is defined as the upper-bound on the test accuracy that any algorithm can reach on this problem, knowing the model and its parameters $\alpha, \lambda, \mu$ and $\rho$. It is of particular interest since it will allow us to check how far the GCNs are from the optimality and how much improvement can one hope for.

The BO performance on this problem has been derived in \cite{cSBM18} and \cite{dz23csbm}. It is expressed as a function of the fixed-point of an algorithm based on approximate message-passing (AMP). In the limit of large degrees $d=\Theta(N)$ this algorithm can be tracked by a few scalar state-evolution (SE) equations that we reproduce in appendix \ref{sec:eqSE}.

\section{Asymptotic characterization of the GCN}
In this section we provide an asymptotic characterization of the performance of the GCNs previously defined. It relies on a finite set of order parameters that satisfy a system of self-consistent, or fixed-point, equations, that we obtain thanks to the replica method in the high-dimensional limit at finite $K$. In a second part, for the continuous GCN, we show how to take the limit $K\to\infty$ for the order parameters and for their self-consistent equations. The continuous GCN is still described by a finite set of order parameters, but these are now continuous functions and the self-consistent equations are integral equations.

Notice that for a quadratic loss function $\ell$ there is an analytical expression for the minimizer $w^*$ of the regularized loss $L_{A,X}$ eq.~\eqref{eq:loss}, given by the regularized least-squares formula. Based on that, a computation of the performance of the GCN with random matrix theory (RMT) is possible. It would not be straightforward in the sense that the convolved features, the weights $w^*$ and the labels $y$ are correlated, and such a computation would have to take in account these correlations. Instead, we prefer to use the replica method, which has already been successfully applied to analyze several architectures of one (learnable) layer neural networks in articles such that \cite{seung1992statistical,aubin20glmReg}. Compared to RMT, the replica method allows us to seamlessly handle the regularized pseudo-inverse of the least-squares and to deal with logistic regression, where no explicit expression for $w^*$ exists.

We compute the average train and test errors and accuracies eqs.~\eqref{eq:erreur} and \eqref{eq:precision} in the high-dimensional limit $N$ and $M$ large. We define the Hamiltonian
\begin{equation}
H(w) = s\sum_{i\in R}\ell(y_ih(w)_i)+r\sum_\nu\gamma(w_\nu) + s'\sum_{i\in R'}\ell(y_ih(w)_i) \label{eq:hamiltonien}
\end{equation}
where $s$ and $s'$ are external fields to probe the observables. The loss of the test samples is in $H$ for the purpose of the analysis; we will take $s'=0$ later and the GCN is minimizing the training loss~\eqref{eq:loss}. The free energy $f$ is defined as
\begin{equation}
Z = \int\mathrm dw\, e^{-\beta H(w)}\ ,\quad f = -\frac{1}{\beta N}\mathbb E\log Z\ .
\end{equation}
$\beta$ is an inverse temperature; we consider the limit $\beta\to\infty$ where the partition function $Z$ concentrates over $w^*$ at $s=1$ and $s'=0$. The train and test errors are then obtained according to
\begin{equation}
E_\mathrm{train} = \frac{1}{\rho}\frac{\partial f}{\partial s}\ ,\quad E_\mathrm{test} = \frac{1}{\rho'}\frac{\partial f}{\partial s'}
\end{equation}
both evaluated at $(s,s')=(1,0)$. One can, in the same manner, compute the average accuracies by introducing the observables $\sum_{i\in\hat R}\delta_{y_i=\sign{h(w)_i}}$ in $H$. To compute $f$ we introduce $n$ replica:
\begin{equation}
\mathbb E\log Z = \mathbb E\frac{\partial Z^n}{\partial n}(n=0) = \left(\frac{\partial}{\partial n}\mathbb EZ^n\right)(n=0)\ .
\label{eq:replica}
\end{equation}
To pursue the computation we need to precise the architecture of the GCN.

\subsection{Discrete GCN}
\label{sec:discreteGCN}
\subsubsection{Asymptotic characterization}
\label{sec:discrAsymptChara}
In this section, we work at finite $K$. We consider only the asymmetric graph. We define the state of the GCN after the $k^\mathrm{th}$ convolution step as
\begin{equation}
h_k = \left(\frac{1}{\sqrt N}\tilde A+c_kI_N\right)h_{k-1}\ ,\quad h_0 = \frac{1}{\sqrt N}Xw\ .
\end{equation}
$h_K=h(w)\in\mathbb R^N$ is the output of the full GCN. We introduce $h_k$ in the replicated partition function $Z^n$ and we integrate over the fluctuations of $A$ and $X$. This couples the variables across the different layers $k=0\ldots K$ and one has to take in account the correlations between the different $h_k$, which will result into order parameters of dimension $K$. One has to keep separate the indices $i\in R$ and $i\notin R$, whether the loss $\ell$ is active or not; consequently the free entropy of the problem will be a linear combination of $\rho$ times a potential with $\ell$ and $(1-\rho)$ times without $\ell$. The limit $N\to\infty$ is taken thanks to Laplace's method. The extremization is done in the space of the replica-symmetric ansatz, which is justified by the convexity of $H$. The detailed computation is given in appendix \ref{sec:appendixDiscreteGCN}.

The outcome of the computation is that this problem is described by a set of twelve order parameters (or summary statistics). They are $\Theta=\{m_w\in\mathbb R, Q_w\in\mathbb R, V_w\in\mathbb R, m\in\mathbb R^K, Q\in\mathbb R^{K\times K}, V\in\mathbb R^{K\times K}\}$ and their conjugates $\hat\Theta=\{\hat m_w\in\mathbb R, \hat Q_w\in\mathbb R, \hat V_w\in\mathbb R, \hat m\in\mathbb R, \hat Q\in\mathbb R^{K\times K}, \hat V\in\mathbb R^{K\times K}\}$, where
\begin{align}
& m_w = \frac{1}{N}u^Tw\ , && m_k = \frac{1}{N}y^Th_k\ ,\\
& Q_w = \frac{1}{N}w^Tw\ , && Q_{k,l} = \frac{1}{N}h^T_kh_l\ ,\\
& V_w = \frac{\beta}{N}\tr(\cov_\beta(w,w))\ , && V_{k,l} = \frac{\beta}{N}\tr(\cov_\beta(h_k,h_l))\ . \label{eq:defV}
\end{align}
$m_w$ and $m_k$ are the magnetizations (or overlaps) between the weights and the hidden variables and between the $k^\mathrm{th}$ layer and the labels; the $Q$s are the self-overlaps (or scalar products) between the different layers; and, writing $\cov_\beta$ for the covariance under the density $e^{-\beta H}$, the $V$s are the covariances between different trainings on the same data, after rescaling by $\beta$.

The order parameters $\Theta$ and $\hat\Theta$ satisfy the property that they extremize the following free entropy $\phi$:
\begin{widetext}
\begin{align}
\phi &= \frac{1}{2}\left(\hat V_wV_w+\hat V_wQ_w-V_w\hat Q_w\right)-\hat m_wm_w+\frac{1}{2}\mathrm{tr}\left(\hat VV+\hat VQ-V\hat Q\right)-\hat m^Tm \\
 &\quad{}+\frac{1}{\alpha}\mathbb E_{u,\varsigma}\left(\log\int\mathrm dw\,e^{\psi_w(w)}\right)+\rho\mathbb E_{y,\xi,\zeta,\chi}\left(\log\int\prod_{k=0}^K\mathrm dh_k e^{\psi_h(h;s)}\right)+(1-\rho)\mathbb E_{y,\xi,\zeta,\chi}\left(\log\int\prod_{k=0}^K\mathrm dh_k e^{\psi_h(h;s')}\right)\ ,  \nonumber
\end{align}
the potentials being
\begin{align}
\psi_w(w) &= -r\gamma(w)-\frac{1}{2}\hat V_ww^2+\left(\sqrt{\hat Q_w}\varsigma+u\hat m_w\right)w \\
\psi_h(h; \bar s) &= -\bar s\ell(yh_K)-\frac{1}{2}h_{<K}^T\hat Vh_{<K}+\left(\xi^T\hat Q^{1/2}+y\hat m^T\right)h_{<K} \label{eq:psiH} \\
&\qquad {}+\log\mathcal N\left(h_0\left|\sqrt\mu ym_w+\sqrt{Q_w}\zeta ; V_w \right.\right) +\log\mathcal N\left(h_{>0}\left|c\odot h_{<K}+\lambda ym+Q^{1/2}\chi ; V \right.\right)\ , \nonumber
\end{align}
\end{widetext}
for $w\in\mathbb R$ and $h\in\mathbb R^{K+1}$, where we introduced the Gaussian random variables $\varsigma\sim\mathcal N(0,1)$, $\xi\sim\mathcal N(0,I_K)$, $\zeta\sim\mathcal N(0,1)$ and $\chi\sim\mathcal N(0,I_K)$, take $y$ Rademacher and $u\sim\mathcal N(0,1)$, where we set $h_{>0}=(h_1,\ldots,h_K)^T$, $h_{<K}=(h_0,\ldots,h_{K-1})^T$ and $c\odot h_{<K}=(c_1h_0,\ldots,c_Kh_{K-1})^T$ and where $\bar s\in\{0,1\}$ controls whether the loss $\ell$ is active or not. We use the notation $\mathcal N(\cdot|m;V)$ for a Gaussian density of mean $m$ and variance $V$. We emphasize that $\psi_w$ and $\psi_h$ are effective potentials taking in account the randomness of the model and that are defined over a finite number of variables, contrary to the initial loss function $H$.

The extremality condition $\nabla_{\Theta,\hat\Theta}\,\phi=0$ can be stated in terms of a system of self-consistent equations that we give here. In the limit $\beta\to\infty$ one has to consider the extremizers of $\psi_w$ and $\psi_h$ defined as
\begin{align}
w^* &= \argmax_w\psi_w(w)\in\mathbb R \\
h^* &= \argmax_h\psi_h(h;\bar s=1)\in\mathbb R^{K+1} \\
h^{'*} &= \argmax_h\psi_h(h;\bar s=0)\in\mathbb R^{K+1}\ .
\end{align}
We also need to introduce $\cov_{\psi_h}(h)$ and $\cov_{\psi_h}(h')$ the covariances of $h$ under the densities $e^{\psi_h(h,\bar s=1)}$ and $e^{\psi_h(h,\bar s=0)}$. In the limit $\beta\to\infty$ they read
\begin{align}
\cov_{\psi_h}(h) &= \nabla\nabla\psi_h(h^*;\bar s=1) \\
\cov_{\psi_h}(h') &= \nabla\nabla\psi_h(h^{'*};\bar s=0)\ ,
\end{align}
$\nabla\nabla$ being the Hessian with respect to $h$. Last, for compactness we introduce the operator $\mathcal P$ that, for a function $g$ in $h$, acts according to
\begin{equation}
\mathcal P(g(h))=\rho g(h^*)+(1-\rho)g(h^{'*})\ .
\end{equation}
For instance $\mathcal P(hh^T)=\rho h^*(h^*)^T+(1-\rho)h^{'*}(h^{'*})^T$ and $\mathcal P(\cov_{\psi_h}(h))=\rho \cov_{\psi_h}(h)+(1-\rho) \cov_{\psi_h}(h')$. Then the extremality condition gives the following self-consistent, or fixed-point, equations on the order parameters:
{\small
\begin{align}
& m_w = \frac{1}{\alpha}\mathbb E_{u,\varsigma}\,uw^* \label{eq:pointFixeDébut} \\
& Q_w=\frac{1}{\alpha}\mathbb E_{u,\varsigma}(w^*)^2 \\
& V_w=\frac{1}{\alpha}\frac{1}{\sqrt{\hat Q_w}}\mathbb E_{u,\varsigma}\,\varsigma w^* \\
& m=\mathbb E_{y,\xi,\zeta,\chi}\,y\mathcal P(h_{<K}) \\
& Q=\mathbb E_{y,\xi,\zeta,\chi}\mathcal P(h_{<K}h_{<K}^T) \\
& V=\mathbb E_{y,\xi,\zeta,\chi}\mathcal P(\cov_{\psi_h}(h_{<K})) \\
& \hat m_w=\frac{\sqrt\mu}{V_w}\mathbb E_{y,\xi,\zeta,\chi}\,y\mathcal P(h_0-\sqrt\mu ym_w) \\
& \hat Q_w=\frac{1}{V_w^2}\mathbb E_{y,\xi,\zeta,\chi}\mathcal P\left((h_0-\sqrt\mu ym_w-\sqrt Q_w\zeta)^2\right) \\
& \hat V_w=\frac{1}{V_w}-\frac{1}{V_w^2}\mathbb E_{y,\xi,\zeta,\chi}\mathcal P(\cov_{\psi_h}(h_0)) \\
& \hat m=\lambda V^{-1}\mathbb E_{y,\xi,\zeta,\chi}\,y\mathcal P(h_{>0}-c\odot h_{<K}-\lambda ym) \\
& \hat Q=V^{-1}\mathbb E_{y,\xi,\zeta,\chi}\mathcal P\left((h_{>0}-c\odot h_{<K}-\lambda ym-Q^{1/2}\chi)^{\otimes 2}\right)V^{-1} \\
& \hat V=V^{-1}-V^{-1}\mathbb E_{y,\xi,\zeta,\chi}\mathcal P(\cov_{\psi_h}(h_{>0}-c\odot h_{<K}))V^{-1} \label{eq:pointFixeFin}
\end{align}
}

Once this system of equations is solved, the expected errors and accuracies can be expressed as
\begin{align}
& E_\mathrm{train} = \mathbb E_{y,\xi,\zeta,\chi} \ell(yh_K^*)\ , && \mathrm{Acc}_\mathrm{train} = \mathbb E_{y,\xi,\zeta,\chi} \delta_{y=\sign(h_K^*)} \\
& E_\mathrm{test} = \mathbb E_{y,\xi,\zeta,\chi} \ell(yh_K^{'*})\ , && \mathrm{Acc}_\mathrm{test} = \mathbb E_{y,\xi,\zeta,\chi} \delta_{y=\sign(h_K^{'*})}\ . \label{eq:pointFixeFinFin}
\end{align}

\begin{figure*}[t]
 \centering
 \includegraphics[width=0.9\linewidth]{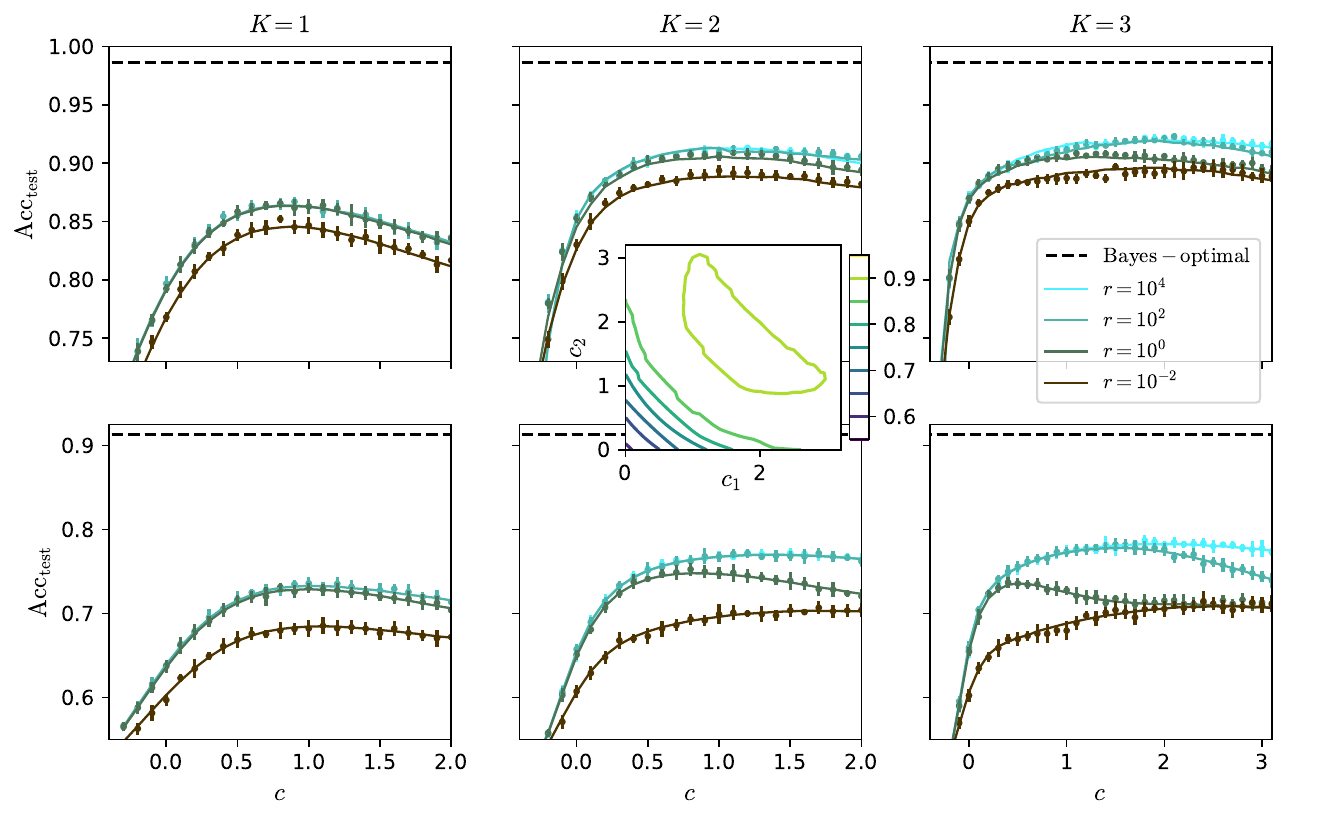}
 \caption{\label{fig:différentsKvsC} Predicted test accuracy $\mathrm{Acc}_\mathrm{test}$ for different values of $K$. \emph{Top:} for $\lambda=1.5$, $\mu=3$ and logistic loss; \emph{bottom:} for $\lambda=1$, $\mu=2$ and quadratic loss; $\alpha=4$ and $\rho=0.1$. We take $c_k=c$ for all $k$. \emph{Inset:} $\mathrm{Acc}_\mathrm{test}$ vs $c_1$ and $c_2$ at $K=2$ and at large $r$. Dots: numerical simulation of the GCN for $N=10^4$ and $d=30$, averaged over ten experiments.}
\end{figure*}

\subsubsection{Analytical solution}
\label{sec:discrAnalyticalSol}
In general the system of self-consistent equations~(\ref{eq:pointFixeDébut}-\ref{eq:pointFixeFin}) has to be solved numerically. The equations are applied iteratively, starting from arbitrary $\Theta$ and $\hat\Theta$, until convergence.

An analytical solution can be computed in some special cases. We consider ridge regression (i.e. quadratic $\ell$) and take $c=0$ no residual connections. Then $\cov_{\psi_h}(h)$, $\cov_{\psi_h'}(h)$, $V$ and $\hat V$ are diagonal. We obtain that
\begin{equation}
\mathrm{Acc}_\mathrm{test} = \frac{1}{2}\left(1+\mathrm{erf}\left(\frac{\lambda q_{y,K-1}}{\sqrt{2}}\right)\right)\ ,\quad q_{y,k}=\frac{m_k}{\sqrt{Q_{k,k}}}\ . \label{eq:accC0Rinf}
\end{equation}
The test accuracy only depends on the angle (or overlap) $q_{y,k}$ between the labels $y$ and the last hidden state $h_{K-1}$ of the GCN. $q_{y,k}$ can easily be computed in the limit $r\to\infty$. In appendix \ref{sec:accDiscrQuadC0rInf} we explicit the equations (\ref{eq:pointFixeDébut}-\ref{eq:pointFixeFinFin}) and give their solution in that limit. In particular we obtain for any $k$
{\small
\begin{align}
m_k &= \frac{\rho}{\alpha r}\left(\mu\lambda^{K+k}+\sum_{l=0}^k\lambda^{K-k+2l}\right) \label{eq:mC0Rinf} \\
Q_{k,k} &= \frac{\rho}{\alpha^2r^2}\left(\alpha\left(1+\rho\mu\lambda^{2K}+\rho\sum_{l=1}^{K}\lambda^{2l}\right)\right. \label{eq:qC0Rinf} \\
&\qquad\left. {}+\sum_{l=0}^{k}\left(1+\rho\sum_{l'=1}^{K-1-l}\lambda^{2l'}+\frac{\alpha^2r^2}{\rho}m_l^2\right) \right) \, .\nonumber
\end{align}
}

\subsubsection{Consequences: going to large $K$ is necessary}
\label{sec:discrConsequences}
We derive consequences from the previous theoretical predictions. We numerically solve eqs.~(\ref{eq:pointFixeDébut}-\ref{eq:pointFixeFin}) for some plausible values of the parameters of the data model. We keep balanced the signals from the graph, $\lambda^2$, and from the features, $\mu^2/\alpha$; we take $\rho=0.1$ to stick to the common case where few train nodes are available. We focus on searching the architecture that maximizes the test accuracy by varying the loss $\ell$, the regularization $r$, the residual connections $c_k$ and $K$. For simplicity we will mostly consider the case where $c_k=c$ for all $k$ and for a given $c$. We compare our theoretical predictions to simulations of the GCN for $N=10^4$ in fig.~\ref{fig:différentsKvsC}; as expected, the predictions are within the statistical errors. Details on the numerics are provided in appendix~\ref{sec:numériques}. We provide the code to run our predictions in the supplementary material.

\cite{dz24gcn} already studies in detail the effect of $\ell$, $r$ and $c$ at $K=1$. It reaches the conclusion that the optimal regularization is $r\to\infty$, that the choice of the loss $\ell$ has little effect and that there is an optimal $c=c^*$ of order one. According to fig.~\ref{fig:différentsKvsC}, it seems that these results can be extrapolated to $K>1$. We indeed observe that, for both the quadratic and the logistic loss, at $K\in\{1, 2, 3\}$, $r\to\infty$ seems optimal. Then the choice of the loss has little effect, because at large $r$ the output $h(w)$ of the network is small and only the behaviour of $\ell$ around 0 matters. Notice that, though $h(w)$ is small and the error $E_\mathrm{train/test}$ is trivially equal to $\ell(0)$, the sign of $h(w)$ is mostly correct and the accuracy $\mathrm{Acc}_\mathrm{train/test}$ is not trivial. Last, according to the inset of fig.~\ref{fig:différentsKvsC} for $K=2$, to take $c_1=c_2$ is optimal and our assumption $c_k=c$ for all $k$ is justified.

To take trainable $c_k$ would degrade the test performances. We show this in fig.~\ref{fig:différentsKvsC_Etrain} in appendix~\ref{sec:supplFig}, where optimizing the train error $E_\mathrm{train}$ over $c_k=c$ trivially leads to $c=+\infty$. Indeed, in this case the graph is discarded and the convolved features are proportional to the features $X$; which, if $\alpha\rho$ is small enough, are separable, and lead to a null train error. Consequently $c_k$ should be treated as a hyperparameter, tuned to maximize $\mathrm{Acc}_\mathrm{test}$, as we do in the rest of the article.

\begin{figure}[t]
 \centering
 \includegraphics[width=\linewidth]{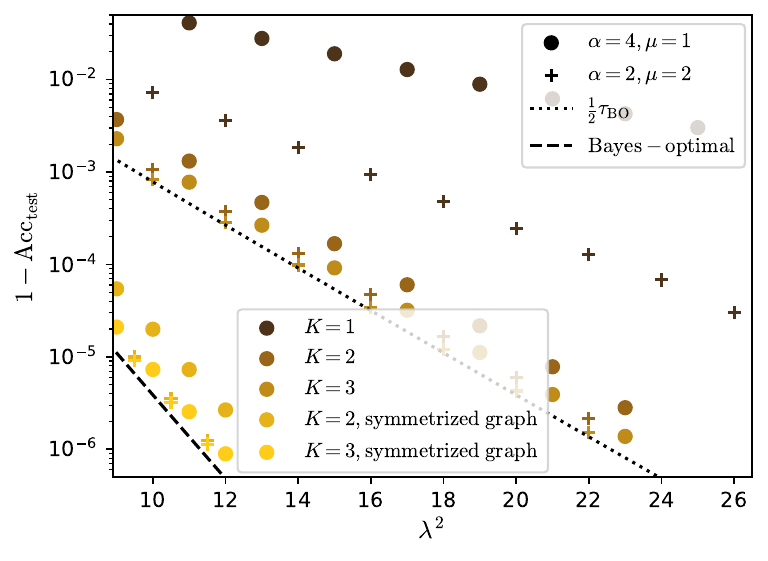}
 \caption{\label{fig:taux} Predicted misclassification error $1-\mathrm{Acc}_\mathrm{test}$ at large $\lambda$ for two strengths of the feature signal. $r=\infty$, $c=c^*$ is optimized by grid search and $\rho=0.1$. The dots are theoretical predictions given by numerically solving the self-consistent equations (\ref{eq:pointFixeDébut}-\ref{eq:pointFixeFin}) simplified in the limit $r\to\infty$. For the symmetrized graph the self-consistent equations are eqs.~(\ref{eq:pointFixeCont_Vqh}-\ref{eq:pointFixeCont_Qh}) in the next part.}
\end{figure}

\paragraph{Finite $K$:}
We focus on the effect of varying the number $K$ of aggregation steps. \cite{dz24gcn} shows that at $K=1$ there is a large gap between the Bayes-optimal test accuracy and the best test accuracy of the GCN. We find that, according to fig.~\ref{fig:différentsKvsC}, for $K\in\{1, 2, 3\}$, to increase $K$ reduces more and more the gap. Thus going to higher depth allows to approach the Bayes-optimality.

This also stands as to the learning rate when the signal $\lambda$ of the graph increases. At $\lambda\to\infty$ the GCN is consistent and correctly predicts the labels of all the test nodes, that is $\mathrm{Acc}_\mathrm{test} \underset{\lambda\to\infty}{\longrightarrow} 1$. The learning rate $\tau>0$ of the GCN is defined as
\begin{equation}
\log(1-\mathrm{Acc}_\mathrm{test}) \underset{\lambda\to\infty}{\sim} -\tau\lambda^2\ .
\end{equation}
As shown in \cite{dz23csbm}, the rate $\tau_\mathrm{BO}$ of the Bayes-optimal test accuracy is
\begin{equation}
\tau_\mathrm{BO}=1\ .
\end{equation}
For $K=1$ \cite{dz24gcn} proves that $\tau\le\tau_\mathrm{BO}/2$ and that $\tau\to\tau_\mathrm{BO}/2$ when the signal from the features $\mu^2/\alpha$ diverges. We obtain that if $K>1$ then $\tau=\tau_\mathrm{BO}/2$ for any signal from the features. This is shown on fig.~\ref{fig:taux}, where for $K=1$ the slope of the residual error varies with $\mu$ and $\alpha$ but does not reach half of the Bayes-optimal slope; while for $K>1$ it does, and the features only contribute with a sub-leading order.

Analytically, taking the limit in eqs.~\eqref{eq:mC0Rinf} and \eqref{eq:qC0Rinf}, at $c=0$ and $r\to\infty$ we have that
\begin{equation}
\underset{\lambda\to\infty}{\mathrm{lim}} q_{y,K-1} \left \{
\begin{array}{rr}
= 1 & \mathrm{if}\,K>1 \\
< 1 & \mathrm{if}\,K=1
\end{array}
\right .
\end{equation}
Since $\log(1-\erf(\lambda q_{y,K-1}/\sqrt{2}))\underset{\lambda\to\infty}{\sim}-\lambda^2q_{y,K-1}^2/2$ we recover the leading behaviour depicted on fig.~\ref{fig:taux}. $c$ has little effect on the rate $\tau$; it only seems to vary the test accuracy by a sub-leading term.

\paragraph{Symmetrization:}
\label{sec:symmetrization}
We found that in order to reach the Bayes-optimal rate one has to further symmetrize the graph, according to eq.~\eqref{eq:symétrisation}, and to perform the convolution steps by applying $\tilde A^\mathrm{s}$ instead of $\tilde A$. Then, as shown on fig.~\ref{fig:taux}, the GCN reaches the BO rate for any $K>1$, at any signal from the features.

The reason of this improvement is the following. The GCN we consider is not able to deal with the asymmetry of the graph and the supplementary information it gives. This is shown on fig.~\ref{fig:symVsAsym} in appendix \ref{sec:supplFig} for different values of $K$ at finite signal, in agreement with \cite{shi2022statistical}. There is little difference in the performance of the simple GCN whether the graph is symmetric or not with same $\lambda$. As to the rates, as shown by the computation in appendix~\ref{sec:eqSE}, a symmetric graph with signal $\lambda$ would lead to a BO rate $\tau_\mathrm{BO}^\mathrm{s}=1/2$, which is the rate the GCN achieves on the asymmetric graph. It is thus better to let the GCN process the symmetrized the graph, which has a higher signal $\lambda^\mathrm{s}=\sqrt 2\lambda$, and which leads to $\tau=1=\tau_\mathrm{BO}$.

Symmetrization is an important step toward the optimality and we will detail the analysis of the GCN on the symmetrized graph in part \ref{sec:continuousGCN}.

\begin{figure}[h]
 \centering
 \includegraphics[width=\linewidth]{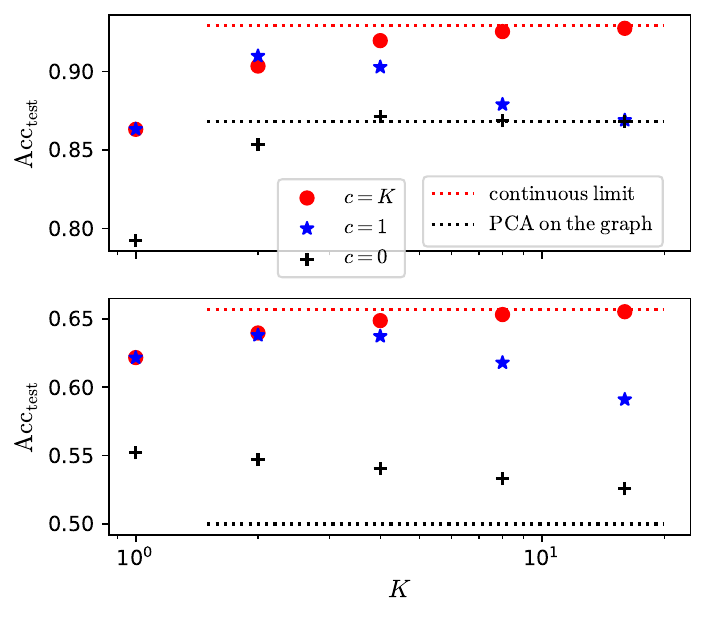}
 \caption{\label{fig:vsKetC} Predicted test accuracy $\mathrm{Acc}_\mathrm{test}$ vs $K$ for different scalings of $c$, at $r=\infty$. \emph{Top:} for $\lambda=1.5$, $\mu=3$; \emph{bottom:} for $\lambda=0.7$, $\mu=1$; $\alpha=4$, $\rho=0.1$. The predictions are given either by the explicit expression eqs.~(\ref{eq:accC0Rinf}-\ref{eq:qC0Rinf}) for $c=0$, either by solving the self-consistent equations (\ref{eq:pointFixeDébut}-\ref{eq:pointFixeFin}) simplified in the limit $r\to\infty$. The performance for the continuous limit are derived and given in the next section \ref{sec:continuousGCN}, while the performance of PCA on the graph are given by eqs.~(\ref{eq:qPCA}-\ref{eq:accPCA}).}
\end{figure}

\paragraph{Large $K$ and scaling of $c$:}
Going to larger $K$ is beneficial and allows the network to approach the Bayes optimality. Yet $K=3$ is not enough to reach it at finite~$\lambda$, and one can ask what happens at larger $K$. An important point is that $c$ has to be well tuned. On fig.~\ref{fig:différentsKvsC} we observe that $c^*$, the optimal $c$, is increasing with $K$. To make this point more precise, on fig.~\ref{fig:vsKetC} we show the predicted test accuracy for larger $K$ for different scalings of $c$. We take $r=\infty$ since it appears to be the optimal regularization. We consider no residual connections, $c=0$; constant residual connections, $c=1$; or growing residual connections, $c\propto K$.

A main observation is that, on fig.~\ref{fig:vsKetC} for $K\to\infty$, $c=0$ or $c=1$ converge to the same limit while $c\propto K$ converge to a different limit, that has higher accuracy.

In the case where $c=0$ or $c=1$ the GCN oversmooths at large $K$. The limit it converges to corresponds to the accuracy of principal component analysis (PCA) on the sole graph; that is, it corresponds to the accuracy of the estimator $\hat y_\mathrm{PCA}=\sign\left(\mathrm{Re}(y_1)\right)$ where $y_1$ is the leading eigenvector of $\tilde A$. The overlap $q_\mathrm{PCA}$ between $y$ and $\hat y_\mathrm{PCA}$ and the accuracy are
\begin{align}
& q_\mathrm{PCA} =\left \{
\begin{array}{cr}
\sqrt{1-\lambda^{-2}} & \mathrm{if}\,\lambda\ge 1 \\
0 & \mathrm{if}\,\lambda\le 1
\end{array} \right. \ , \label{eq:qPCA}\\
& \mathrm{Acc}_\mathrm{test,PCA} = \frac{1}{2}\left(1+\mathrm{erf}\left(\frac{\lambda q_\mathrm{PCA}}{\sqrt{2}}\right)\right)\ . \label{eq:accPCA}
\end{align}
Consequently, if $c$ does not grow, the GCN will oversmooth at large $K$, in the sense that all the information from the features $X$ vanishes. Only the information from the graph remains, that can still be informative if $\lambda>1$. The formula (\ref{eq:qPCA}--\ref{eq:accPCA}) is obtained by taking the limit $K\to\infty$ in eqs.~(\ref{eq:accC0Rinf}--\ref{eq:qC0Rinf}), for $c=0$. For any constant $c$ it can also be recovered by considering the leading eigenvector $y_1$ of $\tilde A$. At large $K$, $(\tilde A/\sqrt N+cI)^K$ is dominated by $y_1$ and the output of the GCN is $h(w)\propto y_1$ for any $w$. Consequently the GCN exactly acts like thresholded PCA on $\tilde A$. The sharp transition at $\lambda=1$ corresponds to the BBP phase transition in the spectrum of $A^g$ and $\tilde A$ \cite{baik05transition}. According to eqs.~(\ref{eq:accC0Rinf}--\ref{eq:qC0Rinf}) the convergence of $q_{y,K-1}$ toward $q_\mathrm{PCA}$ is exponentially fast in $K$ if $\lambda>1$; it is like $1/\sqrt K$, much slower, if $\lambda<1$.

The fact that the oversmoothed features can be informative differs from several previous works where they are fully non-informative, such as \cite{li18oversmoothing,oono20oversmoothing,keriven22oversm}. This is mainly due to the normalization $\tilde A$ of $A$ we use and that these works do not use. It allows to remove the uniform eigenvector $(1, \ldots, 1)^T$, that otherwise dominates $A$ and leads to non-informative features. \cite{wang24GCNmultiConv} emphasizes on this point and compares different ways of normalizing and correcting $A$. This work concludes, as we do, that for a correct rescaling $\tilde A$ of $A$, similar to ours, going to higher $K$ is always beneficial if $\lambda$ is high enough, and that the convergence to the limit is exponentially fast. Yet, at large $K$ it obtains bounds on the test accuracy that do not depend on the features: the network they consider still oversmooths in the precise sense we defined. This can be expected since it does not have residual connections, i.e. $c=0$, that appear to be decisive.

In the case where $c\propto K$ the GCN does not oversmooth and it converges to a continuous limit, obtained as $(cI+\tilde A/\sqrt N)^K\propto (I+t\tilde A/K\sqrt N)^K\to e^{\frac{t\tilde A}{\sqrt N}}$. We study this limit in detail in the next part where we predict the resulting accuracy for all constant ratios $t=c/K$. In general the continuous limit has a better performance than the limit at constant $c$ that relies only on the graph, performing PCA, because it can take in account the features, which bring additional information.

Fig.~\ref{fig:vsKetC} suggests that $\mathrm{Acc}_\mathrm{test}$ is monotonically increasing with $K$ if $c\propto K$ and that the continuous limit is the upper-bound on the performance at any $K$. We will make this point more precise in the next part. Yet we can already see that, for this to be true, one has to correctly tune the ratio $c/K$: for instance if $\lambda$ is small $\tilde A$ mostly contains noise and applying it to $X$ will mostly lower the accuracy. Shortly, if $c/K$ is optimized then $K\to\infty$ is better than any fixed $K$. Consequently the continuous limit is the correct limit to maximize the test accuracy and it is of particular relevance.
\vspace{-2mm}

\subsection{Continuous GCN}
\label{sec:continuousGCN}
\vspace{-2mm}
In this section we present the asymptotic characterization of the continuous GCN, both for the asymmetric graph and for its symmetrization. The continuous GCN is the limit of the discrete GCN when the number of convolution steps $K$ diverges while the residual connections $c$ become large. The order parameters that describe it, as well as the self-consistent equations they follow, can be obtained as the limit of those of the discrete GCN. We give a detailed derivation of how the limit is taken, since it is of independent interest.

The outcome is that the state $h$ of the GCN across the convolutions is described by a set of equations resembling the dynamical mean-field theory. The order parameters of the problem are continuous functions and the self-consistent equations can be expressed by expansion around large regularization $r\to\infty$ as integral equations, that specialize to differential equations in the asymmetric case. The resulting equations can be solved analytically; for asymmetric graphs, the covariance and its conjugate are propagators (or resolvant) of the two-dimensional Klein-Gordon equation. We show numerically that our approach is justified and agrees with simulations. Last we show that going to the continuous limit while symmetrizing the graph corresponds to the optimum of the architecture and allows to approach the Bayes-optimality.

\subsubsection{Asymptotic characterization}
\label{sec:contAsymptChara}
To deal with both cases, asymmetric or symmetrized, we define $(\delta_\mathrm{e}, \tilde A^\mathrm{e}, \lambda^\mathrm{e})\in\{(0, \tilde A, \lambda), (1, \tilde A^\mathrm{s}, \lambda^\mathrm{s})\}$, where we remind that $\tilde A^\mathrm{s}$ is the symmetrized $\tilde A$ with effective signal $\lambda^\mathrm{s}=\sqrt 2\lambda$. In particular $\delta_\mathrm{e}=0$ for the asymmetric and $\delta_\mathrm{e}=1$ for the symmetrized.

The continuous GCN is defined by the output function
\begin{equation}
h(w)=e^{\frac{t}{\sqrt N}\tilde A^\mathrm{e}}\frac{1}{\sqrt N}Xw\ .
\end{equation}
We first derive the free entropy of the discretization of the GCN and then take the continuous limit. The discretization at finite $K$ is
\begin{align}
& h(w) = h_K\ ,\\
& h_{k+1} = \left(I_N+\frac{t}{K\sqrt N}\tilde A^\mathrm{e}\right)h_k\ ,\\
& h_0 = \frac{1}{\sqrt N}Xw\ .
\end{align}
In the case of the asymmetric graph this discretization can be mapped to the discrete GCN of the previous section \ref{sec:appendixDiscreteGCN} as detailed in eq.~\eqref{eq:GCNcontinuBis} and the following paragraph; the free entropy and the order parameters of the two models are the same, up to a rescaling by $c$.

The order parameters of the discretization of the GCN are $m_w\in\mathbb R, Q_w\in\mathbb R, V_w\in\mathbb R, m\in\mathbb R^K, Q_h\in\mathbb R^{K\times K}, V_h\in\mathbb R^{K\times K}$, their conjugates $\hat m_w\in\mathbb R, \hat Q_w\in\mathbb R, \hat V_w\in\mathbb R, \hat m\in\mathbb R, \hat Q_h\in\mathbb R^{K\times K}, \hat V_h\in\mathbb R^{K\times K}$ and the two additional order parameters $Q_{qh}\in\mathbb R^{K\times K}$ and $V_{qh}\in\mathbb R^{K\times K}$ that account for the supplementary correlations the symmetry of the graph induces; $Q_{qh}=V_{qh}=0$ for the asymmetric case.

The free entropy and its derivation are given in appendix \ref{sec:appendixContinuousGCN}. The outcome is that $h$ is described by the effective low-dimensional potential $\psi_h$ over $\mathbb R^{K+1}$ that is
\begin{align}
\psi_h(h;\bar s) &= -\frac{1}{2}h^TGh+h^T\left(B_h+D_{qh}^TG_0^{-1}B\right) \ ; \label{eq:psiHCont}
\end{align}
where
\begin{align}
G &= G_h+D_{qh}^TG_0^{-1}D_{qh}\ , \label{eq:Gcont}\\
G_h &=\left(\begin{smallmatrix}\hat V_h & 0 \\ 0 & \bar s\end{smallmatrix}\right)\ ,\\
G_0 &=\left(\begin{smallmatrix}K^2V_w & 0 \\ 0 & t^2V_h\end{smallmatrix}\right)\ ,\\
D_{qh} &=D-t\left(\begin{smallmatrix} 0 & 0\\ -\mathrm i\delta_\mathrm{e}V_{qh}^T & 0 \end{smallmatrix}\right)
\end{align}
are $(K+1)\times(K+1)$ block matrices;
\begin{align}
D &= K\left(\begin{smallmatrix}1 & & & 0 \\ -1 & \ddots & & \\ & \ddots & \ddots & \\ 0 & & -1 & 1 \end{smallmatrix}\right)
\end{align}
is the $(K+1)\times(K+1)$ discrete derivative;
\begin{align}
B &= \left(\begin{smallmatrix}K\sqrt{Q_w}\chi \\ \mathrm it\left(\hat Q^{1/2}\zeta\right)_q\end{smallmatrix}\right) + y\left(\begin{smallmatrix}K\sqrt\mu m_w \\ \lambda^\mathrm{e}tm\end{smallmatrix}\right)\ ,\\
B_h &= \left(\begin{smallmatrix}\left(\hat Q^{1/2}\zeta\right)_h \\ 0\end{smallmatrix}\right) + y\left(\begin{smallmatrix}\hat m \\ \bar s\end{smallmatrix}\right)\ ,\\
\left(\begin{smallmatrix}(\hat Q^{1/2}\zeta)_q\\ (\hat Q^{1/2}\zeta)_h\end{smallmatrix}\right) &= \left(\begin{smallmatrix}-Q_h & -\delta_\mathrm{e}Q_{qh}^T \\ -\delta_\mathrm{e}Q_{qh} & \hat Q_h\end{smallmatrix}\right)^{1/2}\left(\begin{smallmatrix}\zeta_q\\ \zeta_h\end{smallmatrix}\right)
\end{align}
are vectors of size $K+1$, where $y=\pm 1$ is Rademacher and $\zeta_q\sim\mathcal N(0,I_{K+1})$, $\zeta_h\sim\mathcal N(0,I_{K+1})$ and $\chi\sim\mathcal N(0,1)$ are standard Gaussians. $\bar s$ determines whether the loss is active $\bar s=1$ or not $\bar s=0$. We assumed that $\ell$ is quadratic. Later we will take the limit $r\to\infty$ where $h$ is small and where $\ell$ can effectively be expanded around $0$ as a quadratic potential. Notice that in the case $\delta_\mathrm{e}=0$ we recover the potential $\psi_h$ eq.~\eqref{eq:psiH} of the previous part.

This potential eq.~\eqref{eq:psiHCont} corresponds to a one dimensional interacting chain, involving the positions $h$ and their effective derivative $D_{qh}h$, and with constraints at the two ends, for the loss on $h_K$ and the regularized weights on $h_0$. Its extremizer $h^*$ is
\begin{equation}
h^*=G^{-1}\left(B_h+D_{qh}^TG_0^{-1}B\right)\ .
\end{equation}

The order parameters are determined by the following fixed-point equations, obtained by extremizing the free entropy. As before $\mathcal P$ acts by linearly combining quantities evaluated at $h^*$, taken with $\bar s=1$ and $\bar s=0$ with weights $\rho$ and $1-\rho$.
{\small
\begin{align}
& m_w = \frac{1}{\alpha}\frac{\hat m_w}{r+\hat V_w} \label{eq:pointFixeC_début} \\
& Q_w = \frac{1}{\alpha}\frac{\hat Q_w+\hat m_w^2}{(r+\hat V_w)^2} \\
& V_w = \frac{1}{\alpha}\frac{1}{r+\hat V_w} \\
& \left(\begin{smallmatrix}\hat m_w \\ \hat m \\ m \\ \cdot \end{smallmatrix}\right) = \left(\begin{smallmatrix}K\sqrt\mu & & 0 \\ & \lambda^\mathrm{e}tI_K & \\ 0 & & I_{K+1}\end{smallmatrix}\right)\mathbb E_{y,\xi,\zeta}\,y\mathcal P\left(\begin{smallmatrix}G_0^{-1}(D_{qh}h-B) \\ h\end{smallmatrix}\right) \\
& \begin{multlined}[t] \left(\begin{smallmatrix}\hat Q_w & & & \cdot \\ & \hat Q_h & Q_{qh} & \\ & Q_{qh}^T & Q_h & \\ \cdot & & & \cdot\end{smallmatrix}\right) = \left(\begin{smallmatrix}K & & 0 \\ & tI_K & \\ 0 & & I_{K+1}\end{smallmatrix}\right) \\
\qquad\mathbb E_{y,\xi,\zeta}\mathcal P\left(\left(\begin{smallmatrix}G_0^{-1}(D_{qh}h-B) \\ h\end{smallmatrix}\right)^{\otimes 2}\right) \left(\begin{smallmatrix}K & & 0 \\ & tI_K & \\ 0 & & I_{K+1}\end{smallmatrix}\right) \end{multlined} \\
& \left(\begin{smallmatrix} \cdot & \cdot \\ -\mathrm iV_{qh} & \cdot \end{smallmatrix}\right) = t\mathcal P\left(G_0^{-1}D_{qh}G^{-1}\right) \label{eq:pointFixeC_débutV}\\
& \left(\begin{smallmatrix}V_h & \cdot \\ \cdot & \cdot\end{smallmatrix}\right) = \mathcal P\left(G^{-1}\right) \\
& \left(\begin{smallmatrix}\hat V_w & \cdot \\ \cdot & \hat V_h \end{smallmatrix}\right) = \left(\begin{smallmatrix}K^2 & 0 \\ 0 & t^2I_K \end{smallmatrix}\right) \mathcal P\left(G_0^{-1} -G_0^{-1}D_{qh}G^{-1}D_{qh}^TG_0^{-1}\right) \label{eq:pointFixeC_fin}
\end{align}
}
where $\cdot$ are unspecified elements that pad the vector to the size $2(K+1)$ and the matrices to the size $2(K+1)\times 2(K+1)$ and $(K+1)\times(K+1)$. On $w$ we assumed $l_2$ regularization and obtained the same equations as in part \ref{sec:discreteGCN}.

Once a solution to this system is found the train and test accuracies are expressed as
\begin{align}
\mathrm{Acc}_\mathrm{train/test} = \mathbb E_{y,\zeta,\chi} \delta_{y=\sign(h_K^*)}\ ,
\end{align}
taking $\bar s=1$ or $\bar s=0$.

\subsubsection{Expansion around large regularization $r$ and continuous limit}
\label{sec:développementRlimiteCont}
Solving the above self-consistent equations (\ref{eq:pointFixeC_début}-\ref{eq:pointFixeC_fin}) is difficult as such. One can solve them numerically by repeated updates; but this does not allow to go to large $K$ because of numerical instability. One has to invert $G$ eq.~\eqref{eq:Gcont} and to make sense of the continuous limit of matrix inverts. This is an issue in the sense that, for a generic $K\times K$ matrix $(M)_{ij}$ whose elements vary smoothly with $i$ and $j$ in the limit of large $K$, the elements of its inverse $M^{-1}$ are not necessarly continuous with respect to their indices and can vary with a large magnitude.

Our analysis from the previous part \ref{sec:discreteGCN} gives an insight on how to achieve this. It appears that the limit of large regularization $r\to\infty$ is of particular relevance. In this limit the above system can be solved analytically thanks to an expansion around large $r$. This expansion is natural in the sense that it leads to several simplifications and corresponds to expanding the matrix inverts in Neumann series. Keeping the first terms of the expansion the limit $K\to\infty$ is then well defined. In this section we detail this expansion; we take the continuous limit and, keeping the first constant order, we solve (\ref{eq:pointFixeC_début}-\ref{eq:pointFixeC_fin}).

In the limit of large regularization $h$ and $w$ are of order $1/r$; the parameters $m_w$, $m$, $V_w$ and $V$ are of order $1/r$ and $Q_w$ and $Q$ are of order $1/r^2$, while all their conjugates, $Q_{qh}$ and $V_{qh}$ are of order one. Consequently we have $G_0^{-1}\sim r\gg G_h\sim 1$ and we expand $G^{-1}$ around $G_0$:
\begin{align}
G^{-1} &= D_{qh}^{-1}G_0D_{qh}^{-1,T}\sum_{a\ge 0}\left(-G_hD_{qh}^{-1}G_0D_{qh}^{-1,T}\right)^a\ .
\end{align}

\paragraph{Constant order:}
We detail how to solve the self-consistent equations~(\ref{eq:pointFixeC_début}-\ref{eq:pointFixeC_fin}) taking the continuous limit $K\to\infty$ at the constant order in $1/r$. As we will show later, truncating $G^{-1}$ to the constant order gives predictions that are close to the simulations at finite $r$, even for $r\approx 1$ if $t$ is not too large. Considering higher orders is feasible but more challenging and we will only provide insights on how to pursue the computation.

The truncated expansion gives, starting from the variances:
\begin{align}
\left(\begin{smallmatrix} \cdot & \cdot \\ -\mathrm iV_{qh} & \cdot \end{smallmatrix}\right) &= tD_{qh}^{-1,T}\ , \label{eq:pointFixeCont_Vqh} \\
\left(\begin{smallmatrix}V_h & \cdot \\ \cdot & \cdot \end{smallmatrix}\right) &= D_{qh}^{-1}\left(\begin{smallmatrix}K^2V_w & 0 \\ 0 & t^2V_h\end{smallmatrix}\right)D_{qh}^{-1,T} ,\label{eq:pointFixeCont_Vh} \\
\left(\begin{smallmatrix}\hat V_w & \cdot \\ \cdot & \hat V_h \end{smallmatrix}\right) &= \left(\begin{smallmatrix}K^2 & 0 \\ 0 & t^2I_K \end{smallmatrix}\right) D_{qh}^{-1,T}\left(\begin{smallmatrix}\hat V_h & 0 \\ 0 & \rho\end{smallmatrix}\right)D_{qh}^{-1}\ .\label{eq:pointFixeCont_VCh}
\end{align}
We kept the order $a=0$ for $V_{qh}$ and $V_h$, and the orders $a\le 1$ for $\hat V_w$ and $\hat V_h$. We expand $h^* \approx D_{qh}^{-1}G_0D_{qh}^{-1,T}B_h+D_{qh}^{-1}B$ keeping the order $a=0$ and obtain the remaining self-consistent equations
\begin{align}
\left(\begin{smallmatrix}\hat m_w \\ \hat m \end{smallmatrix}\right) &= \left(\begin{smallmatrix}K\sqrt\mu & 0 \\ 0 & \lambda^\mathrm{e}t I_K \cdot\end{smallmatrix}\right)D_{qh}^{-1,T}\left(\begin{smallmatrix}\hat m \\ \rho \end{smallmatrix}\right) \label{eq:pointFixeCont_mC} \\
\left(\begin{smallmatrix}m \\ \cdot \end{smallmatrix}\right) &= D_{qh}^{-1}G_0D_{qh}^{-1,T} \left(\begin{smallmatrix}\hat m \\ \rho \end{smallmatrix}\right) + D_{qh}^{-1}\left(\begin{smallmatrix}K\sqrt\mu m_w \\ \lambda^\mathrm{e}tm \end{smallmatrix}\right)
\end{align}
\begin{widetext}
{\small
\begin{align}
\left(\begin{smallmatrix}\hat Q_w & \cdot \\ \cdot & \hat Q_h \end{smallmatrix}\right) &= \left(\begin{smallmatrix} K & 0 \\ 0 & tI_K \end{smallmatrix}\right)D_{qh}^{-1,T}\left(\left(\begin{smallmatrix} \hat Q_h & 0 \\ 0 & 0 \end{smallmatrix}\right) + \rho\left(\begin{smallmatrix}\hat m\\ 1 \end{smallmatrix}\right)^{\otimes 2}+(1-\rho)\left(\begin{smallmatrix}\hat m\\ 0 \end{smallmatrix}\right)^{\otimes 2}\right)D_{qh}^{-1}\left(\begin{smallmatrix} K & 0 \\ 0 & tI_K \end{smallmatrix}\right) \\
\left(\begin{smallmatrix}\cdot & \cdot \\ -\mathrm iQ_{qh} & \cdot \end{smallmatrix}\right) &= tD_{qh}^{-1,T}\left[ \left( t\delta_\mathrm{e}\left(\begin{smallmatrix}0 & -\mathrm iQ_{qh} \\ 0 & 0 \end{smallmatrix}\right) + \left(\begin{smallmatrix}\hat m \\ \rho \end{smallmatrix}\right)\left(\begin{smallmatrix}K\sqrt\mu m_w \\ \lambda^\mathrm{e}tm \end{smallmatrix}\right)^T\right)D_{qh}^{-1,T} + \left(\left(\begin{smallmatrix}\hat Q_h & 0 \\ 0 & 0 \end{smallmatrix}\right) + \rho\left(\begin{smallmatrix}\hat m\\ 1 \end{smallmatrix}\right)^{\otimes 2}+(1-\rho)\left(\begin{smallmatrix}\hat m\\ 0 \end{smallmatrix}\right)^{\otimes 2}\right)D_{qh}^{-1}G_0D_{qh}^{-1,T} \right] \\
\left(\begin{smallmatrix}Q_h & \cdot \\ \cdot & \cdot \end{smallmatrix}\right) &= D_{qh}^{-1}G_0D_{qh}^{-1,T}\left(\left(\begin{smallmatrix}\hat Q_h & 0 \\ 0 & 0 \end{smallmatrix}\right) + \rho\left(\begin{smallmatrix}\hat m\\ 1 \end{smallmatrix}\right)^{\otimes 2}+(1-\rho)\left(\begin{smallmatrix}\hat m\\ 0 \end{smallmatrix}\right)^{\otimes 2}\right)D_{qh}^{-1}G_0D_{qh}^{-1,T}
+ D_{qh}^{-1}\left(\left(\begin{smallmatrix}K^2Q_w & 0 \\ 0 & t^2Q_h \end{smallmatrix}\right)+\left(\begin{smallmatrix}K\sqrt\mu m_w \\ \lambda^\mathrm{e}tm \end{smallmatrix}\right)^{\otimes 2}\right)D_{qh}^{-1,T} \nonumber \\
&{}+D_{qh}^{-1}G_0D_{qh}^{-1,T}\left(t\delta_\mathrm{e}\left(\begin{smallmatrix}0 & -\mathrm iQ_{qh} \\ 0 & 0 \end{smallmatrix}\right) + \left(\begin{smallmatrix}\hat m \\ \rho \end{smallmatrix}\right)\left(\begin{smallmatrix}K\sqrt\mu m_w \\ \lambda^\mathrm{e}tm \end{smallmatrix}\right)^T\right)D_{qh}^{-1,T} + D_{qh}^{-1}\left(t\delta_\mathrm{e}\left(\begin{smallmatrix}0 & 0 \\ -\mathrm iQ_{qh}^T & 0 \end{smallmatrix}\right) + \left(\begin{smallmatrix}K\sqrt\mu m_w \\ \lambda^\mathrm{e}tm \end{smallmatrix}\right)\left(\begin{smallmatrix}\hat m \\ \rho \end{smallmatrix}\right)^T\right)D_{qh}^{-1}G_0D_{qh}^{-1,T} \label{eq:pointFixeCont_Qh}
\end{align}
}
\end{widetext}
We see that all these self-consistent equations (\ref{eq:pointFixeCont_Vh}-\ref{eq:pointFixeCont_Qh}) are vectorial or matricial equations of the form $x=\lambda^\mathrm{e}tD_{qh}^{-1}x$ or $X=t^2D_{qh}^{-1}XD_{qh}^{-1,T}$, over $x$ or $X$, plus inhomogenuous terms and boundary conditions at 0 or $(0,0)$. The equations are recursive in the sense that each equation only depends on the previous ones and they can be solved one by one. It is thus enough to compute the resolvants of these two equations. Last eq.~\eqref{eq:pointFixeCont_Vqh} shows how to invert $D_{qh}$ and express $D_{qh}^{-1}$. These different properties make the system of self-consistent equations easily solvable, provided one can compute $D_{qh}$ and the two resolvants. This furthermore highlights the relevance of the $r\to\infty$ limit.

We take the continuous limit $K\to\infty$. We translate the above self-consistent equations into functional equations. Thanks to the expansion around large $r$ we have a well defined limit, that does not involve any matrix inverse. We set $x=k/K$ and $z=l/K$ continuous indices ranging from 0 to 1. We extend the vectors and the matrices by continuity to match the correct dimensions. We apply the following rescaling to obtain quantities that are independent of $K$ in that limit:
\begin{align}
& \hat m\to K\hat m\ , && \hat Q_h\to K^2\hat Q_h\ , && \hat V_h\to K^2\hat V_h\ ,\\
& Q_{qh}\to KQ_{qh}\ , && V_{qh}\to KV_{qh}\ .
\end{align}

We first compute the effective derivative $D_{qh}=D-t\left(\begin{smallmatrix} 0 & 0\\ -\mathrm i\delta_\mathrm{e}V_{qh}^T & 0 \end{smallmatrix}\right)$ and its inverse. In the asymmetric case we have $D_{qh}=D$, the usual derivative. In the symmetric case we have $D_{qh}=D-tV_{qh}^T$ where $V_{qh}$ satisfies eq.~\eqref{eq:pointFixeCont_Vqh} which reads
\begin{align}
& \partial_zV_{qh}(x,z) + \delta(z)V_{qh}(x,z) = \\
&\qquad\qquad t\delta(z-x)+t\int_0^1\mathrm dx'\,V_{qh}(x,x')V_{qh}(x',z)\ , \nonumber
\end{align}
where we multiplied both sides by $D_{qh}^T$ and took $V_{qh}(x,z)$ for $-\mathrm iV_{qh}$. The solution to this integro-differential equation is
\begin{align}
V_{qh}(x,z) &= \theta(z-x)\frac{I_1(2t(z-x))}{z-x}
\end{align}
with $\theta$ the step function and $I_\nu$ the modified Bessel function of the second kind of order $\nu$. Consequently we obtain the effective inverse derivative
\begin{align}
D_{qh}^{-1}(x,z) &= D_{qh}^{-1,T}(z,x) = \left \{\begin{array}{cc}
\theta(x-z) & \mathrm{if}\,\delta_\mathrm{e}=0 \\
\frac{1}{t}V_{qh}(z,x) & \mathrm{if}\,\delta_\mathrm{e}=1
\end{array}\right.\ . \label{eq:DqhCont}
\end{align}
\vspace{-4mm}

We then define the resolvants (or propagators) $\varphi$ and $\Phi$ of the integral equations as
\begin{align}
D_{qh}\varphi(x) &= \lambda^\mathrm{e}t\varphi(x)+\delta(x)\ ,\\
D_{qh}\Phi(x,z)D_{qh}^T &= t^2\Phi(x,z)+\delta(x,z)\ .\label{eq:resPhi}
\end{align}
Notice that in the asymmetric case, $D_{qh}=\partial_x$, $D_{qh}^T=\partial_z$ and $\Phi$ is the propagator of the two-dimensional Klein-Gordon equation up to a change of variables. The resolvants can be expressed as
\begin{align}
\varphi(x) &= \left \{\begin{array}{cc}
e^{\lambda^\mathrm{e} tx} & \mathrm{if}\,\delta_\mathrm{e}=0 \\
\sum_{\nu>0}^\infty\nu(\lambda^\mathrm{e})^{\nu-1}\frac{I_\nu(2tx)}{tx} & \mathrm{if}\,\delta_\mathrm{e}=1
\end{array}\right .\ ,\\
\Phi(x,z) &= \left \{\begin{array}{cc}
I_0(2t\sqrt{xz}) & \mathrm{if}\,\delta_\mathrm{e}=0 \\
\frac{I_1(2t(x+z))}{t(x+z)} & \mathrm{if}\,\delta_\mathrm{e}=1
\end{array}\right .\ .
\end{align}

We obtain the solution of the self-consistent equations by convolving $\varphi$ or $\Phi$ with the non-homogenuous terms. We flip $\hat m$ along it axis to match the vectorial equation with boundary condition at $x=0$; we do the same for $\hat V_h$ and $\hat Q_h$ along there two axes, and for $Q_{qh}$ along its first axis. This gives the following expressions for the order parameters:
{\small
\begin{align}
& V_w = \frac{1}{r\alpha} \\
& V_h(x,z) = V_w\Phi(x,z) \\
& \hat V_h(1-x,1-z) = t^2\rho\Phi(x,z) \\
& \hat V_w = t^{-2}\hat V_h(0,0) \\
& \hat m(1-x) = \rho\lambda^\mathrm{e}t\varphi(x) \\
& \hat m_w = \sqrt\mu\frac{1}{\lambda^\mathrm{e}t}\hat m(0) \\
& m_w = \frac{\hat m_w}{r\alpha} \\
& m(x) = (1+\mu)\frac{m_w}{\sqrt\mu}\varphi(x) \\
& \quad {}+\frac{t}{\lambda^\mathrm{e}}\int_0^x\mathrm dx'\int_0^1\mathrm dx''\,\varphi(x-x')V_h(x',x'')\hat m(x'') \nonumber\\
& \hat Q_w = t^{-2}\hat Q_h(0,0) \\
& Q_w = \frac{\hat Q_w+\hat m_w^2}{r^2\alpha}
\end{align}
}
\begin{widetext}
{\small
\begin{align}
& \hat Q_h(1-x, 1-z) = t^2\int_{0^-,0^-}^{x,z}\mathrm dx'\mathrm dz'\ \Phi(x-x',z-z')\left[\mathcal P(\hat m^{\otimes 2})(1-x',1-z')\right] \\
& Q_{qh}(1-x,z) = t\int_{0^-,0^-}^{x,z}\mathrm dx'\mathrm dz'\ \Phi(x-x',z-z')\Bigg{[}\mathcal P(\hat m)(1-x')(\lambda^\mathrm{e}tm(z')+\sqrt\mu m_w\delta(z')) \\
&\qquad \left. {}+\int_{0,0^-}^{1^+,1}\mathrm dx''\mathrm dz''\,\left(\hat Q_h(1-x',x'')+\mathcal P(\hat m^{\otimes 2})(1-x',x'')\right)D_{qh}^{-1}(x'',z'')G_0(z'',z')\right] \nonumber
\end{align}
\begin{align}
& Q_h(x,z) = \int_{0^-,0^-}^{x,z}\mathrm dx'\mathrm dz'\ \Phi(x-x',z-z')\Bigg{[}\hat Q_w\delta(x',z')+(\lambda^\mathrm{e}tm(x')+\sqrt\mu m_w\delta(x'))(\lambda^\mathrm{e}tm(z')+\sqrt\mu m_w\delta(z')) \\
& \qquad {}+\int_{0^-,0}^{1,1^+}\mathrm dx''\mathrm dx'''\,G_0(x',x'')D_{qh}^{-1,T}(x'',x''')\left(t\delta_\mathrm{e}Q_{qh}(x''',z')+\mathcal P(\hat m)(x''')(\lambda^\mathrm{e}tm(z')+\sqrt\mu m_w\delta(z'))\right) \nonumber \\
& \qquad {}+\int_{0,0^-}^{1^+,1}\mathrm dz'''\mathrm dz''\,\left(t\delta_\mathrm{e}Q_{qh}(z''',x')+(\lambda^\mathrm{e}tm(x')+\sqrt\mu m_w\delta(x'))\mathcal P(\hat m)(z''')\right)D_{qh}^{-1}(z''',z'')G_0(z'',z') \nonumber \\
& \quad\left. {}+\int_{0^-,0,0,0^-}^{1,1^+,1^+,1}\mathrm dx''\mathrm dx'''\mathrm dz'''\mathrm dz''\,G_0(x',x'')D_{qh}^{-1,T}(x'',x''')\left(\hat Q_h(x''',z''')+\mathcal P(\hat m^{\otimes 2})(x''',z''')\right)D_{qh}^{-1}(z''',z'')G_0(z'',z')\right] \ ; \nonumber
\end{align}
}
\end{widetext}
where we set
\begin{align}
\mathcal P(\hat m)(x) &= \hat m(x)+\rho\delta(1-x)\ ,\\
\mathcal P(\hat m^{\otimes 2})(x,z) &= \rho\left(\hat m(x)+\delta(1-x)\right)\left(\hat m(z)+\delta(1-z)\right) \nonumber\\
&\quad{}+(1-\rho)\hat m(x)\hat m(z)\ ,\\
G_0(x,z) &= t^2V_h(x,z)+V_w\delta(x,z)
\end{align}
and take $Q_{qh}(x,z)$ for $-\mathrm iQ_{qh}$. The accuracies are, with $\bar s=1$ for train and $\bar s=0$ for test:
{\small
\begin{align}
& \mathrm{Acc}_\mathrm{train/test} = \label{eq:precisionContinue}\\
& \quad\frac{1}{2}\left(1+\mathrm{erf}\left(\frac{m(1)+(\bar s-\rho)V_h(1,1)}{\sqrt 2\sqrt{Q_h(1,1)-m(1)^2-\rho(1-\rho)V_h(1,1)^2}}\right)\right)\ .\nonumber
\end{align}
}

Notice that we fully solved the model, in a certain limit, by giving an explicit expression of the performance of the GCN. This is an uncommon result in the sense that, in several works analyzing the performance of neural networks in a high-dimensional limit, the performance are only expressed as the function of the self-consistent of a system of equations similar to ours (\ref{eq:pointFixeC_début}-\ref{eq:pointFixeC_fin}). These systems have to be solved numerically, which may be unsatisfactory for the understanding of the studied models.

So far, we dealt with infinite regularization $r$ keeping only the first constant order. The predicted accuracy \eqref{eq:precisionContinue} does not depend on~$r$. We briefly show how to pursue the computation at any order in appendix \ref{sec:appOrdresSuivants}, by a perturbative approach with expansion in powers of $1/r$.

\paragraph{Interpretation in terms of dynamical mean-field theory:}
\label{sec:lienDMFT}
The order parameters $V_h$, $V_{qh}$ and $\hat V_h$ come from the replica computation and were introduced as the covariances between $h$ and its conjugate $q$. Their values are determined by extremizing the free entropy of the problem. In the above lines we derived that $V_h(x,z)\propto\Phi(x,z)$ is the forward propagator, from the weights to the loss, while $\hat V_h(x,z)\propto\Phi(1-x,1-z)$ is the backward propagator, from the loss to the weights.

In this section we state an equivalence between these order parameters and the correlation and response functions of the dynamical process followed by $h$.

We introduce the tilting field $\eta(x)\in\mathbb R^N$ and the tilted Hamiltonian as
\begin{align}
& \frac{\mathrm dh}{\mathrm dx}(x) = \frac{t}{\sqrt N}\tilde A^\mathrm{e}h(x)+\eta(x)+\delta(x)\frac{1}{\sqrt N}Xw\ ,\\
& h(x) = \int_{0^-}^x\mathrm dx'e^{(x-x')\frac{t}{\sqrt N}\tilde A^\mathrm{e}}\left(\eta(x')+\delta(x')\frac{1}{\sqrt N}Xw\right)\ ,\\
& H(\eta) = \frac{1}{2}(y-h(1))^TR(y-h(1))+\frac{r}{2}w^Tw\ ,
\end{align}
where $R\in\mathbb R^{N\times N}$ diagonal accounts for the train and test nodes. We write $\langle \cdot\rangle_\beta$ the expectation under the density $e^{-\beta H(\eta)}/Z$ (normalized only at $\eta=0$).

Then we have
\begin{align}
V_h(x,z) &= \frac{\beta}{N}\tr\left[\langle h(x)h(z)^T\rangle_\beta-\langle h(x)\rangle_\beta\langle h(z)^T\rangle_\beta\right]|_{\eta=0}\ , \\
V_{qh}(x,z) &= \frac{t}{N}\tr\frac{\partial}{\partial\eta(z)}\langle h(x)\rangle_\beta|_{\eta=0}\ , \\
\hat V_h(x,z) &= \frac{t^2}{\beta^2N}\tr\frac{\partial^2}{\partial\eta(x)\partial\eta(z)}\langle 1\rangle_\beta|_{\eta=0}\ ;
\end{align}
that is to say $V_h$ is the correlation function, $V_{qh}\approx tD_{qh}^{-1,T}$ is the response function and $\hat V_h$ is the correlation function of the responses of $h$. We prove these equalities at the constant order in $r$ using random matrix theory in the appendix \ref{sec:appLienDMFT}.

\subsubsection{Consequences}
\label{sec:continuConsequences}
\begin{figure}[t]
 \centering
 \includegraphics[width=\linewidth]{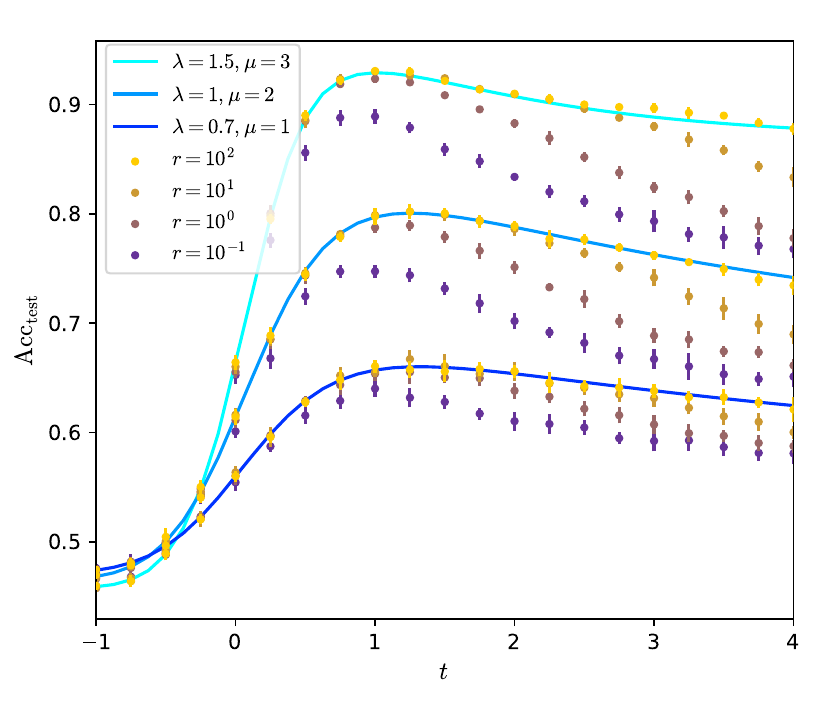}
 \caption{\label{fig:GCNcontinu_vsTetR} Predicted test accuracy $\mathrm{Acc}_\mathrm{test}$ of the continuous GCN on the asymmetric graph, at $r=\infty$. $\alpha=4$ and $\rho=0.1$. The performance of the continuous GCN are given by eq.~\eqref{eq:precisionContinue}. Dots: numerical simulation of the continuous GCN for $N=10^4$ and $d=30$, trained with quadratic loss, averaged over ten experiments.}
\end{figure}

\paragraph{Convergences:}
\label{sec:convergences}
We compare our predictions to numerical simulations of the continuous GCN for $N=10^4$ and $N=7\times 10^3$ in fig.~\ref{fig:GCNcontinu_vsTetR} and figs.~\ref{fig:GCNcontinu_vsTetRbis}, \ref{fig:GCNcontinu_vsTetR_sym} and \ref{fig:GCNcontinu_vsTetR_symBis} in appendix \ref{sec:supplFig}. The predicted test accuracies are well within the statistical errors. On these figures we can observe the convergence of $\mathrm{Acc}_\mathrm{test}$ with respect to $r$. The interversion of the two limits $r\to\infty$ and $K\to\infty$ we did to obtain \eqref{eq:precisionContinue} seems valid. Indeed on the figures we simulate the continuous GCN with $e^{\frac{t\tilde A}{\sqrt N}}$ or $e^{\frac{t\tilde A^\mathrm{s}}{\sqrt N}}$ and take $r\to\infty$ after the continuous limit $K\to\infty$; and we observe that the simulated accuracies converge well toward the predicted ones. To keep only the constant order in $1/r$ gives a good approximation of the continuous GCN. Indeed, the convergence with respect to $1/r$ can be fast: at $t\lessapprox 1$ not too large, $r\gtrapprox 1$ is enough to reach the continuous limit.

The convergence with respect to $K\to\infty$, taken after $r\to\infty$, is depicted in fig.~\ref{fig:GCNcontinu_vsTetK} and fig.~\ref{fig:GCNcontinu_vsTetKbis} in appendix \ref{sec:supplFig}. Again the continuous limit enjoyes good convergence properties since $K\gtrapprox 16$ is enough if $t$ is not too large.
\begin{figure}[h]
 \centering
 \includegraphics[width=\linewidth]{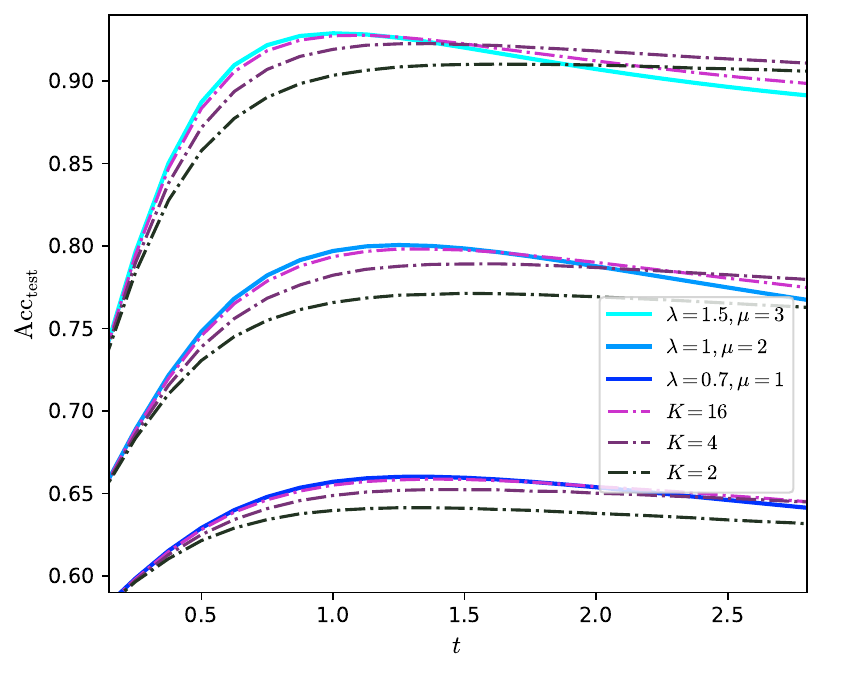}
 \caption{\label{fig:GCNcontinu_vsTetK} Predicted test accuracy $\mathrm{Acc}_\mathrm{test}$ of the continuous GCN and of its discrete counterpart with depth $K$ on the asymmetric graph, at $r=\infty$. $\alpha=1$ and $\rho=0.1$. The performance of the continuous GCN are given by eq.~\eqref{eq:precisionContinue} while for the discrete GCN they are given by numerically solving the fixed-point equations (\ref{eq:pointFixeCont_Vh}-\ref{eq:pointFixeCont_Qh}).}
\end{figure}

To summarize, figs.~\ref{fig:GCNcontinu_vsTetR}, \ref{fig:GCNcontinu_vsTetK} and appendix \ref{sec:supplFig} validate our method that consists in deriving the self-consistent equations at finite $K$ with replica, expanding them with respect to $1/r$, taking the continuous limit $K\to\infty$ and then solving the integral equations.

\paragraph{Optimal diffusion time $t^*$:}
\label{sec:tOpt}
We observe on the previous figures that there is an optimal diffusion time $t^*$ that maximizes $\mathrm{Acc}_\mathrm{test}$. Though we are able to solve the self-consistent equations and to obtain an explicit and analytical expression \eqref{eq:precisionContinue}, it is hard to analyze it in order to evaluate $t^*$. We have to consider further limiting cases or to compute $t^*$ numerically. The derivation of the following equations is detailed in appendix \ref{sec:contCasLimites}.

We first consider the case $t\to 0$. Expanding \eqref{eq:precisionContinue} to the first order in $t$ we obtain
\begin{equation}
\mathrm{Acc}_\mathrm{test} \underset{t\to 0}{=} \frac{1}{2}\left(1+\mathrm{erf}\left(\frac{1}{\sqrt 2}\sqrt\frac{\rho}{\alpha}\frac{\mu+\lambda^{\mathrm e} t(2+\mu)}{\sqrt{1+\rho\mu}}\right)\right)+o(t)\ . \label{eq:accTpetit}
\end{equation}
This expression shows in particular that $t^*>0$, i.e. some diffusion on the graph is always beneficial compared to no diffusion, as long as $\lambda t>0$ i.e. the diffusion is done forward if the graph is homophilic $\lambda>0$ and backward if it is heterophilic $\lambda<0$. We recover the result of \cite{keriven22oversm} for the discrete case in a slightly different setting. This holds even if the features of the graph are not informative $\mu=0$. Notice the explicit invariance by the change $(\lambda,t)\to(-\lambda,-t)$ in the potential~\eqref{eq:psiHCont} and in \eqref{eq:accTpetit}, which allows us to focus on $\lambda\ge 0$.
The case $t=0$ no diffusion corresponds to performing ridge regression on the Gaussian mixture $X$ alone. Such a model has been studied in \cite{mignacco20gaussMixt}; we checked we obtain the same expression as theirs at large regularization.

We now consider the case $t\to+\infty$ and $\lambda\ge 0$. Taking the limit in \eqref{eq:precisionContinue} we obtain
\begin{equation}
\mathrm{Acc}_\mathrm{test} \underset{t\to\infty}{\longrightarrow} \frac{1}{2}\left(1+\mathrm{erf}\left(\frac{\lambda^\mathrm{e} q_\mathrm{PCA}}{\sqrt 2}\right)\right)\ , \label{eq:accTgrand}
\end{equation}
where $q_\mathrm{PCA}$ is the same as for the discrete GCN, defined in eq.~\eqref{eq:qPCA}. This shows that the continuous GCN will oversmooth at large diffusion times. Thus, if the features are informative, if $\mu^2/\alpha>0$, the optimal diffusion time should be finite, $t^*<+\infty$. The continuous GCN does exactly as does the discrete GCN at $K\to\infty$ if $c$ is fixed. This is not surprising because of the mapping $c=K/t$: taking $t$ large is equivalent to take $c$ small with respect to $K$. $e^{\frac{t}{\sqrt N}\tilde A}$ is dominated by the same leading eigenvector $y_1$.

These two limits show that at finite time $t$ the GCN avoids oversmoothing and interpolates between an estimator that is only function of the features at $t=0$ and an estimator only function of the graph at $t=\infty$. $t$ has to be fine-tuned to reach the best trade-off $t^*$ and the optimal performance. 

In the insets of fig.~\ref{fig:GCNcontinu_vsL} and fig.~\ref{fig:GCNcontinu_vsLbis} in appendix \ref{sec:supplFig} we show how $t^*$ depends on $\lambda$. In particular, $t^*$ is finite for any $\lambda$: some diffusion is always beneficial but too much diffusion leads to oversmoothing. We have $t^*\underset{\lambda\to 0}{\longrightarrow}0$. This is expected since if $\lambda=0$ then $A$ is not informative and any diffusion $t>0$ would degrade the performance. The non-monotonicity of $t^*$ with respect to $\lambda$ is less expected and we do not have a clear interpretation for it. Last $t^*$ decreases when the feature signal $\mu^2/\alpha$ increases: the more informative $X$ the less needed diffusion is.
\begin{figure}[t]
 \centering
 \includegraphics[width=\linewidth]{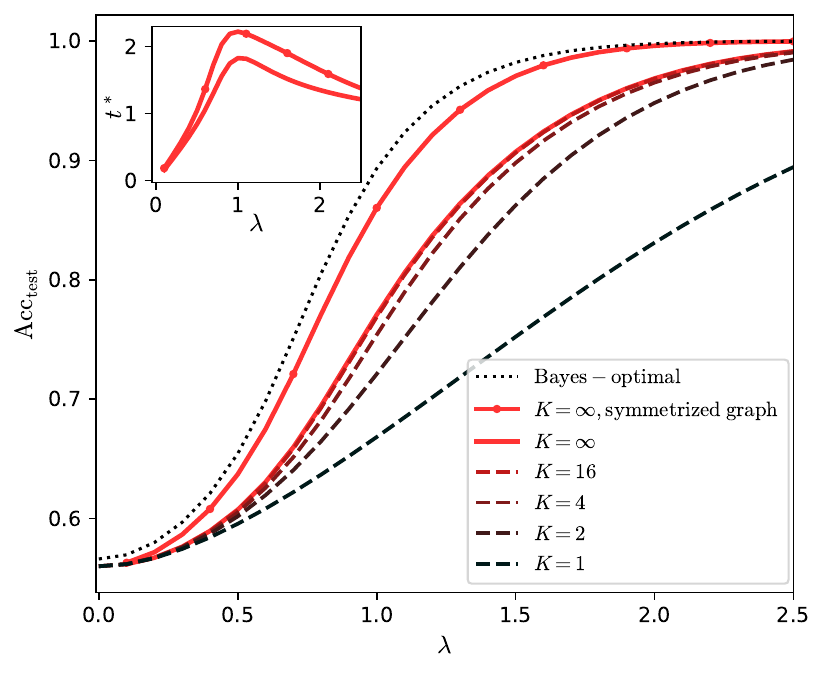}
 \caption{\label{fig:GCNcontinu_vsL} Predicted test accuracy $\mathrm{Acc}_\mathrm{test}$ of the continuous GCN and of its discrete counterpart with depth $K$, at optimal times $t^*$ and $r=\infty$. $\alpha=4$, $\mu=1$ and $\rho=0.1$. The performance of the continuous GCN $K=\infty$ are given by eq.~\eqref{eq:precisionContinue} while for its discretization at finite $K$ they are given by numerically solving eqs.~(\ref{eq:pointFixeCont_Vqh}-\ref{eq:pointFixeCont_Qh}). \emph{Inset:} $t^*$ the maximizer at $K=\infty$.}
\end{figure}

\paragraph{Optimality of the continuous limit:}
\label{sec:optimalitéCont}
A major result is that, at $t=t^*$, the continuous GCN is better than any fixed-$K$ GCN. Taking the continuous limit of the simple GCN is the way to reach its optimal performance.
This was suggested by fig.~\ref{fig:vsKetC} in the previous part; we show this more precisely in fig.~\ref{fig:GCNcontinu_vsL} and fig.~\ref{fig:GCNcontinu_vsLbis} in appendix~\ref{sec:supplFig}. We compare the continuous GCN to its discretization at different depths $K$ for several configurations $\alpha, \lambda, \mu$ and $\rho$ of the data model. The result is that at $t^*$ the test accuracy appears to be always an increasing function of $K$, and that its value at $K\to\infty$ and $t^*$ is a upper-bound for all $K$ and $t$.

Additionally, if the GCN is run on the symmetrized graph it can approach the Bayes-optimality and almost close the gap that \cite{dz24gcn} describes, as shown by figs.~\ref{fig:GCNcontinu_vsL}, \ref{fig:GCNcontinu_vsLbis} and \ref{fig:BOvsGCNcontinu} right. For all the considered $\lambda$ and $\mu$ the GCN is less than a few percents of accuracy far from the optimality.

However we shall precise this statement: the GCN approaches the Bayes-optimality only for a certain range of the parameters of the CSBM, as exemplified by figs.~\ref{fig:GCNcontinu_vsLbis} and \ref{fig:BOvsGCNcontinu} left. In these figures, the GCN is far from the Bayes-optimality when $\lambda$ is small but $\mu$ is large. In this regime we have $\mathrm{snr}_\mathrm{CSBM}>1$; even at $\rho=0$ information can be retrieved on the labels and the problem is closer to an unsupervised classification of the sole features~$X$. On $X$ the GCN acts as a supervised classifier, and as long as $\rho\neq 1$ it cannot catch all information. As previously highlighted by \cite{dz23csbm} the comparison with the Bayes-optimality is more relevant at $\mathrm{snr}_\mathrm{CSBM}<1$ where supervision is necessary. Then, as shown by figs.~\ref{fig:GCNcontinu_vsL}, \ref{fig:GCNcontinu_vsLbis} and \ref{fig:BOvsGCNcontinu} the symmetrized continuous GCN is close to the Bayes-optimality. The GCN is also able to close the gap in the region where $\lambda$ is large because, as we saw, it can perform unsupervised PCA on $A$.

\section{Conclusion}
In this article we derived the performance of a simple GCN trained for node classification in a semi-supervised way on data generated by the CSBM in the high-dimensional limit. We first studied a discrete network with a finite number $K$ of convolution steps. We showed the importance of going to large $K$ to approach the Bayes-optimality, while scaling accordingly the residual connections $c$ of the network to avoid oversmoothing. The resulting limit is a continuous GCN.

In a second part we were able to explicitly derive the performance of the continuous GCN. We highlighted the importance of the double limit $r, K\to\infty$, which allows to reach the optimal architecture and which can be analyzed thanks to an expansion in powers of $1/r$. In is an interesting question for future work whether this approach could allow the study of fully-connected large-depth neural networks.

Though the continuous GCN can be close to the Bayes-optimality, it has to better handle the features, especially when they are the main source of information.

\section*{Acknowledgments}
We acknowledge useful discussions with J. Zavatone-Veth, F. Zamponi and V. Erba.
This work is supported by the Swiss National Science Foundation under grant SNSF SMArtNet (grant number 212049).

\newpage

\appendix

\medskip

\input{appendix_discr.tex}

\medskip

\input{appendix_sym.tex}

\medskip

\section{State-evolution equations for the Bayes-optimal performance}
\label{sec:eqSE}
The Bayes-optimal (BO) performance for semi-supervised classification on the binary CSBM can be computed thanks to the following iterative state-evolution equations, that have been derived in \cite{cSBM18,dz23csbm}.

The equations have been derived for a symmetric graph. We map the asymmetric $\tilde A$ to a symmetric matrix by the symmetrization $(\tilde A+\tilde A^T)/\sqrt 2$. Thus the BO performance on $A$ asymmetric are the BO performance on $A$ symmetrized and effective signal $\lambda^\mathrm{s}=\sqrt 2\lambda$.

Let $m_y^0$ and $m_u^0$ be the initial condition. The state-evolution equations are
\begin{align}
& m_u^{t+1} = \frac{\mu m_y^t}{1+\mu m_y^t} \\
& m^t = \frac{\mu}{\alpha}m_u^t+(\lambda^\mathrm{s})^2m_y^{t-1} \\
& m_y^t = \rho+(1-\rho)\mathbb E_{W}\left[\tanh\left(m^t+\sqrt{m^t}W\right)\right]
\end{align}
where $W$ is a standard scalar Gaussian. These equations are iterated until convergence to a fixed-point $(m,m_y,m_u)$. Then the BO test accuracy is
\begin{equation}
\mathrm{Acc}_\mathrm{test} = \frac{1}{2}(1+\mathrm{erf}\sqrt{m/2})\ . \label{eq:precBOcsbm}
\end{equation}
In the large $\lambda$ limit we have $m_y\to 1$ and
\begin{equation}
\log(1-\mathrm{Acc}_\mathrm{test}) \underset{\lambda\to\infty}{\sim} -\lambda^2\ .
\end{equation}

\medskip 

\section{Details on the numerics}
\label{sec:numériques}
For the discrete GCN, the system of fixed-point equations (\ref{eq:pointFixeDébut}--\ref{eq:pointFixeFin}) is solved by iterating it until convergence. The iterations are stable up to $K\approx 4$ and no damping is necessary. The integration over $(\xi, \zeta, \chi)$ is done by Hermite quadrature (quadratic loss) or Monte-Carlo sampling (logistic loss) over about $10^6$ samples. For the quadratic loss $h^*$ has to be computed by Newton's method. Then the whole computation takes around one minute on a single CPU.

For the continuous GCN the equation \eqref{eq:precisionContinue} is evaluated by a trapezoidal integration scheme with a hundred of discretization points. In the nested integrals of $Q(1,1)$, $\hat Q$ can be evaluated only once at each discretization point. The whole computation takes a few seconds.

We provide the code to evaluate our predictions in the supplementary material.

\newpage

\begin{widetext}
\section{Supplementary figures}
\label{sec:supplFig}
In this section we provide the supplementary figures of part \ref{sec:continuConsequences}. They show the convergence to the continuous limit with respect to $K$ and $r$, and that the continuous limit can be close to the optimality. We also provide the supplementary figures of part \ref{sec:discrConsequences}, that compare the GCN on symmetric and asymmetric graphs, and that show the train error versus the residual connection strength.

\vspace{-2mm}
\subsection*{Asymmetric graph}
The following figures support the discussion of part \ref{sec:convergences} for the asymmetric graph. They compare the theoretical predictions for the continuous GCN to numerical simulations of the trained network. They show the convergence towards the limit $r\to\infty$ and the optimality of the continuous GCN over its discretization at finite $K$.
\begin{figure}[th]
 \centering
 \includegraphics[width=0.48\linewidth]{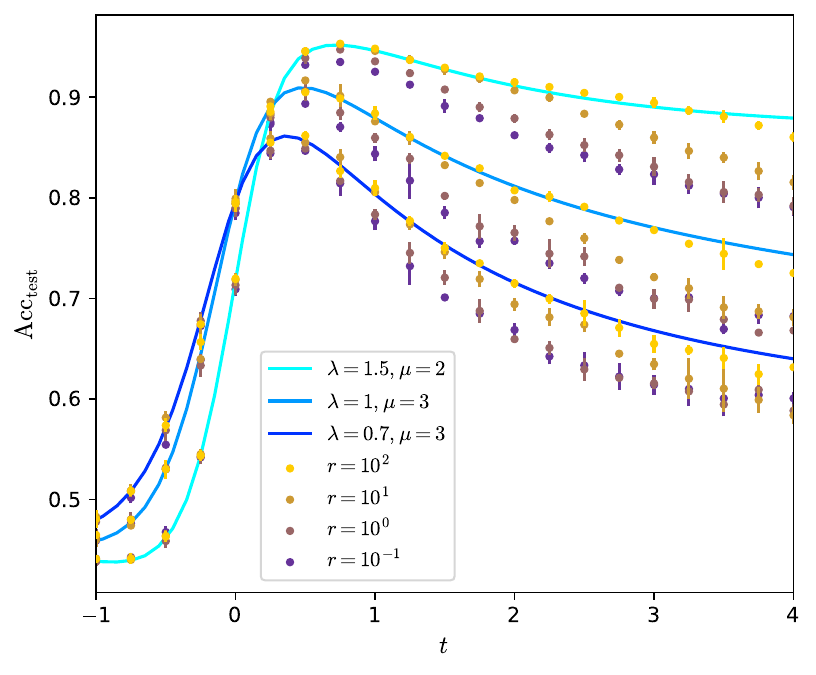}
 \includegraphics[width=0.48\linewidth]{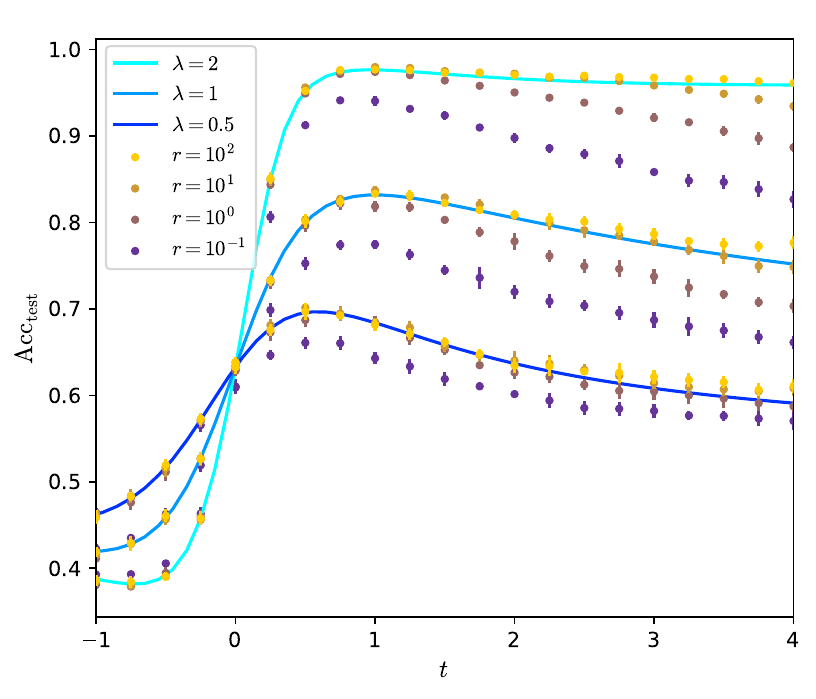}
 \caption{\label{fig:GCNcontinu_vsTetRbis} Predicted test accuracy $\mathrm{Acc}_\mathrm{test}$ of the continuous GCN, at $r=\infty$. \emph{Left:} for $\alpha=1$ and $\rho=0.1$; \emph{right:} for $\alpha=2$, $\mu=1$ and $\rho=0.3$. The performance of the continuous GCN are given by eq.~\eqref{eq:precisionContinue}. Dots: numerical simulation of the continuous GCN for $N=7\times 10^3$ and $d=30$, trained with quadratic loss, averaged over ten experiments.}
\end{figure}
\vspace{-4mm}
\begin{figure}[th]
 \centering
 \includegraphics[width=0.48\linewidth]{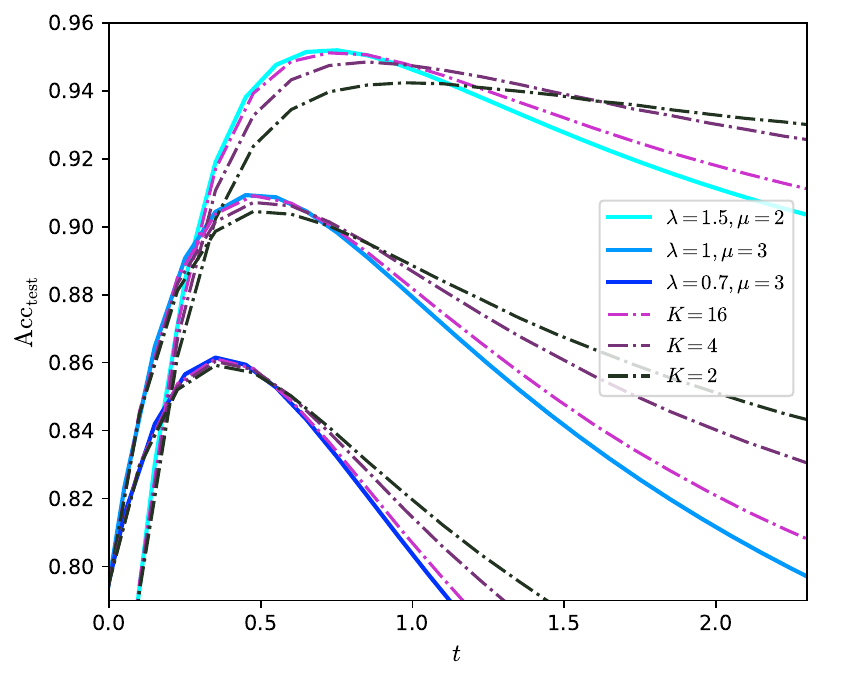}
 \includegraphics[width=0.48\linewidth]{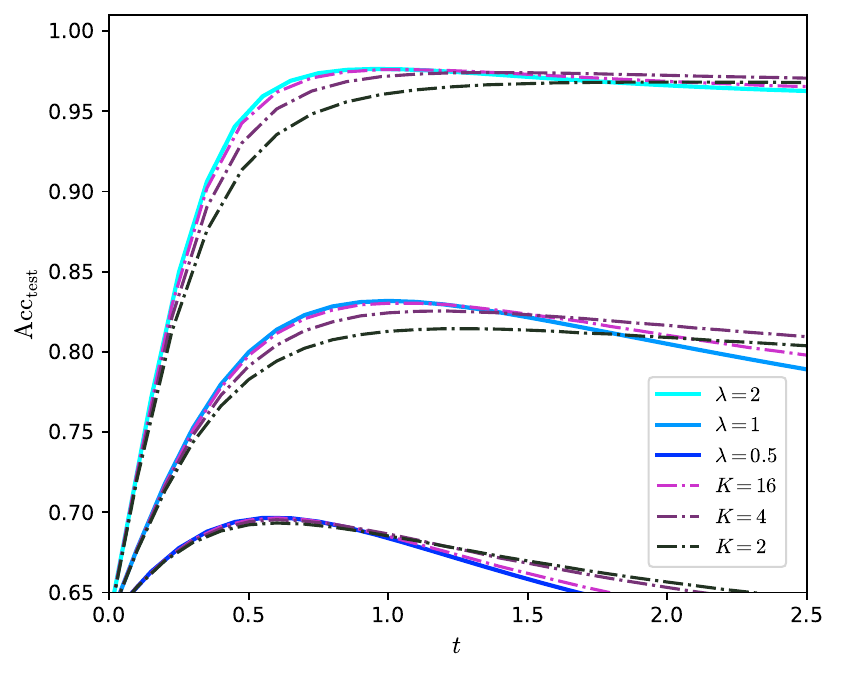}
 \caption{\label{fig:GCNcontinu_vsTetKbis} Predicted test accuracy $\mathrm{Acc}_\mathrm{test}$ of the continuous GCN, at $r=\infty$. \emph{Left:} for $\alpha=1$ and $\rho=0.1$; \emph{right:} for $\alpha=2$, $\mu=1$ and $\rho=0.3$. The performance of the continuous GCN are given by eq.~\eqref{eq:precisionContinue} while for its discretization at finite $K$ they are given by numerically solving the fixed-point equations (\ref{eq:pointFixeCont_Vqh}-\ref{eq:pointFixeCont_Qh}).}
\end{figure}

\newpage

\subsection*{Symmetrized graph}
The following figures support the discussion of part \ref{sec:convergences} for the symmetrized graph. They compare the theoretical predictions for the continuous GCN to numerical simulations of the trained network. They show the convergence towards the limit $r\to\infty$.
\begin{figure}[th]
 \centering
 \includegraphics[width=0.48\linewidth]{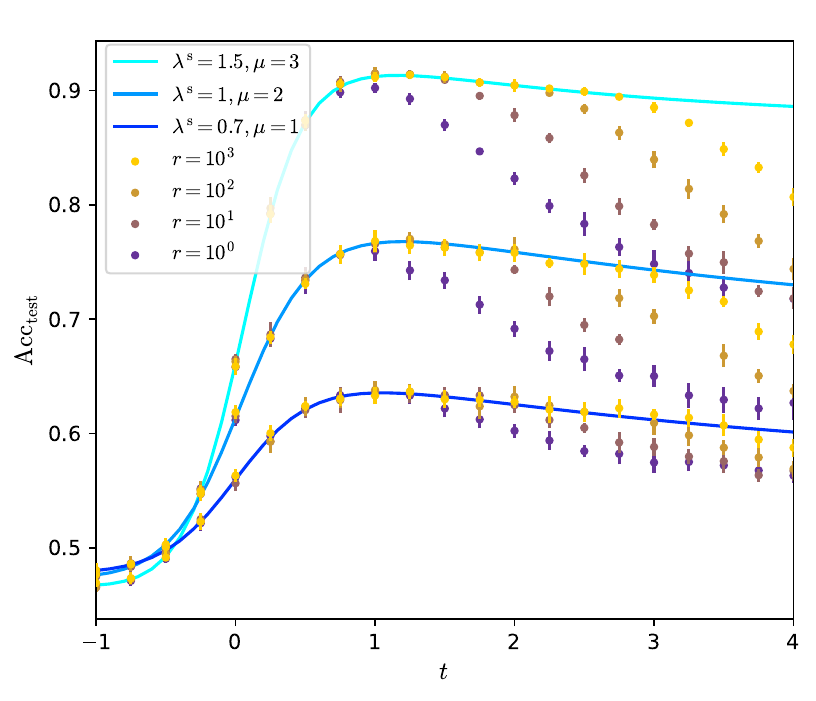}
 \caption{\label{fig:GCNcontinu_vsTetR_sym} Predicted test accuracy $\mathrm{Acc}_\mathrm{test}$ of the continuous GCN, at $r=\infty$ for a symmetrized graph. $\alpha=4$, $\rho=0.1$. We remind that $\lambda^\mathrm{s}=\sqrt 2\lambda$. The performance of the continuous GCN are given by eq.~\eqref{eq:precisionContinue}. Dots: numerical simulation of the continuous GCN for $N=10^4$ and $d=30$, trained with quadratic loss, averaged over ten experiments.}
\end{figure}
\begin{figure}[th]
 \centering
 \includegraphics[width=0.48\linewidth]{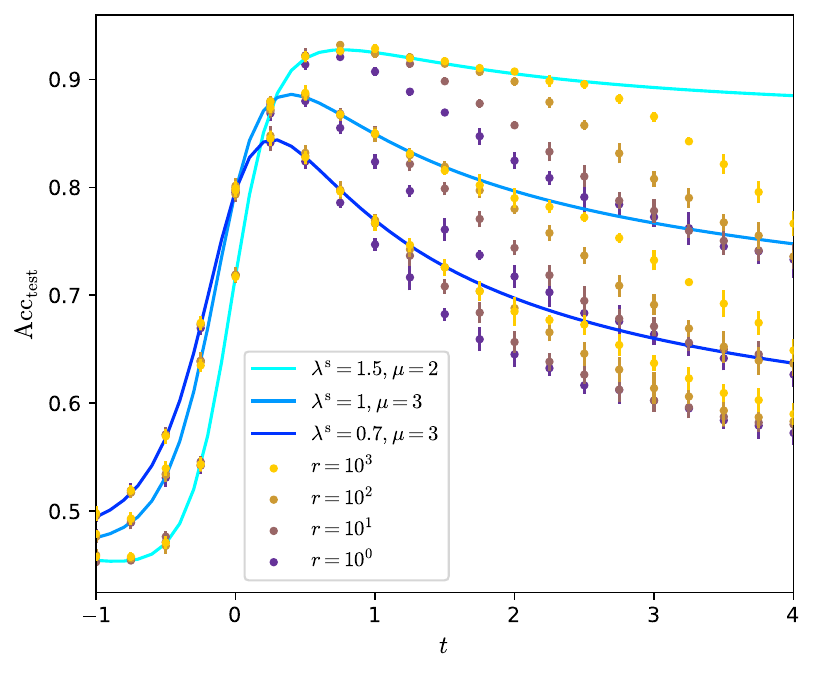}
 \includegraphics[width=0.48\linewidth]{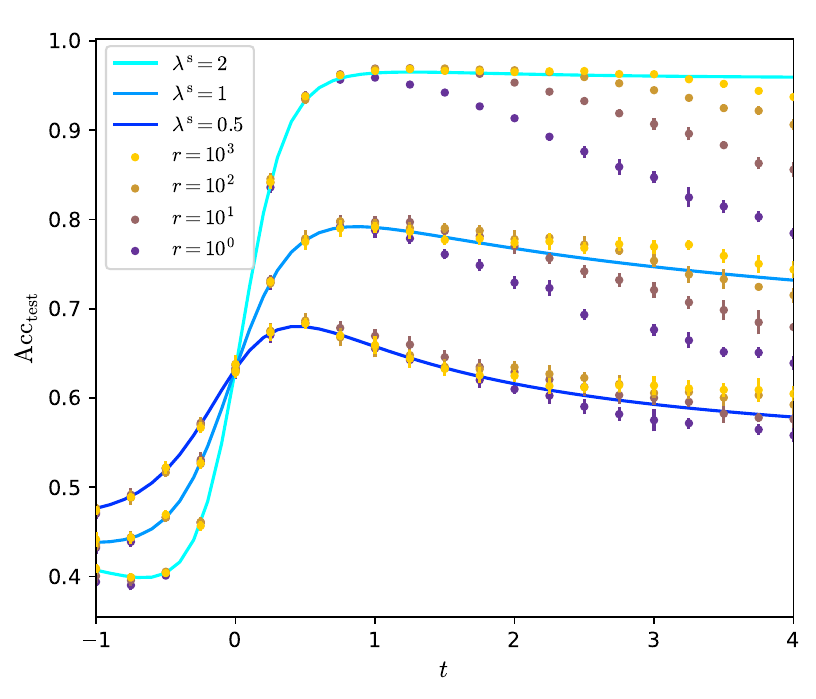}
 \caption{\label{fig:GCNcontinu_vsTetR_symBis} Predicted test accuracy $\mathrm{Acc}_\mathrm{test}$ of the continuous GCN, at $r=\infty$ for a symmetrized graph. \emph{Left:} for $\alpha=1$ and $\rho=0.1$; \emph{right:} for $\alpha=2$, $\mu=1$ and $\rho=0.3$. We remind that $\lambda^\mathrm{s}=\sqrt 2\lambda$. The performance of the continuous GCN are given by eq.~\eqref{eq:precisionContinue}. Dots: numerical simulation of the continuous GCN for $N=7\times 10^3$ and $d=30$, trained with quadratic loss, averaged over ten experiments.}
\end{figure}

\newpage

\subsection*{Comparison with optimality}
The following figures support the discussion of parts \ref{sec:tOpt} and \ref{sec:optimalitéCont}. They show how the optimal diffusion time $t^*$ varies with respect to the parameters of the model and they compare the performance of the optimal continuous GCN and its discrete counterpart to the Bayes-optimality.
\begin{figure}[th]
 \centering
 \includegraphics[width=0.48\linewidth]{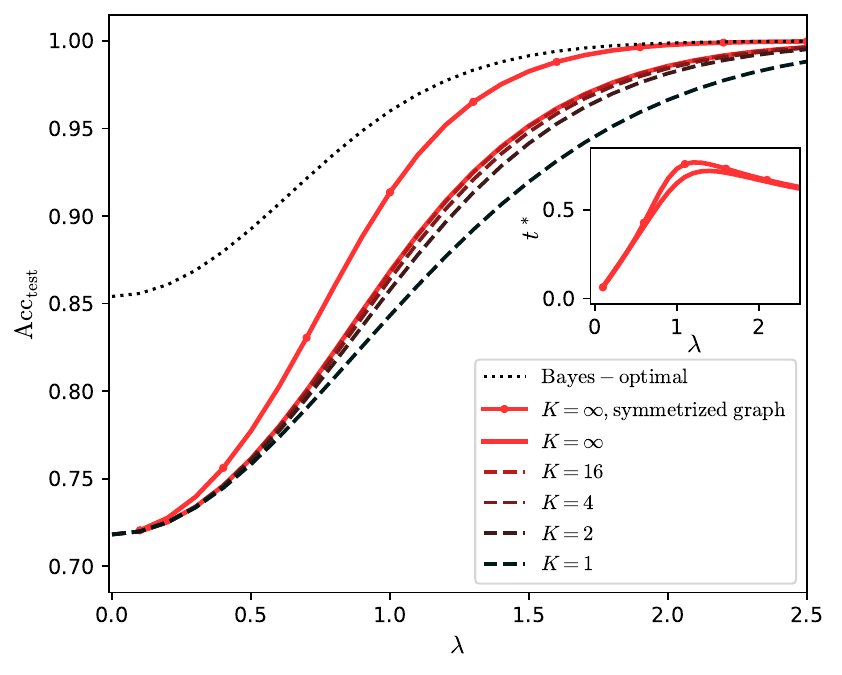}
 \includegraphics[width=0.48\linewidth]{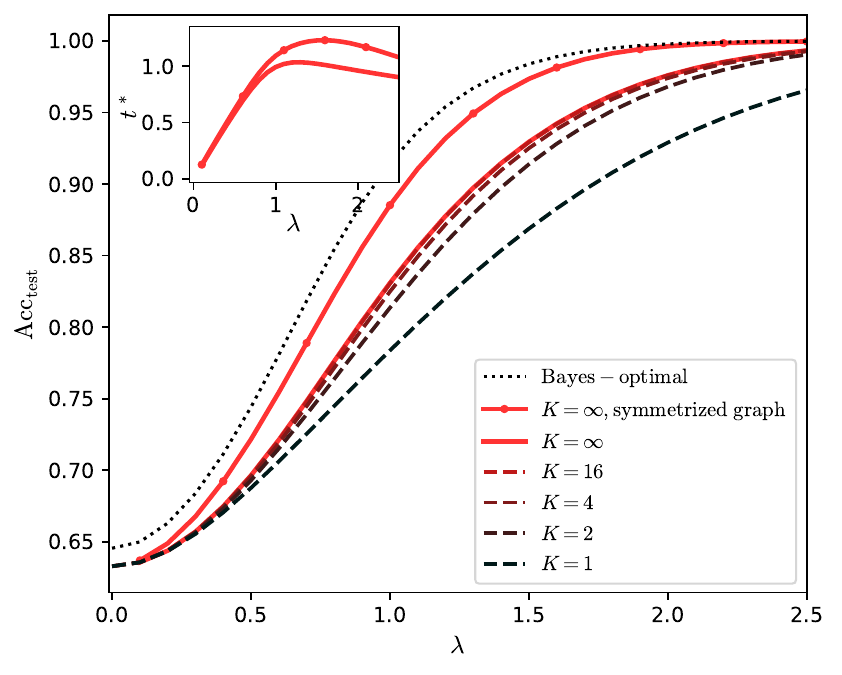}
 \caption{\label{fig:GCNcontinu_vsLbis} Predicted test accuracy $\mathrm{Acc}_\mathrm{test}$ of the continuous GCN and of its discrete counterpart with depth $K$, at optimal times $t^*$ and $r=\infty$. \emph{Left:} for $\alpha=1$, $\mu=2$ and $\rho=0.1$; \emph{right:} for $\alpha=2$, $\mu=1$ and $\rho=0.3$. The performance of the continuous GCN $K=\infty$ are given by eq.~\eqref{eq:precisionContinue} while for its discretization at finite $K$ they are given by numerically solving the fixed-point equations (\ref{eq:pointFixeCont_Vqh}-\ref{eq:pointFixeCont_Qh}). \emph{Inset:} $t^*$ the maximizer at $K=\infty$.}
\end{figure}

\begin{figure}[th]
 \centering
 \includegraphics[width=0.48\linewidth]{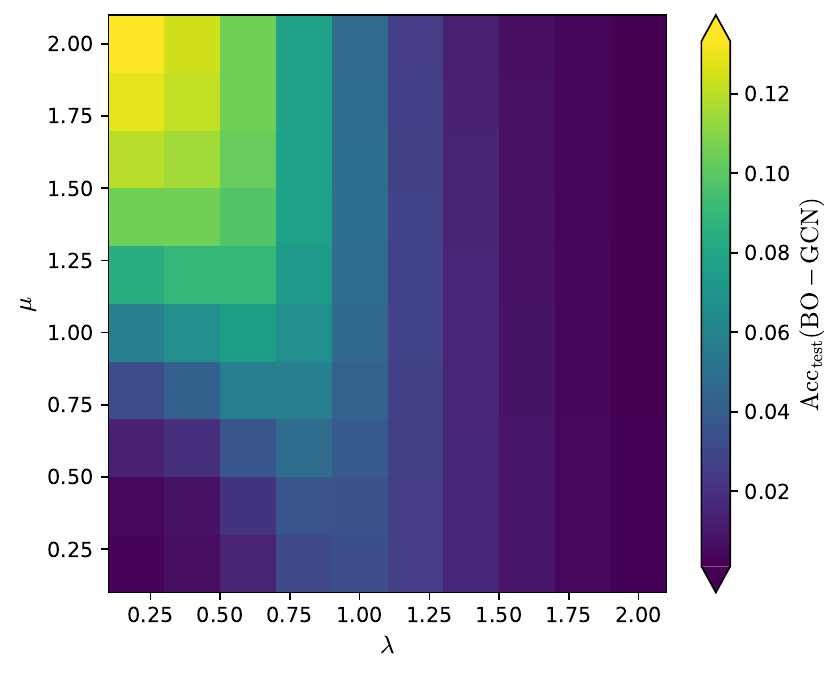}
 \includegraphics[width=0.48\linewidth]{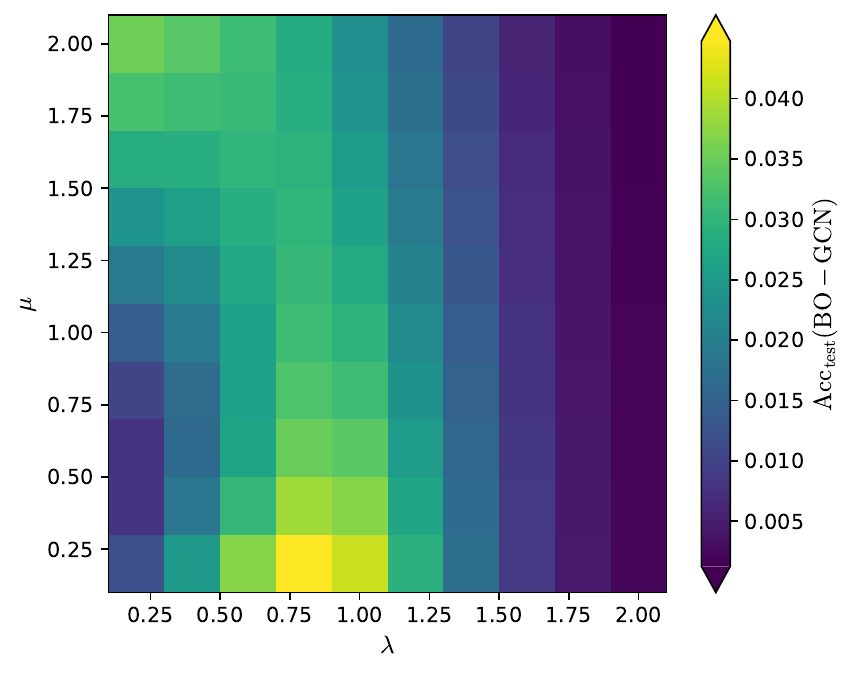}
 \caption{\label{fig:BOvsGCNcontinu} Gap to the Bayes-optimality. Predicted difference between the Bayes-optimal test accuracy and the test accuracy of the continuous GCN at optimal time $t^*$ and $r=\infty$, vs the two signals $\lambda$ and $\mu$. \emph{Left:} for $\alpha=1$ and $\rho=0.1$; \emph{right:} for $\alpha=2$ and $\rho=0.3$. The performance of the continuous GCN are given by eq.~\eqref{eq:precisionContinue}.}
\end{figure}

\newpage

\subsection*{Comparison between symmetric and asymmetric graphs}
\vspace{-3mm}
The following figure supports the claim of part \ref{sec:symmetrization}, that the performance of the GCN depends little whether the graph is symmetric or not at same $\lambda$, and that it is not able to deal with the supplementary information the asymmetry gives.
\vspace{-3mm}
\begin{figure}[th]
 \centering
 \includegraphics[width=\linewidth]{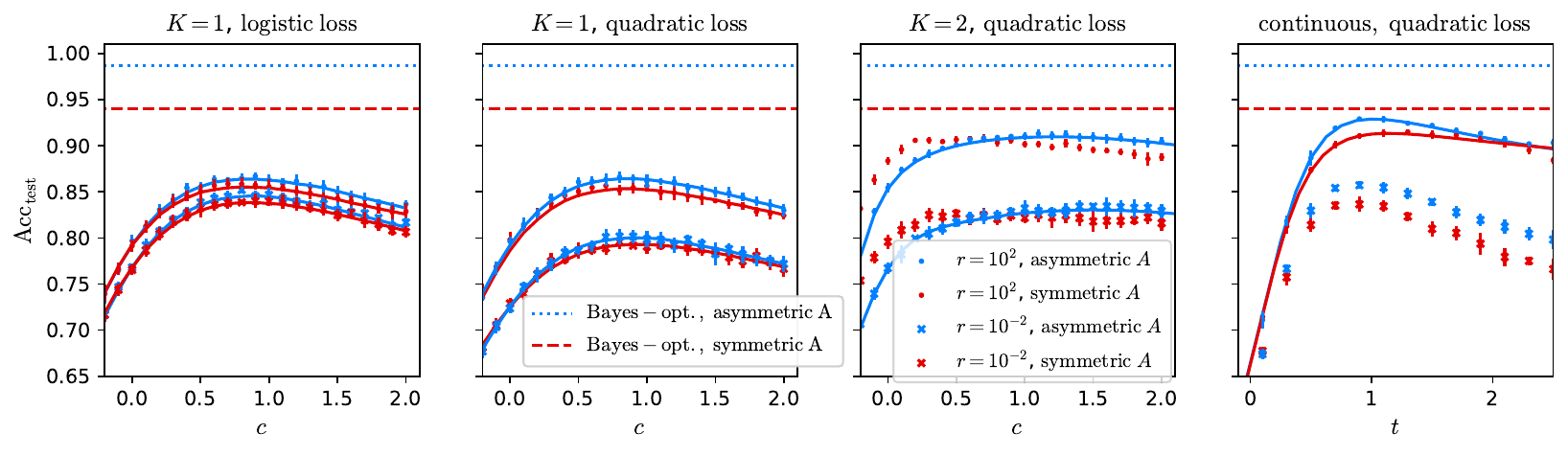}
\vspace{-4mm}
 \caption{\label{fig:symVsAsym} Test accuracy of the GCN, on asymmetric $A$ and its symmetric counterpart, obtained by equaling $A_{ij}$ with $A_{ji}$ for all $i<j$. $\alpha=4$, $\lambda=1.5$, $\mu=3$ and $\rho=0.1$. Lines: predictions. Dots: numerical simulation of the GCN for $N=10^4$ and $d=30$, averaged over ten experiments.}
\end{figure}

\vspace{-8mm}
\subsection*{Train error}
\vspace{-3mm}
The following figure displays the train error $E_\mathrm{train}$ eq. \ref{eq:erreur} vs the self-loop intensity $c$, in the same settings as fig.~\ref{fig:différentsKvsC} of part \ref{sec:discreteGCN}. It shows in particular that to treat $c$ as a parameter trained to minimize the train error would degrade the performance, since it would lead to $c\to\infty$. As a consequence, $c$ should be treated as a hyperparameter, tuned to maximize the test accuracy, as done in the main part of the article.
\vspace{-3mm}
\begin{figure}[th]
 \centering
 \includegraphics[width=0.9\linewidth]{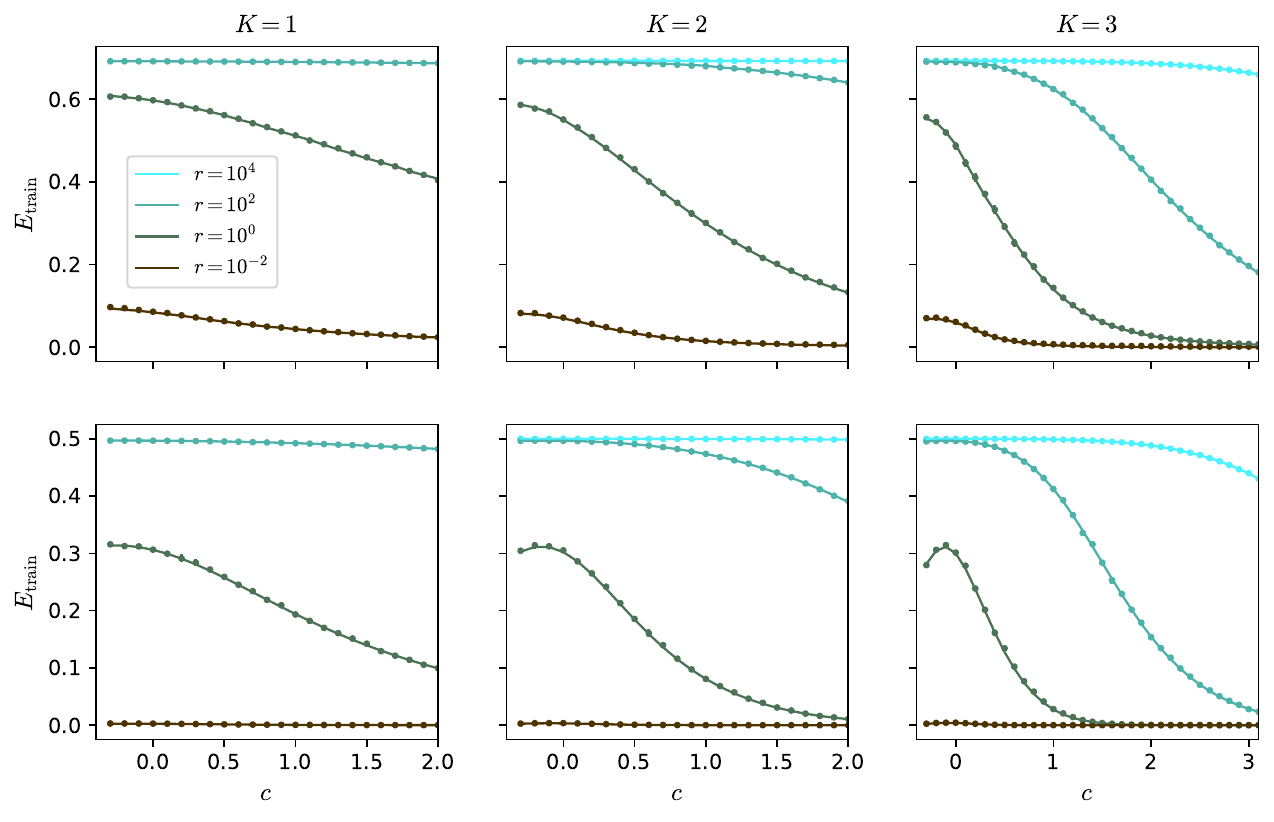}
\vspace{-4mm}
 \caption{\label{fig:différentsKvsC_Etrain} Predicted train error $E_\mathrm{test}$ for different values of $K$. \emph{Top:} for $\lambda=1.5$, $\mu=3$ and logistic loss; \emph{bottom:} for $\lambda=1$, $\mu=2$ and quadratic loss; $\alpha=4$ and $\rho=0.1$. We take $c_k=c$ for all $k$. Dots: numerical simulation of the GCN for $N=10^4$ and $d=30$, averaged over ten experiments.}
\end{figure}

\end{widetext}

\newpage

\bibliography{main}

\end{document}

%% file: appendix_discr.tex
\begin{widetext}
\section{Asymptotic characterisation of the discrete GCN}
\label{sec:appendixDiscreteGCN}
In this part we compute the free energy of the discrete finite-$K$ GCN using replica. We derive the fixed-point equations for the order parameters of the problem and the asymptotic characterization of the errors and accuracies in function of the order parameters. We consider only the asymmetric graph $\tilde A$; the symmetrized case $\tilde A^\mathrm{s}$ is analyzed in the following section \ref{sec:appendixContinuousGCN} together with the continuous GCN.

The free energy of the problem is $-\beta Nf = \partial_n\mathbb E_{u, \Xi, W, y}Z^n(n=0)$ where the partition function is
\begin{align}
Z &= \int\prod_\nu^M\mathrm dw_\nu e^{-\beta r\gamma(w_\nu)} e^{-\beta s\sum_{i\in R}\ell(y_ih(w)_i)-\beta s'\sum_{i\in R'}\ell(y_ih(w)_i)}\ .
\end{align}
To lighten the notations we take $\rho'=1-\rho$ i.e. the test set is the whole complementary of the train set. This does not change the result since the performances do not depend on the size of the test set.

We recall that $\tilde A$ admits the following Gaussian equivalent:
\begin{equation}
\tilde A \approx A^\mathrm{g}=\frac{\lambda}{\sqrt N}yy^T+\Xi\ ,\quad \Xi_{ij}\sim\mathcal N(0,1)\ .
\end{equation}
$\tilde A$ can be approximated by $A^\mathrm{g}$ with a vanishing change in the free energy $f$.

\subsection{Derivation of the free energy}
We define the intermediate states of the GCN as
\begin{equation}
h_k = \left(\frac{1}{\sqrt N}\tilde A+c_kI_N\right)h_{k-1}\ ,\quad h_0 = \frac{1}{\sqrt N}Xw\ .
\end{equation}
We introduce them in $Z$ thanks to Dirac deltas. The expectation of the replicated partition function is
\begin{align}
\mathbb EZ^n \propto &\, \mathbb E_{u, \Xi, W, y}\int \prod_a^n\prod_\nu^M\mathrm dw_\nu^ae^{-\beta r\gamma(w_\nu^a)}\prod_a^n\prod_i^N\prod_{k=0}^K\mathrm dh_{i,k}^a\mathrm dq_{i,k}^a e^{-\beta s\sum_{a,i\in R}\ell(y_ih_{i,K}^a)-\beta s'\sum_{a,i\in R'}\ell(y_ih_{i,K}^a)} \\
 & \quad\quad e^{\sum_{a,i}\sum_{k=1}^K\mathrm iq_{i,k}^a\left(h_{i,k}^a-\frac{1}{\sqrt N}\sum_j(\frac{\lambda}{\sqrt N}y_iy_j+\Xi_{ij})h_{j,k-1}^a-c_kh_{i,k-1}^a\right)+\sum_{a,i}\mathrm iq_{i,0}^a\left(h_{i,0}^a-\frac{1}{\sqrt N}\sum_\nu\left(\sqrt\frac{\mu}{N}y_iu_\nu+W_{i\nu}\right)w_\nu^a\right)} \nonumber \\
 = &\, \mathbb E_{u, y}\int \prod_{a,\nu}\mathrm dw_\nu^ae^{-\beta r\gamma(w_\nu^a)}\prod_{a,i,k}\mathrm dh_{i,k}^a e^{-\beta s\sum_{a,i\in R}\ell(y_ih_{i,K}^a)-\beta s'\sum_{a,i\in R'}\ell(y_ih_{i,K}^a)} \nonumber \\
 & \quad\quad \prod_i\mathcal N\left(h_{i,>0} \left| c\odot h_{i,<K}+y_i\frac{\lambda}{N}\sum_jy_jh_{j,<K} ; \tilde Q \right.\right)\prod_i\mathcal N\left(h_{i,0} \left| y_i\frac{\sqrt\mu}{N}\sum_\nu u_\nu w_\nu ; \frac{1}{N}\sum_\nu w_\nu w_\nu^T \right. \right) \ . \nonumber
\end{align}
$\mathcal N(\cdot|m;V)$ is the Gaussian density of mean $m$ and covariance $V$. We integrated over the random fluctuations $\Xi$ and $W$ and then over the $q$s. We collected the replica in vectors of size $n$ and assembled them as
\begin{align}
& h_{i,>0} = \left(\begin{smallmatrix}h_{i,1} \\ \vdots \\ h_{i,K}\end{smallmatrix}\right) \in\mathbb R^{nK}, \quad h_{i,<K} = \left(\begin{smallmatrix}h_{i,0} \\ \vdots \\ h_{i,K-1}\end{smallmatrix}\right) \in\mathbb R^{nK}, \quad c\odot h_{i,<K} = \left(\begin{smallmatrix}c_1h_{i,0} \\ \vdots \\ c_Kh_{i,K-1}\end{smallmatrix}\right), \\
& \tilde Q_{k,l} = \frac{1}{N}\sum_jh_{j,k}h_{j,l}^T\ ,\quad \tilde Q = \left(\begin{smallmatrix}\tilde Q_{0,0} & \ldots & \tilde Q_{0,K-1} \\ \vdots & & \vdots \\ \tilde Q_{K-1,0} & \ldots & \tilde Q_{K-1,K-1}\end{smallmatrix}\right) \in\mathbb R^{nK\times nK}\ .
\end{align} 
We introduce the order parameters
\begin{align}
& m_w^a = \frac{1}{N}\sum_\nu u_\nu w_\nu^a\ , \quad Q_w^{ab} = \frac{1}{N}\sum_\nu w_\nu^a w_\nu^b\ , \\
& m_k^a = \frac{1}{N}\sum_jy_jk_{j,k}^a\ , \quad Q_k^{ab} = (\tilde Q_{k,k})_{a,b} = \frac{1}{N}\sum_jh_{j,k}^ah_{j,k}^b\ , \quad Q_{k,l}^{ab} = (\tilde Q_{k,l})_{a,b}=\frac{1}{N}\sum_jh_{j,k}^ah_{j,l}^b\ .
\end{align} 
$m_k$ is the magnetization (or overlap) between the $k^\mathrm{th}$ layer and the labels; $m_w$ is the magnetization between the weights and the hidden variables and the $Q$s are the self-overlaps across the different layers. In the following we write $\tilde Q$ for the matrix with elements $(\tilde Q)_{ak,bl}=Q_{k,l}^{ab}$. We introduce these quantities thanks to new Dirac deltas. This allows us to factorize the spacial $i$ and $\nu$ indices.
\begin{align}
\mathbb EZ^n \propto &\, \int \prod_a\prod_{k=0}^{K-1}\mathrm d\hat m_k^a\mathrm dm_k^ae^{N\hat m_k^am_k^a} \prod_a\mathrm d\hat m_w^a\mathrm dm_w^ae^{N\hat m_w^am_w^a} \prod_{a\le b}\prod_{k=0}^{K-1}\mathrm d\hat Q_k^{ab}\mathrm dQ_k^{ab}e^{N\hat Q_k^{ab}Q_k^{ab}} \prod_{a,b}\prod_{k<l}^{K-1}\mathrm d\hat Q_{k,l}^{ab}\mathrm dQ_{k,l}^{ab}e^{N\hat Q_{k,l}^{ab}Q_{k,l}^{ab}} \\
& \prod_{a\le b}d\hat Q_w^{ab}\mathrm dQ_w^{ab}e^{N\hat Q_w^{ab}Q_w^{ab}} \left[\mathbb E_u\int \prod_a\mathrm dw^ae^{\psi_w^{(n)}(w)} \right]^\frac{N}{\alpha}
\left[\mathbb E_y\int \prod_{a,k}\mathrm dh_k^a e^{\psi_h^{(n)}(h; s)} \right]^{\rho N}
\left[\mathbb E_y\int \prod_{a,k}\mathrm dh_k^a e^{\psi_h^{(n)}(h; s')} \right]^{(1-\rho)N} \nonumber
\end{align}
where we defined the two potentials
\begin{align}
& \psi_w^{(n)}(w) = -\beta r\sum_a\gamma(w^a)-\sum_{a\le b}\hat Q_w^{ab}w^aw^b-\sum_a\hat m_w^auw^a \\
& \psi_h^{(n)}(h; \bar s) = -\beta\bar s\sum_a\ell(yh_K^a)-\sum_{a\le b}\sum_{k=0}^{K-1}\hat Q_k^{ab}h_k^ah_k^b -\sum_{a,b}\sum_{k<l}^{K-1}\hat Q_{k,l}^{ab}h_k^ah_l^b-\sum_a\sum_{k=0}^{K-1}\hat m_k^ayh_k^a \nonumber \\
& \qquad\qquad{}+\log \mathcal N\left(h_{>0} \left|c\odot h_{<K}+\lambda ym_{<K} ; \tilde Q \right.\right) +\log \mathcal N\left(h_0 \left| \sqrt\mu ym_w ; Q_w \right. \right)\ .
\end{align}
We leverage the replica-symmetric ansatz. It is justified by the convexity of the Hamiltonian $H$. We assume that for all $a$ and $b$
\begin{align}
& m_k^a = m_k\ , && \hat m_k^a = -\hat m_k\ , && m_w^a = m_w\ , && \hat m_w^a = -\hat m_w\ , \\
& Q_k^{ab} = Q_kJ+V_kI\ , && \hat Q_k^{ab} = -\hat Q_kJ+\frac{1}{2}(\hat V_k+\hat Q_k)I\ , && Q_w^{ab} = Q_wJ+V_wI\ , && \hat Q_w^{ab} = -\hat Q_wJ+\frac{1}{2}(\hat V_w+\hat Q_w)I\ , \\
& Q_{k,l}^{ab} = Q_{k,l}J+V_{k,l}I\ , && \hat Q_{k,l}^{ab} = -\hat Q_{k,l}J+\hat V_{k,l}I\ .
\end{align}
$I$ is the $n\times n$ identity and $J$ is the $n\times n$ matrix filled with ones. We introduce the $K\times K$ symmetric matrices $Q$ and $V$, filled with $(Q_k)_{0\leq k\leq K-1}$ and $(V_k)_{0\leq k\leq K-1}$ on the diagonal, and $(Q_{k,l})_{0\leq k<l\leq K-1}$ and $(V_{k,l})_{0\leq k<l\leq K-1}$ off the diagonal, such that $\tilde Q$ can be written in terms of Kronecker products as
\begin{align}
\tilde Q=Q\otimes J+V\otimes I\ .
\end{align}
The entropic terms of $\psi_w^{(n)}$ and $\psi_h^{(n)}$ can be computed. Since we will take $n=0$ we discard subleading terms in $n$. We obtain
\begin{align}
& \sum_a\hat m_w^am_w^a = n\hat m_wm_w\ , \quad
\sum_{a\leq b}\hat Q_w^{ab}Q_w^{ab} = \frac{n}{2}(\hat V_wV_w+\hat V_wQ_w-V_w\hat Q_w)\ ,\\
& \sum_a\hat m_k^am_k^a = n\hat m_km_k\ , \quad
\sum_{a\leq b}\hat Q_k^{ab}Q_k^{ab} = \frac{n}{2}(\hat V_kV_k+\hat V_kQ_k-V_k\hat Q_k)\ , \quad \sum_{a,b}\hat Q_{k,l}^{ab}Q_{k,l}^{ab} = n(\hat V_{k,l}V_{k,l}+\hat V_{k,l}Q_{k,l}-V_{k,l}\hat Q_{k,l})\ .
\end{align}
The Gaussian densities can be explicited, keeping again the main order in $n$ and using the formula for a rank-1 update to a matrix (Sherman-Morrison formula):
\begin{align}
& Q_w^{-1} = \frac{1}{V_w}I-\frac{Q_w}{V_w^2}J\ ,\quad \log\det Q_w = n\frac{Q_w}{V_w}+n\log V_w\ ,\\
& \tilde Q^{-1} = V^{-1}\otimes I-(V^{-1}QV^{-1})\otimes J\ ,\quad \log\det\tilde Q = n\tr(V^{-1}Q)+n\log V\ .
\end{align}
Then we can factorize the replica by introducing random Gaussian variables:
\begin{align}
\int \prod_a\mathrm dw^ae^{\psi_w^{(n)}(w)} &= \int\prod_a\mathrm dw^ae^{\sum_a\log P_W(w^a)+\frac{1}{2}\hat Q_ww^TJw-\frac{1}{2}\hat V_ww^Tw+u\hat m_w^Tw} = \mathbb E_\varsigma\left(\int\mathrm dwe^{\psi_w(w)}\right)^n
\end{align}
where $\varsigma\sim\mathcal N(0,1)$ and the potential is
\begin{align}
\psi_w(w) = \log P_W(w)-\frac{1}{2}\hat V_ww^2+\left(\sqrt{\hat Q_w}\varsigma+u\hat m_w\right)w\ ;
\end{align}
and samely
\begin{align}
\int \prod_{a,k}\mathrm dh_k^ae^{\psi_h^{(n)}(h;\bar s)} &= \int\prod_{a,k}\mathrm dh_k^ae^{-\beta\bar s\sum_a \ell(yk_K^a)+\sum_{k=0}^{K-1}\left(\frac{1}{2}\hat Q_kh_k^TJh_k-\frac{1}{2}\hat V_kh_k^Th_k+y\hat m_k^Th_k\right)+\sum_{k<l}^{K-1}(\hat Q_{k,l}h_k^TJh_l-\hat V_{k,l}h_k^Th_l)} \\
& e^{-\frac{1}{2}(h_0-\sqrt\mu ym_w)^T\left(\frac{1}{V_w}I-\frac{Q_w}{V_w^2}J\right)(h_0-\sqrt\mu ym_w)-\frac{1}{2}n\frac{Q_w}{V_w}-\frac{1}{2}n\log V_w} \nonumber \\
& e^{-\frac{1}{2}\sum_{k,l}^K(h_k-c_kh_{k-1}-\lambda ym_{k-1})^T\left({(V^{-1})_{k-1,l-1}}I-(V^{-1}QV^{-1})_{k-1,l-1}J\right)(h_l-c_lh_{l-1}-\lambda ym_{l-1})-\frac{n}{2}\tr(V^{-1}Q)-\frac{n}{2}\log\det V} \nonumber \\
&= \mathbb E_{\xi,\chi,\zeta}\left(\int\prod_{k=0}^K\mathrm dh_ke^{\psi_h(h; \bar s)}\right)^n
\end{align}
where $\xi\sim\mathcal N(0,I_K)$, $\chi\sim\mathcal N(0,I_K)$, $\zeta\sim\mathcal N(0,1)$ and the potential is
\begin{align}
\psi_h(h; \bar s) &= -\beta\bar s\ell(yh_K)-\frac{1}{2}h_{<K}^T\hat Vh_{<K}+\left(\xi^T\hat Q^{1/2}+y\hat m^T\right)h_{<K} \\
&\quad\quad {}+\log\mathcal N\left(h_0\left|\sqrt\mu ym_w+\sqrt{Q_w}\zeta ; V_w \right.\right) +\log\mathcal N\left(h_{>0}\left|c\odot h_{<K}+\lambda ym+Q^{1/2}\chi ; V \right.\right)\ ; \nonumber
\end{align}
where $h_{>0}=(h_1,\ldots,h_K)\in\mathbb R^K$, $h_{<K}=(h_0,\ldots,h_{K-1})\in\mathbb R^K$, $c\odot h_{<K}=(c_1h_0,\ldots,c_Kh_{K-1})$, $\hat m=(\hat m_0,\ldots,\hat m_{K-1})\in\mathbb R^K$, $m=(m_0,\ldots,m_{K-1})\in\mathbb R^K$, $\hat Q$ and $\hat V$ are the $K\times K$ symmetric matrix filled with $(\hat Q_k)_{0\leq k\leq K-1}$ and $(\hat V_k)_{0\leq k\leq K-1}$ on the diagonal, and $(\hat Q_{k,l})_{0\leq k<l\leq K-1}$ and $(\hat V_{k,l})_{0\leq k<l\leq K-1}$ off the diagonal. We used that $\mathbb E_\zeta e^{-\frac{n}{2}\frac{Q_w}{V_w}\zeta^2}=e^{-\frac{n}{2}\frac{Q_w}{V_w}}$ in the limit $n\to 0$ to factorize $\sqrt{Q_w}\zeta$ and the same for $Q^{1/2}\chi$.

We pursue the computation:
\begin{align}
\mathbb EZ^n \propto &\, \int \mathrm d\hat m_w\mathrm dm_we^{Nn\hat m_wm_w}\mathrm d\hat Q_w\mathrm dQ_w\mathrm d\hat V_w\mathrm dV_we^{N\frac{n}{2}(\hat V_wV_w+\hat V_wQ_w-V_w\hat Q_w)}\prod_{k=0}^{K-1}\mathrm d\hat m_k\mathrm dm_ke^{Nn\hat m^Tm} \\
& \quad \prod_{k=0}^{K-1}\mathrm d\hat Q_k\mathrm dQ_k\mathrm d\hat V_k\mathrm dV_k\prod_{k<l}^{K-1}\mathrm d\hat Q_{k,l}\mathrm dQ_{k,l}\mathrm d\hat V_{k,l}\mathrm dV_{k,l}e^{N\frac{n}{2}\mathrm{tr}\left(\hat VV+\hat VQ-V\hat Q\right)} \nonumber \\
& \quad \left[\mathbb E_{u,\varsigma}\left(\int\mathrm dwe^{\psi_w(w)}\right)^n \right]^{N/\alpha}
\left[\mathbb E_{y,\xi,\chi,\zeta}\left(\int\prod_{k=0}^K\mathrm dh_ke^{\psi_h(h; s)}\right)^n \right]^{\rho N}
\left[\mathbb E_{y,\xi,\chi,\zeta}\left(\int\prod_{k=0}^K\mathrm dh_ke^{\psi_h(h; s')}\right)^n \right]^{(1-\rho)N} \nonumber \\
 :=&\ \int \mathrm d\Theta\mathrm d\hat\Theta e^{N\phi^{(n)}(\Theta,\hat\Theta)}\ .
\end{align}
where $\Theta=\{m_w, Q_w, V_w, m, Q, V\}$ and $\hat\Theta=\{\hat m_w, \hat Q_w, \hat V_w, \hat m, \hat Q, \hat V\}$ are the sets of the order parameters. We can now take the limit $N\to\infty$ thanks to Laplace's method.
\begin{align}
-\beta f\propto & \frac{1}{N}\frac{\partial}{\partial n}(n=0) \int\mathrm d\Theta\mathrm d\hat\Theta\,e^{N\phi^{(n)}(\Theta, \hat\Theta)} \\
= & \extr_{\Theta, \hat\Theta}\frac{\partial}{\partial n}(n=0)\phi^{(n)}(\Theta, \hat\Theta) \\
:= & \extr_{\Theta, \hat\Theta}\phi(\Theta, \hat\Theta)\ ,
\end{align}
where we extremize the following free entropy $\phi$:
\begin{align}
\phi &= \frac{1}{2}\left(\hat V_wV_w+\hat V_wQ_w-V_w\hat Q_w\right)-\hat m_wm_w+\frac{1}{2}\mathrm{tr}\left(\hat VV+\hat VQ-V\hat Q\right)-\hat m^Tm \\
 &\quad{}+\frac{1}{\alpha}\mathbb E_{u,\xi}\left(\log\int\mathrm dw\,e^{\psi_w(w)}\right)+\rho\mathbb E_{y,\xi,\zeta,\chi}\left(\log\int\prod_{k=0}^K\mathrm dh_k e^{\psi_h(h;s)}\right)+(1-\rho)\mathbb E_{y,\xi,\zeta,\chi}\left(\log\int\prod_{k=0}^K\mathrm dh_k e^{\psi_h(h;s')}\right)\ .  \nonumber
\end{align}

We take the limit $\beta\to\infty$. Later we will differentiate $\phi$ with respect to the order parameters or to $\bar s$ and these derivatives will simplify in that limit.
We introduce the measures
\begin{align}
\mathrm dP_w = \frac{\mathrm dw\,e^{\psi_w(w)}}{\int\mathrm dw\,e^{\psi_w(w)}} \quad,\quad
\mathrm dP_h = \frac{\prod_{k=0}^K\mathrm dh_k\,e^{\psi_h(h;\bar s=1)}}{\int\prod_{k=0}^K\mathrm dh_k\,e^{\psi_h(h;\bar s=1)}} \quad,\quad
\mathrm dP_h' = \frac{\prod_{k=0}^K\mathrm dh_k\,e^{\psi_h(h;\bar s=0)}}{\int\prod_{k=0}^K\mathrm dh_k\,e^{\psi_h(h;\bar s=0)}}\ .
\end{align}
We have to rescale the order parameters not to obtain a degenerated solution when $\beta\to\infty$ (we recall that, in $\psi_w$, $\log P_W(w)\propto\beta$). We take
\begin{align}
& \hat m_w\to\beta\hat m_w\ ,&& \hat Q_w\to\beta^2\hat Q_w\ ,&& \hat V_w\to\beta\hat V_w\ ,&& V_w\to\beta^{-1}V_w \\
& \hat m\to\beta\hat m\ ,&& \hat Q\to\beta^2\hat Q\ ,&& \hat V\to\beta\hat V\ ,&& V\to\beta^{-1}V
\end{align}
So we obtain that $f=-\phi$. Then $\mathrm dP_w$, $\mathrm dP_h$ and $\mathrm dP_h'$ are picked around their maximum and can be approximated by Gaussian measures. We define
\begin{align}
w^* = \argmax_w\psi_w(w)\ ,\quad h^* = \argmax_h\psi_h(h;\bar s=1)\ ,\quad h^{'*} = \argmax_h\psi_h(h;\bar s=0)\ .
\end{align}
Then we have the expected value of a function $g$ in $h$ $\mathbb E_{P_h}g(h)=g(h^*)$ and the covariance $\cov_{P_h}(h)=-\frac{1}{2}(\nabla\nabla\psi_h(h^*))^{-1}$ with $\nabla\nabla$ the Hessian; and similarly for $\mathrm dP_w$ and $\mathrm dP_h'$.

Last we compute the expected errors and accuracies. 
We differentiate the free energy $f$ with respect to $s$ and $s'$ to obtain that
\begin{align}
E_\mathrm{train} = \mathbb E_{y,\xi,\zeta,\chi}\ell(yh_K^*)\ ,\quad E_\mathrm{test} = \mathbb E_{y,\xi,\zeta,\chi}\ell(yh_K^{'*})\ . \label{eqApp:erreursM}
\end{align}
Augmenting $H$ with the observable $\frac{1}{|\hat R|}\sum_{i\in\hat R}\delta_{y_i=\sign h(w)_i}$ and following the same steps gives the expected accuracies
\begin{align}
\mathrm{Acc}_\mathrm{train} = \mathbb E_{y,\xi,\zeta,\chi}\delta_{y=\sign(h_K^*)}\ ,\quad \mathrm{Acc}_\mathrm{test} = \mathbb E_{y,\xi,\zeta,\chi}\delta_{y=\sign(h_K^{'*})}\ . \label{eqApp:précisionsM}
\end{align}

\subsection{Self-consistent equations}
The two above formula \eqref{eqApp:erreursM} and \eqref{eqApp:précisionsM} are valid only at the values of the order parameters that extremize the free entropy. We seek the extremizer of $\phi$. The extremality condition $\nabla_{\Theta, \hat\Theta}\phi=0$ gives the following self-consistent equations:
\begin{align}
& m_w = \frac{1}{\alpha}\mathbb E_{u,\varsigma}\,uw^* \quad\quad m=\mathbb E_{y,\xi,\zeta,\chi}\,y\left(\rho h_{<K}^*+(1-\rho)h_{<K}^{'*}\right) \\
& Q_w=\frac{1}{\alpha}\mathbb E_{u,\varsigma}(w^*)^2 \quad\quad Q=\mathbb E_{y,\xi,\zeta,\chi}\left(\rho (h_{<K}^*)^{\otimes 2}+(1-\rho)(h_{<K}^{'*})^{\otimes 2}\right) \\
& V_w=\frac{1}{\alpha}\frac{1}{\sqrt{\hat Q_w}}\mathbb E_{u,\varsigma}\,\varsigma w^* \quad\quad V=\mathbb E_{y,\xi,\zeta,\chi}\left(\rho\cov_{P_h}(h_{<K})+(1-\rho)\cov_{P_h'}(h_{<K})\right) \\
& \hat m_w=\frac{\sqrt\mu}{V_w}\mathbb E_{y,\xi,\zeta,\chi}\,y\left(\rho(h^*_0-\sqrt\mu ym_w)+(1-\rho)(h_0^{'*}-\sqrt\mu ym_w)\right) \\
& \hat Q_w=\frac{1}{V_w^2}\mathbb E_{y,\xi,\zeta,\chi}\left(\rho(h_0^*-\sqrt\mu ym_w-\sqrt Q_w\zeta)^2+(1-\rho)(h_0^{'*}-\sqrt\mu ym_w-\sqrt Q_w\zeta)^2\right) \\
& \hat V_w=\frac{1}{V_w}-\frac{1}{V_w^2}\mathbb E_{y,\xi,\zeta,\chi}\left(\rho\cov_{P_h}(h_0)+(1-\rho)\cov_{P_h}(h_0)\right) \\
& \hat m=\lambda V^{-1}\mathbb E_{y,\xi,\zeta,\chi}\,y\left(\rho(h_{>0}^*-c\odot h_{<K}^*-\lambda ym)+(1-\rho)(h_{>0}^{'*}-c\odot h_{<K}^{'*}-\lambda ym)\right) \\
& \hat Q=V^{-1}\mathbb E_{y,\xi,\zeta,\chi}\left(\rho(h_{>0}^*-c\odot h_{<K}^*-\lambda ym-Q^{1/2}\chi)^{\otimes 2}+(1-\rho)(h_{>0}^{'*}-c\odot h_{<K}^{'*}-\lambda ym-Q^{1/2}\chi)^{\otimes 2}\right)V^{-1} \\
& \hat V=V^{-1}-V^{-1}\mathbb E_{y,\xi,\zeta,\chi}\left(\rho\cov_{P_h}(h_{>0}-c\odot h_{<K})+(1-\rho)\cov_{P_h'}(h_{>0}-c\odot h_{<K})\right)V^{-1}
\end{align}
We introduced the covariance $\cov_P(x)=\mathbb E_P(xx^T)-\mathbb E_P(x)\mathbb E_P(x^T)$ and the tensorial product $x^{\otimes 2}=xx^T$. We used Stein's lemma to simplify the differentials of $Q^{1/2}$ and $\hat Q^{1/2}$ and to transform the expression of $\hat V_w$ into a more accurate expression for numerical computation in terms of covariance. We used the identities $2x^TQ^{1/2}\frac{\partial Q^{1/2}}{\partial E_{k,l}}x=x^TE_{k,l}x$ and $-x^TV\frac{\partial V^{-1}}{\partial E_{k,l}}Vx=x^TE_{k,l}x$ for any element matrix $E_{k,l}$ and for any vector $x$. We have also that $\nabla_V\log\det V=V^{-1}$, considering its comatrix. Last we kept the first order in $\beta$ with the approximations $Q+V\approx Q$ and $\hat Q-\hat V\approx\hat Q$.

These self-consistent equations are reproduced in the main part \ref{sec:discrAsymptChara}.

\subsection{Solution for ridge regression}
\label{sec:accDiscrQuadC0rInf}
We take quadratic $\ell$ and $\gamma$. Moreover we assume there is no residual connections $c=0$; this simplifies largely the analysis in the sense that the covariances of $h$ under $P_h$ or $P_h'$ become diagonal. We have
\begin{align}
&\cov_{P_h}(h) = \mathrm{diag}\left( \frac{V_w}{1+V_w\hat V_0}, \frac{V_0}{1+V_0\hat V_1}, \ldots, \frac{V_{K-2}}{1+V_{K-2}\hat V_{K-1}}, \frac{V_{K-1}}{1+V_{K-1}}\right) \\
&\cov_{P_h'}(h) = \mathrm{diag}\left( \frac{V_w}{1+V_w\hat V_0}, \frac{V_0}{1+V_0\hat V_1}, \ldots, \frac{V_{K-2}}{1+V_{K-2}\hat V_{K-1}}, V_{K-1}\right) \\
&h^* = \cov_{P_h}(h)\left(
\begin{pmatrix}
\hat Q^{1/2}\xi+y\hat m \\ y
\end{pmatrix}
+\begin{pmatrix}
\frac{1}{V_w}(\sqrt\mu ym_w+\sqrt{Q_w}\zeta) \\ V^{-1}\left(\lambda ym+Q^{1/2}\chi\right)
\end{pmatrix}\right) \\
&h^{'*} = \cov_{P_h'}(h)\left(
\begin{pmatrix}
\hat Q^{1/2}\xi+y\hat m \\ 0
\end{pmatrix}
+\begin{pmatrix}
\frac{1}{V_w}(\sqrt\mu ym_w+\sqrt{Q_w}\zeta) \\ V^{-1}\left(\lambda ym+Q^{1/2}\chi\right)
\end{pmatrix}\right)
\end{align}
where diag means the diagonal matrix with the given diagonal. We packed elements into block vectors of size $K+1$. The self-consistent equations can be explicited:
\begin{align}
& m_w = \frac{1}{\alpha}\frac{\hat m_w}{r+\hat V_w} &&\quad V_w = \frac{1}{\alpha}\frac{1}{r+\hat V_w} &&\quad Q_w = \frac{1}{\alpha}\frac{\hat Q_w+\hat m_w^2}{(r+\hat V_w)^2} \\
& m = V\left(\hat m+\left(\begin{smallmatrix}\sqrt\mu\frac{m_w}{V_w} \\ \lambda V_{<K-1}^{-1}m_{<K-1}\end{smallmatrix}\right)\right)
&&\quad V = \mathrm{diag}\left(\begin{smallmatrix} \frac{V_w}{1+V_w\hat V_0}, \frac{V_0}{1+V_0\hat V_1}, \ldots, \frac{V_{K-2}}{1+V_{K-2}\hat V_{K-1}} \end{smallmatrix}\right)
&&  \\
& \hat m_w =\frac{\sqrt\mu}{V_w}(m_0-\sqrt\mu m_w) &&\quad \hat V_w = \frac{\hat V_0}{1+V_w\hat V_0}
&& \\
& \hat m = \lambda\hat V\left(\left(\begin{smallmatrix} \hat V_{>0}^{-1}\hat m_{>0} \\ 1\end{smallmatrix}\right)-\lambda m\right)
&&\quad \hat V = \mathrm{diag}\left(\begin{smallmatrix} \frac{\hat V_1}{1+V_0\hat V_1},\ldots,\frac{\hat V_{K-1}}{1+V_{K-2}\hat V_{K-1}},\frac{\rho}{1+V_{K-1}}\end{smallmatrix}\right)
&&
\end{align}
and
\begin{align}
& \hat Q_w = \frac{V_0^2}{V_w^2}\hat Q_{0,0}+\left(\frac{V_0}{V_w}-1\right)^2\frac{Q_w}{V_w^2}+\frac{\hat m_w^2}{\mu} \\
& Q = V\left(\hat Q+\left(\begin{smallmatrix}\frac{Q_w}{V_w^2} & 0 \\ 0 & V_{<K-1}^{-1}Q_{<K-1}V_{<K-1}^{-1}\end{smallmatrix}\right)\right)V+m^{\otimes 2} \\
& \hat Q = \hat V\left(\left(\begin{smallmatrix}\hat V_{>0}^{-1}\hat Q_{>0}\hat V_{>0}^{-1} & 0 \\ 0 & 0\end{smallmatrix}\right)
+\rho\left(\begin{smallmatrix}I_{K-1} & 0 \\ 0 & \rho^{-1}\end{smallmatrix}\right)Q\left(\begin{smallmatrix}I_{K-1} & 0 \\ 0 & \rho^{-1}\end{smallmatrix}\right)
+(1-\rho)\left(\begin{smallmatrix}I_{K-1} & 0 \\ 0 & 0\end{smallmatrix}\right)Q\left(\begin{smallmatrix}I_{K-1} & 0 \\ 0 & 0\end{smallmatrix}\right) \right)\hat V \nonumber \\
& \qquad {}+ \rho\left(\begin{smallmatrix} \frac{1}{\lambda}\hat m_{<K-1} \\ \frac{1}{\lambda\rho}\hat m_{K-1}\end{smallmatrix}\right)^{\otimes 2} + (1-\rho)\left(\begin{smallmatrix} \frac{1}{\lambda}\hat m_{<K-1} \\ 0\end{smallmatrix}\right)^{\otimes 2}
\end{align}
We used the notations $m_{<K-1}=(m_{k})_{0\leq k<K-1}$, $\hat m_{<K-1}=(\hat m_{k})_{0\leq k<K-1}$, $\hat m_{>0}=(\hat m_{k})_{0<k\leq K-1}$, $Q_{<K-1}=(Q_{k,l})_{0\leq k,l<K-1}$, $Q_{>0}=(Q_{k,l})_{0< k,l\leq K-1}$, $V_{<K-1}=(V_{k,l})_{0\leq k,l<K-1}$ and $V_{>0}=(V_{k,l})_{0< k,l\leq K-1}$.
We simplified the equations by combining the expressions of $V$, $\hat V$, $m$ and $\hat m$: the above system of equations is equivalent to the generic equations only at the fixed-point. The expected losses and accuracies are
\begin{align}
& E_\mathrm{train} = \frac{1}{2\rho}\hat Q_{K-1,K-1} && E_\mathrm{test} = \frac{1}{2\rho}(1+V_{K-1,K-1})^2\hat Q_{K-1,K-1} \\
& \mathrm{Acc}_\mathrm{train} = \frac{1}{2}\left(1+\mathrm{erf}\left(\frac{V_{K-1,K-1}+\lambda m_{K-1}}{\sqrt{2Q_{K-1,K-1}}}\right)\right) && \mathrm{Acc}_\mathrm{test} = \frac{1}{2}\left(1+\mathrm{erf}\left(\frac{\lambda m_{K-1}}{\sqrt{2Q_{K-1,K-1}}}\right)\right) \ .
\end{align}

To obtain a simple solution we take the limit $r\to\infty$. The solution of this system is then
\begin{align}
& m_w = \frac{\rho\sqrt\mu}{\alpha r}\lambda^K &&\quad V_w = \frac{1}{\alpha r} &&\quad Q_w = \frac{1}{\alpha r^2}\left(\rho+\rho^2\mu\lambda^{2K}+\rho^2\sum_{l=1}^{K}\lambda^{2l}\right) \\
& m_k = \frac{\rho}{\alpha r}\left(\mu\lambda^{K+k}+\sum_{l=0}^k\lambda^{K-k+2l}\right)
&&\quad V_{k,k} = \frac{1}{\alpha r} &&\quad \\
& \hat m_w = \rho\sqrt\mu\lambda^K &&\quad \hat V_w = \rho &&\quad \hat Q_w = \rho+\rho^2\sum_{l=1}^{K}\lambda^{2l} \\
& \hat m_k = \rho\lambda^{K-k}
&&\quad \hat V_{k,k} = \rho &&\quad \hat Q_{k,k} = \rho+\rho^2\sum_{l=1}^{K-1-k}\lambda^{2l}
\end{align}
and
\begin{align}
& Q_{k,k} = \frac{\rho}{\alpha^2r^2}\left(\alpha\left(1+\rho\mu\lambda^{2K}+\rho\sum_{l=1}^{K}\lambda^{2l}\right) + \sum_{m=0}^{k}\left(1+\rho\sum_{l=1}^{K-1-m}\lambda^{2l}+\rho\left(\mu\lambda^{K+m}+\sum_{l=0}^m\lambda^{K-m+2l}\right)^2\right) \right)
\end{align}
We did not precise the off-diagonal parts of $Q$ and $\hat Q$ since they do not enter in the computation of the losses and accuracies. The expressions for $m$ and $Q$ are reproduced in the main part \ref{sec:discrAnalyticalSol}.

%% file: appendix_sym.tex
\section{Asymptotic characterization of the continuous GCN, for asymmetric and symmetrized graphs}
\label{sec:appendixContinuousGCN}
In this part we derive the asymptotic characterization of the continuous GCN for both the asymmetric and symmetrized graphs $\tilde A$ and $\tilde A^\mathrm{s}$. As shown in the main section \ref{sec:continuousGCN} this architecture is particularly relevant since it can be close to the Bayes-optimality.

We start by discretizing the GCN and deriving its free energy and the self-consistent equations on its order parameters. Then we take the continuous limit $K\to\infty$, jointly with an expansion around large regularization $r$. The derivation of the free energy and of the self-consistent equations follows the same steps as in the previous section \ref{sec:appendixDiscreteGCN}; in particular for the asymmetric case the expressions are identical up to the point where the continuous limit is taken.

To deal with both cases, asymmetric or symmetrized, we define $(\delta_\mathrm{e}, \tilde A^\mathrm{e}, A^\mathrm{g,e}, \lambda^\mathrm{e}, \Xi^\mathrm{e})\in\{(0, \tilde A, A^\mathrm{g}, \lambda, \Xi), (1, \tilde A^\mathrm{s}, A^\mathrm{g,s}, \lambda^\mathrm{s}, \Xi^\mathrm{s})\}$. In particular $\delta_\mathrm{e}=0$ for the asymmetric and $\delta_\mathrm{e}=1$ for the symmetrized. We remind that $\tilde A^\mathrm{s}$ is the symmetrized $\tilde A$ with effective signal $\lambda^\mathrm{s}=\sqrt 2\lambda$. $\tilde A^\mathrm{e}$ admits the following Gaussian equivalent \cite{lesieur2017constrained,cSBM18,shi2022statistical}:
\begin{equation}
\tilde A^\mathrm{e} \approx A^\mathrm{g,e}=\frac{\lambda^\mathrm{e}}{\sqrt N}yy^T+\Xi^\mathrm{e} \ ,
\end{equation}
with $(\Xi)_{ij}$ i.i.d. for all $i$ and $j$ while $\Xi^\mathrm{s}$ is taken from the Gaussian orthogonal ensemble.

\subsection{Derivation of the free energy}
The continuous GCN is defined by the output function
\begin{equation}
h(w)=e^{\frac{t}{\sqrt N}\tilde A^\mathrm{e}}\frac{1}{\sqrt N}Xw\ .
\end{equation}
Its discretization at finite $K$ is
\begin{equation}
h(w) = h_K\ ,\qquad h_k = \left(I_N+\frac{t}{\sqrt N}\tilde A^\mathrm{e}\right)h_{k-1}\ ,\qquad h_0 = \frac{1}{\sqrt N}Xw\ .
\end{equation}
It can be mapped to the discrete GCN of the previous section \ref{sec:appendixDiscreteGCN} by taking $c=t/K$.

The free energy is $-\beta Nf = \partial_n\mathbb E_{u, \Xi^\mathrm{e}, W, y}Z^n(n=0)$ where the partition function is
\begin{align}
Z &= \int\prod_\nu^M\mathrm dw_\nu e^{-\beta r\gamma(w_\nu)} e^{-\beta s\sum_{i\in R}\ell(y_ih(w)_i)-\beta s'\sum_{i\in R'}\ell(y_ih(w)_i)}\ .
\end{align}
The expectation of the replicated partition function is
\begin{align}
& \mathbb EZ^n \propto \, \mathbb E_{u, \Xi^\mathrm{e}, W, y}\int \prod_a^n\prod_\nu^M\mathrm dw_\nu^ae^{-\beta r\gamma(w_\nu^a)}\prod_a^n\prod_i^N\prod_{k=0}^K\mathrm dh_{i,k}^a\mathrm dq_{i,k}^a e^{-\beta s\sum_{a,i\in R}\ell(y_ih_{i,K}^a)-\beta s'\sum_{a,i\in R'}\ell(y_ih_{i,K}^a)} \nonumber \\
 & \qquad\qquad e^{\sum_{a,i}\sum_{k=1}^K\mathrm iq_{i,k}^a\left(\frac{K}{t}h_{i,k}^a-\frac{1}{\sqrt N}\sum_j\left(\sqrt N\frac{K}{t}\delta_{i,j}+\frac{\lambda^\mathrm{e}}{\sqrt N}y_iy_j+\Xi^\mathrm{e}_{ij}\right)h_{j,k-1}^a\right)+\sum_{a,i}\mathrm iq_{i,0}^a\left(h_{i,0}^a-\frac{1}{\sqrt N}\sum_\nu\left(\sqrt\frac{\mu}{N}y_ju_\nu+W_{j\nu}\right)w_\nu^a\right)} \\
 & = \mathbb E_{u, y}\int \prod_{a,\nu}\mathrm dw_\nu^ae^{-\beta r\gamma(w_\nu^a)}\prod_{a,i,k}\mathrm dh_{i,k}^a\mathrm dq_{i,k}^a e^{-\beta s\sum_{a,i\in R}\ell(y_ih_{i,K}^a)-\beta s'\sum_{a,i\in R'}\ell(y_ih_{i,K}^a)+\mathrm i\sum_{a,i,k>0}q_{i,k}^a\left(\frac{K}{t}(h_{i,k}^a-h_{i,k-1}^a)-\frac{\lambda^\mathrm{e}}{\sqrt N}y_i\sum_jy_jh_{j,k-1}^a\right)} \nonumber \\
 & \qquad e^{-\frac{1}{2N}\sum_{i,j}\sum_{a,b}\sum_{k>0,l>0}(q_{i,k}^ah_{j,k-1}^aq_{i,l}^bh_{j,l-1}^b + \delta_\mathrm{e}q_{i,k}^ah_{j,k-1}^aq_{j,l}^bh_{i,l-1}^b)-\mathrm i\sum_{a,i}\frac{\sqrt\mu}{N}y_iq_{i,0}^a\sum_\nu u_\nu w_\nu^a-\frac{1}{2N}\sum_{i,\nu,a,b}q_{i,0}^aq_{i,0}^bw_\nu^aw_\nu^b}\ .
\end{align}
Compared to part \ref{sec:appendixDiscreteGCN}, because of the symmetry the expectation over $\Xi^\mathrm{s}$ gives an additional cross-term. We symmetrized $\sum_{i<j}$ by neglecting the diagonal terms. We introduce new order parameters between $h$ and its conjugate $q$. We set for all $a$ and $b$ and for $0<k\le K$ and $0<l\le K$
\begin{align}
& m_w^a = \frac{1}{N}\sum_\nu u_\nu w_\nu^a\ , \quad Q_w^{ab} = \frac{1}{N}\sum_\nu w_\nu^a w_\nu^b\ , \\
& m_k^a = \frac{1}{N}\sum_jy_jh_{j,k-1}^a\ , \quad Q_{h,kl}^{ab} = \frac{1}{N}\sum_jh_{j,k-1}^ah_{j,l-1}^b\ ,\\
& Q_{q,kl}^{ab} =\frac{1}{N}\sum_jq_{j,k}^aq_{j,l}^b\ ,\quad Q_{qh,kl}^{ab} =\frac{1}{N}\sum_jq_{j,k}^ah_{j,l-1}^b\ .
\end{align} 
We introduce these quantities via $\delta$-Dirac functions. Their conjugates are $\hat m_w^a$, $\hat Q_w^{ab}$, $\hat V_w^{ab}$, $\hat m^a$, $\hat Q^{ab}$ and $\hat V^{ab}$. We factorize the $\nu$ and $i$ indices.
We leverage the replica-symmetric ansatz. We assume that for all $a$ and $b$
{\small
\begin{equation}
m_w^a = m_w\ ,\qquad \hat m_w^a = -\hat m_w\ ,\qquad m_k^a = m_k\ ,\qquad \hat m_k^a = -\hat m_k
\end{equation}
and
\begin{align}
& Q_w^{ab} = Q_w+V_w\delta_{a,b}\ , && \hat Q_w^{ab} = -\hat Q_w+\frac{1}{2}(\hat V_w+\hat Q_w)\delta_{a,b}\ ,  && \\
& Q_{h,kl}^{ab} = Q_{h,kl}+V_{h,kl}\delta_{a,b}\ , && \hat Q_{h,kk}^{ab} = -\hat Q_{h,kk}+\frac{1}{2}(\hat V_{h,kk}+\hat Q_{h,kk})\delta_{a,b}\ , && \hat Q_{h,kl}^{ab} = -\hat Q_{h,kl}+\hat V_{h,kl}\delta_{a,b}\ ,\\
& Q_{q,kl}^{ab} = Q_{q,kl}+V_{q,kl}\delta_{a,b}\ , && \hat Q_{q,kk}^{ab} = -\hat Q_{q,kk}+\frac{1}{2}(\hat V_{q,kk}+\hat Q_{q,kk})\delta_{a,b}\ , && \hat Q_{q,kl}^{ab} = -\hat Q_{q,kl}+\hat V_{q,kl}\delta_{a,b}\ , \\
& Q_{qh,kl}^{ab} = Q_{qh,kl}+V_{qh,kl}\delta_{a,b}\ , && \hat Q_{qh,kk}^{ab} = -\hat Q_{qh,kk}+\hat V_{qh,kk}\delta_{a,b}\ , && \hat Q_{qh,kl}^{ab} = -\hat Q_{qh,kl}+\hat V_{qh,kl}\delta_{a,b}\ .
\end{align}
}
$\delta_{a,b}$ is a Kronecker delta between $a$ and $b$. $Q_h$, $Q_q$, $Q_{qh}$, $V_h$, $V_q$, $V_{qh}$, and their conjugates, written with a hat, are $K\times K$ matrices that we pack into the following $2K\times 2K$ symmetric block matrices:
\begin{align}
& Q=\left(\begin{smallmatrix}Q_q & Q_{qh} \\ Q_{qh}^T & Q_h\end{smallmatrix}\right)\ , && V=\left(\begin{smallmatrix}V_q & V_{qh} \\ V_{qh}^T & V_h\end{smallmatrix}\right)\ ,\\
& \hat Q=\left(\begin{smallmatrix}\hat Q_q & \hat Q_{qh} \\ \hat Q_{qh}^T & \hat Q_h\end{smallmatrix}\right)\ , && \hat V=\left(\begin{smallmatrix}\hat V_q & \hat V_{qh} \\ \hat V_{qh}^T & \hat V_h\end{smallmatrix}\right)\ .
\end{align}
We obtain that
\begin{align}
\mathbb EZ^n \propto &\, \int \mathrm d\hat Q_w\mathrm d\hat V_w\mathrm dQ_w\mathrm dV_w\mathrm d\hat Q\mathrm d\hat V\mathrm dQ\mathrm dVe^{\frac{nN}{2}(\hat V_wV_w+\hat V_wQ_w-V_w\hat Q_w+\mathrm{tr}(\hat VV+\hat VQ-V\hat Q)-\mathrm{tr}(V_qV_h+V_qQ_h+V_hQ_q+\delta_\mathrm{e}V_{qh}^2+2\delta_\mathrm{e}V_{qh}Q_{qh}))} \nonumber \\
& \quad\quad \mathrm d\hat m_w\mathrm dm_w\mathrm d\hat m_\sigma\mathrm dm_\sigma e^{-nN(\hat m_wm_w+\hat m_\sigma m_\sigma)} \left[\mathbb E_u\int\prod_a\mathrm dw^a\,e^{\psi_w^{(n)}(w)} \right]^{N/\alpha} \nonumber \\
 & \quad\quad \left[\mathbb E_y\int\prod_{a,k}\mathrm dh_k^a\mathrm dq_k^a e^{\psi_h^{(n)}(h,q;s)} \right]^{\rho N}
\left[\mathbb E_y\int\prod_{a,k}\mathrm dh_k^a\mathrm dq_k^a e^{\psi_h^{(n)}(h,q;s')} \right]^{(1-\rho)N} \\
 :=&\ \int \mathrm d\Theta\mathrm d\hat\Theta e^{N\phi^{(n)}(\Theta,\hat\Theta)}\ ,
\end{align}
with $\Theta=\{m_w, Q_w, V_w, m, Q, V\}$ and $\hat\Theta=\{\hat m_w, \hat Q_w, \hat V_w, \hat m, \hat Q, \hat V\}$ the sets of order parameters and
\begin{align}
\psi_w^{(n)}(w) &= -\beta r\sum_a\gamma(w^a)-\frac{1}{2}\hat V_w\sum_a(w^a)^2+\hat Q_w\sum_{a,b}w^aw^b+u\hat m_w\sum_aw^a \\
\psi_h^{(n)}(h,q;\bar s) &= -\beta\bar s\sum_a\ell(yh_K^a)-\frac{1}{2}V_w\sum_a(q_0^a)^2+Q_w\sum_{a,b}q_0^aq_0^b-\frac{1}{2}\sum_a
\left(\begin{smallmatrix}q_{>0}^a \\ h_{<K}^a\end{smallmatrix}\right)^T
\hat V\left(\begin{smallmatrix}q_{>0}^a \\ h_{<K}^a\end{smallmatrix}\right)
+\sum_{a,b}\left(\begin{smallmatrix}q_{>0}^a \\ h_{<K}^a\end{smallmatrix}\right)^T\hat Q\left(\begin{smallmatrix}q_{>0}^b \\ h_{<K}^b\end{smallmatrix}\right) \nonumber \\
&\qquad{}+y\hat m^T\sum_ah_{<K}^a+\mathrm{i}\sum_a(q_{>0}^a)^T\left(\frac{K}{t}(h_{>0}^a-h_{<K}^a)-\lambda^\mathrm{e} ym^a\right)-\mathrm i\sqrt\mu ym_w\sum_aq_0^a
\end{align}
$u$ is a scalar standard Gaussian and $y$ is a scalar Rademacher variable. We use the notation $q_{>0}^a\in\mathbb R^K$ for $(q_k^a)_{k>0}$ and similarly as to $h_{>0}^a$ and $h_{<K}^a=(h_k^a)_{k<K}$. We packed them into vectors of size $2K$.

We take the limit $N\to\infty$ thanks to Laplace's method.
\begin{align}
-\beta f\propto & \frac{1}{N}\frac{\partial}{\partial n}(n=0) \int\mathrm d\Theta\mathrm d\hat\Theta\,e^{N\phi^{(n)}(\Theta, \hat\Theta)} \\
= & \extr_{\Theta, \hat\Theta}\frac{\partial}{\partial n}(n=0)\phi^{(n)}(\Theta, \hat\Theta) \\
:= & \extr_{\Theta, \hat\Theta}\phi(\Theta, \hat\Theta)\ ,
\end{align}
where we extremize the following free entropy $\phi$:
\begin{align}
\phi =& \frac{1}{2}(V_w\hat V_w+\hat V_wQ_w-V_w\hat Q_w)+\frac{1}{2}\tr(V_q\hat V_q+\hat V_qQ_q-V_q\hat Q_q)+\frac{1}{2}\tr(V_h\hat V_h+\hat V_hQ_h-V_h\hat Q_h) \\
&{}+\tr(V_{qh}\hat V_{qh}^T+\hat V_{qh}Q_{qh}^T-V_{qh}\hat Q_{qh}^T)-\frac{1}{2}\tr(V_qV_h+V_qQ_h+Q_qV_h+\delta_\mathrm{e}V_{qh}^2+2\delta_\mathrm{e}V_{qh}Q_{qh})-m_w\hat m_w-m^T\hat m \nonumber \\
&{}+\mathbb E_{u,\varsigma}\int\mathrm dw\,e^{\psi_w(w)}+\rho\mathbb E_{y,\zeta,\chi}\int\mathrm dq\mathrm dh\,e^{\psi_{qh}(q,h;s)}+(1-\rho)\mathbb E_{y,\zeta,\chi}\int\mathrm dq\mathrm dh\,e^{\psi_{qh}(q,h;s')}\ . \nonumber
\end{align}
We factorized the replica and took the derivative with respect to $n$ by introducing independent standard Gaussian random variables $\varsigma\in\mathbb R$, $\zeta=\left(\begin{smallmatrix}\zeta_q\\ \zeta_h\end{smallmatrix}\right)\in\mathbb R^{2K}$ and $\chi\in\mathbb R$. The potentials are
\begin{align}
\psi_w(w) =& -\beta r\gamma(w)-\frac{1}{2}\hat V_ww^2+\left(\sqrt{\hat Q_w}\varsigma+u\hat m_w\right)w \\
\psi_{qh}(q,h;\bar s) =& -\beta\bar s\ell(yh_K)-\frac{1}{2}V_wq_0^2-\frac{1}{2}\left(\begin{smallmatrix}q_{>0}\\ h_{<K}\end{smallmatrix}\right)^T
\hat V \left(\begin{smallmatrix}q_{>0}\\ h_{<K}\end{smallmatrix}\right)
+\left(\begin{smallmatrix}q_{>0}\\ h_{<K}\end{smallmatrix}\right)^T
\hat Q^{1/2}\left(\begin{smallmatrix}\zeta_q\\ \zeta_h\end{smallmatrix}\right)\\
&{}+yh_{<K}^T\hat m+\mathrm iq^T\left(\left(\begin{smallmatrix}1/K & \\ & I/t\end{smallmatrix}\right)Dh
-\left(\begin{smallmatrix}y\sqrt\mu m_w+\sqrt{Q_w}\chi\\ y\lambda^\mathrm{e} m\end{smallmatrix}\right)\right) \nonumber
\end{align}

We already extremize $\phi$ with respect to $Q$ and $V$ to obtain the following equalities:
\begin{align}
& \hat V_q=V_h\ ,&& V_q=\hat V_h\ ,&& \hat V_{qh}=\delta_\mathrm{e}V_{qh}^T\ ,\\
& \hat Q_q=-Q_h\ ,&& Q_q=-\hat Q_h\ ,&& \hat Q_{qh}=-\delta_\mathrm{e}Q_{qh}^T\ .
\end{align}
In particular this shows that in the asymmetric case where $\delta_\mathrm{e}=0$ one has $\hat V_{qh}=\hat Q_{qh}=0$ and as a consequence $V_{qh}=Q_{qh}=0$; and we recover the potential $\psi_h$ previously derived in part \ref{sec:appendixDiscreteGCN}.

We assume that $\ell$ is quadratic so $\psi_{qh}$ can be written as the following quadratic potential. Later we will take the limit $r\to\infty$ where $h$ is small and where $\ell$ can effectively be expanded around $0$ as a quadratic potential.
\begin{align}
\psi_{qh}(q,h;\bar s) &= -\frac{1}{2}\left(\begin{smallmatrix}q\\ h\end{smallmatrix}\right)^T \left(\begin{smallmatrix}G_q & -\mathrm iG_{qh}\\ -\mathrm iG_{qh}^T & G_h\end{smallmatrix}\right) \left(\begin{smallmatrix}q\\ h\end{smallmatrix}\right) +\left(\begin{smallmatrix}q\\ h\end{smallmatrix}\right)^T\left(\begin{smallmatrix}-\mathrm iB_q\\ B_h\end{smallmatrix}\right)
\end{align}
with
\begin{align}
& G_q = \left(\begin{smallmatrix}V_w & 0 \\ 0 & \hat V_q\end{smallmatrix}\right)\ , \qquad G_h = \left(\begin{smallmatrix}\hat V_h & 0 \\ 0 & \beta\bar s\end{smallmatrix}\right)\ , \qquad G_{qh} = \left(\begin{smallmatrix}1/K & 0 \\ 0 & I_K/t\end{smallmatrix}\right)D+\left(\begin{smallmatrix} 0 & 0\\ \mathrm i\hat V_{qh}& 0 \end{smallmatrix}\right)\ , \qquad D = K\left(\begin{smallmatrix}1 & & & 0 \\ -1 & \ddots & & \\ & \ddots & \ddots & \\ 0 & & -1 & 1 \end{smallmatrix}\right)\ , \\
& B_q = \left(\begin{smallmatrix}\sqrt{Q_w}\chi \\ \mathrm i\left(\hat Q^{1/2}\zeta\right)_q\end{smallmatrix}\right) + y\left(\begin{smallmatrix}\sqrt\mu m_w \\ \lambda^\mathrm{e} m\end{smallmatrix}\right)\ , \qquad B_h = \left(\begin{smallmatrix}\left(\hat Q^{1/2}\zeta\right)_h \\ 0\end{smallmatrix}\right) + y\left(\begin{smallmatrix}\hat m \\ \beta\bar s\end{smallmatrix}\right)\ , \qquad \left(\begin{smallmatrix}(\hat Q^{1/2}\zeta)_q\\ (\hat Q^{1/2}\zeta)_h\end{smallmatrix}\right) = \left(\begin{smallmatrix}\hat Q_q & \hat Q_{qh}\\ \hat Q_{qh}^T & \hat Q_h\end{smallmatrix}\right)^{1/2}\left(\begin{smallmatrix}\zeta_q\\ \zeta_h\end{smallmatrix}\right)\ .
\end{align}
$G_q$, $G_h$, $G_{qh}$ and $D$ are in $\mathbb R^{(K+1)\times(K+1)}$. $D$ is the discrete derivative. $B_q$, $B_h$ and $\left(\begin{smallmatrix}q\\ h\end{smallmatrix}\right)$ are in $\mathbb R^{2(K+1)}$.
We can marginalize $e^{\psi_{qh}}$ over $q$:
\begin{align}
& \int\mathrm dq\mathrm dh\,e^{\psi_{qh}(q,h;\bar s)} = \int\mathrm dh\,e^{\psi_h(h;\bar s)} \\
& \psi_h(h;\bar s) = -\frac{1}{2}h^TG_hh+h^TB_h-\frac{1}{2}\log\det G_q -\frac{1}{2}(G_{qh}h-B_q)^TG_q^{-1}(G_{qh}h-B_q) \\
&\quad = -\frac{1}{2}h^TGh+h^T\left(B_h+D_{qh}^TG_0^{-1}B\right)-\frac{1}{2}\log\det G_q\ ,\label{eqApp:psiHCont}
\end{align}
where we set
\begin{align}
G &= G_h+D_{qh}^TG_0^{-1}D_{qh}\ , \\
G_0 &=\left(\begin{smallmatrix}K^2V_w & 0 \\ 0 & t^2V_h\end{smallmatrix}\right)\ ,\\
D_{qh} &=D-t\left(\begin{smallmatrix} 0 & 0\\ -\mathrm i\delta_\mathrm{e}V_{qh}^T & 0 \end{smallmatrix}\right)\ ,\\
B &= \left(\begin{smallmatrix}K\sqrt{Q_w}\chi \\ \mathrm it\left(\hat Q^{1/2}\zeta\right)_q\end{smallmatrix}\right) + y\left(\begin{smallmatrix}K\sqrt\mu m_w \\ \lambda^\mathrm{e}tm\end{smallmatrix}\right)\ .
\end{align}
Eq. \eqref{eqApp:psiHCont} is the potential eq.~\eqref{eq:psiHCont} given in the main part, up to a term independent of $h$.

We take the limit $\beta\to\infty$. As before we introduce the measures $\mathrm dP_w$, $\mathrm dP_{qh}$ and $\mathrm dP_{qh}'$, $\mathrm dP_h$ and $\mathrm dP_h'$ whose unnormalized densities are $e^{\psi_\mathrm{w}(w)}$, $e^{\psi_{qh}(h,q;s)}$, $e^{\psi_{qh}(h,q;s')}$, $e^{\psi_h(h;s)}$ and $e^{\psi_h(h;s')}$.
We use Laplace's method to evaluate them. We have to rescale the order parameters not to obtain a degenerated solution. We take
\begin{align}
& m_w\to m_w\ , && Q_w\to Q_w\ , && V_w\to V_w/\beta\ ,\\
& \hat m_w\to\beta\hat m_w\ , && \hat Q_w\to\beta^2\hat Q_w\ , && \hat V_w\to\beta\hat V_w\ ,\\
& m\to m\ , && Q_h\to Q_h\ , && V_h\to V_h/\beta\ ,\\
& \hat m\to\beta\hat m\ , && \hat Q_h\to\beta^2\hat Q_h\ , && \hat V_h\to\beta\hat V_h\ ,\\
& && Q_{qh}\to\beta Q_{qh}\ , && V_{qh}\to V_{qh}\ .
\end{align}
We take this scaling for $Q_{qh}$ and $V_{qh}$ because we want $D_{qh}$ and $B$ to be of order one while $G$, $B_h$ and $G_0^{-1}$ to be of order $\beta$. Taking the matrix square root we obtain the block-wise scaling
\begin{equation}
\hat Q^{1/2}\to\left(\begin{smallmatrix}1 & 1 \\ 1 & \beta\end{smallmatrix}\right)\odot\hat Q^{1/2}\ ,
\end{equation}
which does give $(\hat Q^{1/2}\zeta)_q$ of order one and $(\hat Q^{1/2}\zeta)_h$ of order $\beta$.
As a consequence we obtain that $f=-\phi$ and that $P_w$, $P_h$ and $P_h'$ are peaked around their respective maximum $w^*$, $h^*$ and $h^{'*}$, and that they can be approximated by Gaussian measures. Notice that $P_{qh}$ is not peaked as to its $q$ variable, which has to be integrated over all its range, which leads to the marginale $P_h$ and the potential $\psi_h$ eq.~\eqref{eqApp:psiHCont}.

Last, differentiating the free energy $f$ with respect to $s$ and $s'$ we obtain the expected errors and accuracies:
\begin{align}
& E_\mathrm{train} = \mathbb E_{y,\zeta,\xi}\ell(yh_K^*)\ , && \mathrm{Acc}_\mathrm{train} = \mathbb E_{y,\zeta,\xi}\delta_{y=\sign(h_K^*)}\ ,\\
& E_\mathrm{test} = \mathbb E_{y,\zeta,\xi}\ell(yh_K^{'*})\ , && \mathrm{Acc}_\mathrm{test} = \mathbb E_{y,\zeta,\xi}\delta_{y=\sign(h_K^{'*})}\ .
\end{align}

\subsection{Self-consistent equations}
The extremality condition $\nabla_{\Theta,\hat\Theta}\phi$ gives the following self-consistent equations on the order parameters. $\mathcal P$ is the operator that acts by linearly combining quantities evaluated at $h^*$, taken with $\bar s=1$ and $\bar s=0$ with weights $\rho$ and $1-\rho$, according to $\mathcal P(g(h))=\rho g(h^*)+(1-\rho)g(h^{'*})$. We assume $l_2$ regularization, i.e. $\gamma(w)=w^2/2$.
{\small
\begin{align}
& m_w = \frac{1}{\alpha}\frac{\hat m_w}{r+\hat V_w} \label{eqApp:pointFixeC_début} \\
& Q_w = \frac{1}{\alpha}\frac{\hat Q_w+\hat m_w^2}{(r+\hat V_w)^2} \\
& V_w = \frac{1}{\alpha}\frac{1}{r+\hat V_w} \\
& \left(\begin{smallmatrix}\hat m_w \\ \hat m \\ m \\ \cdot \end{smallmatrix}\right) = \left(\begin{smallmatrix}K\sqrt\mu & & 0 \\ & \lambda^\mathrm{e}tI_K & \\ 0 & & I_{K+1}\end{smallmatrix}\right)\mathbb E_{y,\xi,\zeta}\,y\mathcal P\left(\begin{smallmatrix}G_0^{-1}(D_{qh}h-B) \\ h\end{smallmatrix}\right) \label{eqApp:pointFixeSym_m} \\
& \left(\begin{smallmatrix}\hat Q_w & & & \cdot \\ & \hat Q_h & Q_{qh} & \\ & Q_{qh}^T & Q_h & \\ \cdot & & & \cdot\end{smallmatrix}\right) = \left(\begin{smallmatrix}K & & 0 \\ & tI_K & \\ 0 & & I_{K+1}\end{smallmatrix}\right) \mathbb E_{y,\xi,\zeta}\mathcal P\left(\left(\begin{smallmatrix}G_0^{-1}(D_{qh}h-B) \\ h\end{smallmatrix}\right)^{\otimes 2}\right) \left(\begin{smallmatrix}K & & 0 \\ & tI_K & \\ 0 & & I_{K+1}\end{smallmatrix}\right) \label{eqApp:pointFixeSym_Q} \\
&\left(\begin{smallmatrix}\hat V_w & & & \cdot \\ & \hat V_h & V_{qh} & \\ & V_{qh}^T & V_h& \\ \cdot & & & \cdot\end{smallmatrix}\right)
= \mathcal P\left(\cov_{\psi_{qh}}\left(\begin{smallmatrix}q\\ h\end{smallmatrix}\right)\right) \label{eqApp:pointFixeSym_V}
\end{align}
}
We use the notation $\cdot$ for unspecified padding to reach vectors of size $2(K+1)$ and matrices of size $2(K+1)\times 2(K+1)$.

The extremizer $h^*$ of $\psi_h$ is
\begin{align}
h^*=G^{-1}\left(B_h+D_{qh}^TG_0^{-1}B\right)\ .
\end{align}
It has to be plugged in to the fixed-point equations (\ref{eqApp:pointFixeSym_m}-\ref{eqApp:pointFixeSym_Q}) and the expectation over the disorder has to be taken.

As to the variances eq.~\eqref{eqApp:pointFixeSym_V}, we have $\cov_{\psi_{qh}}\left(\begin{smallmatrix}q\\ h\end{smallmatrix}\right) = \left(\begin{smallmatrix}G_q & -\mathrm iG_{qh}\\ -\mathrm iG_{qh}^T & G_h\end{smallmatrix}\right)^{-1}$ and using Schur's complement on $G_q$ invertible, one obtains
\begin{align}
& \left(\begin{smallmatrix} \cdot & \cdot \\ -\mathrm iV_{qh} & \cdot \end{smallmatrix}\right) = t\mathcal P\left(G_0^{-1}D_{qh}G^{-1}\right) \label{eqApp:pointFixeC_Vqh} \\
& \left(\begin{smallmatrix}V_h & \cdot \\ \cdot & \cdot\end{smallmatrix}\right) = \mathcal P\left(G^{-1}\right) \\
& \left(\begin{smallmatrix}\hat V_w & \cdot \\ \cdot & \hat V_h \end{smallmatrix}\right) = \left(\begin{smallmatrix}K^2 & 0 \\ 0 & t^2I_K \end{smallmatrix}\right) \mathcal P\left(G_0^{-1} -G_0^{-1}D_{qh}G^{-1}D_{qh}^TG_0^{-1}\right) \label{eqApp:pointFixeC_VC}
\end{align}

The continuation of the computation and how to solve these equations is detailed in the main part \ref{sec:développementRlimiteCont}.

\subsection{Solution in the continuous limit at large $r$}
We report the final values of the order parameters, given in the main part \ref{sec:contAsymptChara}. We set $x=k/K$ and $z=l/K$ continuous indices ranging from 0 to 1. We define the resolvants
\begin{align}
\varphi(x) &= \left \{\begin{array}{cc}
e^{\lambda^\mathrm{e} tx} & \mathrm{if}\,\delta_\mathrm{e}=0 \\
\sum_{\nu>0}^\infty\nu(\lambda^\mathrm{e})^{\nu-1}\frac{I_\nu(2tx)}{tx} & \mathrm{if}\,\delta_\mathrm{e}=1
\end{array}\right .\ ,\\
\Phi(x,z) &= \left \{\begin{array}{cc}
I_0(2t\sqrt{xz}) & \mathrm{if}\,\delta_\mathrm{e}=0 \\
\frac{I_1(2t(x+z))}{t(x+z)} & \mathrm{if}\,\delta_\mathrm{e}=1
\end{array}\right .\ ,
\end{align}
with $I_\nu$ the modified Bessel function of the second kind of order $\nu$. The effective inverse derivative is
\begin{align}
V_{qh}(x,z) &= \theta(z-x)(z-x)^{-1}I_1(2t(z-x))\ ,\\
D_{qh}^{-1}(x,z) &= D_{qh}^{-1,T}(z,x) = \left \{\begin{array}{cc}
\theta(x-z) & \mathrm{if}\,\delta_\mathrm{e}=0 \\
\frac{1}{t}V_{qh}(z,x) & \mathrm{if}\,\delta_\mathrm{e}=1
\end{array}\right.\ ,
\end{align}
with $\theta$ the step function.

The solution to the fixed-point equations, in the continuous limit $K\to\infty$, at first constant order in $1/r$, is
{\small
\begin{align}
& V_w = \frac{1}{r\alpha} \\
& V_h(x,z) = V_w\Phi(x,z) \\
& \hat V_h(1-x,1-z) = t^2\rho\Phi(x,z) \\
& \hat V_w = t^{-2}\hat V_h(0,0) \\
& \hat m(1-x) = \rho\lambda^\mathrm{e}t\varphi(x) \\
& \hat m_w = \sqrt\mu\frac{1}{\lambda^\mathrm{e}t}\hat m(0) \\
& m_w = \frac{\hat m_w}{r\alpha} \\
& m(x) = (1+\mu)\frac{m_w}{\sqrt\mu}\varphi(x) +\frac{t}{\lambda^\mathrm{e}}\int_0^x\mathrm dx'\int_0^1\mathrm dx''\,\varphi(x-x')V_h(x',x'')\hat m(x'') \\
& \hat Q_w = t^{-2}\hat Q_h(0,0) \\
& Q_w = \frac{\hat Q_w+\hat m_w^2}{r^2\alpha} \\
& \hat Q_h(1-x, 1-z) = t^2\int_{0^-,0^-}^{x,z}\mathrm dx'\mathrm dz'\ \Phi(x-x',z-z')\left[\mathcal P(\hat m^{\otimes 2})(1-x',1-z')\right] \\
& Q_{qh}(1-x,z) = t\int_{0^-,0^-}^{x,z}\mathrm dx'\mathrm dz'\ \Phi(x-x',z-z')\Bigg{[}\mathcal P(\hat m)(1-x')(\lambda^\mathrm{e}tm(z')+\sqrt\mu m_w\delta(z')) \\
&\qquad \left. {}+\int_{0,0^-}^{1^+,1}\mathrm dx''\mathrm dz''\,\left(\hat Q_h(1-x',x'')+\mathcal P(\hat m^{\otimes 2})(1-x',x'')\right)D_{qh}^{-1}(x'',z'')G_0(z'',z')\right] \nonumber \\
& Q_h(x,z) = \int_{0^-,0^-}^{x,z}\mathrm dx'\mathrm dz'\ \Phi(x-x',z-z')\Bigg{[}\hat Q_w\delta(x',z')+(\lambda^\mathrm{e}tm(x')+\sqrt\mu m_w\delta(x'))(\lambda^\mathrm{e}tm(z')+\sqrt\mu m_w\delta(z')) \\
& \qquad {}+\int_{0^-,0}^{1,1^+}\mathrm dx''\mathrm dx'''\,G_0(x',x'')D_{qh}^{-1,T}(x'',x''')\left(t\delta_\mathrm{e}Q_{qh}(x''',z')+\mathcal P(\hat m)(x''')(\lambda^\mathrm{e}tm(z')+\sqrt\mu m_w\delta(z'))\right) \nonumber \\
& \qquad {}+\int_{0,0^-}^{1^+,1}\mathrm dz'''\mathrm dz''\,\left(t\delta_\mathrm{e}Q_{qh}(z''',x')+(\lambda^\mathrm{e}tm(x')+\sqrt\mu m_w\delta(x'))\mathcal P(\hat m)(z''')\right)D_{qh}^{-1}(z''',z'')G_0(z'',z') \nonumber \\
& \quad\left. {}+\int_{0^-,0,0,0^-}^{1,1^+,1^+,1}\mathrm dx''\mathrm dx'''\mathrm dz'''\mathrm dz''\,G_0(x',x'')D_{qh}^{-1,T}(x'',x''')\left(\hat Q_h(x''',z''')+\mathcal P(\hat m^{\otimes 2})(x''',z''')\right)D_{qh}^{-1}(z''',z'')G_0(z'',z')\right] \ ; \nonumber
\end{align}
}
where we set
\begin{align}
\mathcal P(\hat m)(x) &= \hat m(x)+\rho\delta(1-x)\ ,\\
\mathcal P(\hat m^{\otimes 2})(x,z) &= \rho\left(\hat m(x)+\delta(1-x)\right)\left(\hat m(z)+\delta(1-z)\right)+(1-\rho)\hat m(x)\hat m(z)\ ,\\
G_0(x,z) &= t^2V_h(x,z)+V_w\delta(x,z)\ .
\end{align}

The test and train accuracies are
\begin{align}
\mathrm{Acc}_\mathrm{test} &= \mathbb E_{y,\xi,\zeta,\chi} \delta_{y=\sign(h^{'*}(1))} \\
&=\mathbb E_{\xi,\zeta,\chi} \delta_{0<\sqrt\mu m_w+K\int_0^1\mathrm dx\,V(1,x)\hat m(x)+\lambda t\int_0^1\mathrm dx\,m(x)+\sqrt{Q_w}\zeta+K\int_0^1\mathrm dx\mathrm dz\,V(1,x)\hat Q^{1/2}(x,z)\xi(z)+t\int_0^1\mathrm dx\mathrm dz\,Q^{1/2}(x,z)\chi(z)} \\
&= \frac{1}{2}\left(1+\mathrm{erf}\left(\frac{\sqrt\mu m_w+K\int_0^1\mathrm dx\,V(1,x)\hat m(x)+\lambda t\int_0^1\mathrm dx\,m(x)}{\sqrt 2\sqrt{Q_w+K^2\int_0^1\mathrm dx\mathrm dz\,V(1,x)\hat Q(x,z)V(z,1)+t^2\int_0^1\mathrm dx\mathrm dz\,Q(x,z)}}\right)\right) \\
&= \frac{1}{2}\left(1+\mathrm{erf}\left(\frac{m(1)-\rho V(1,1)}{\sqrt 2\sqrt{Q(1,1)-m(1)^2-\rho(1-\rho)V(1,1)^2}}\right)\right) \label{eqApp:précisionGCNcontinu}
\end{align}
and
\begin{align}
\mathrm{Acc}_\mathrm{train} &= \mathbb E_{y,\xi,\zeta,\chi} \delta_{y=\sign(h^{*}(1))} \\
&= \mathbb E_{y,\xi,\zeta,\chi} \delta_{y=\sign(h^{'*}(1)+V(1,1)y)} \\
&= \frac{1}{2}\left(1+\mathrm{erf}\left(\frac{m(1)+(1-\rho)V(1,1)}{\sqrt 2\sqrt{Q(1,1)-m(1)^2-\rho(1-\rho)V(1,1)^2}}\right)\right)
\end{align}
To obtain the last expressions we integrated $m$ and $Q$ by parts thanks to the self-consistent conditions they satisfy.
\end{widetext}

\subsection{Higher orders in $1/r$: how to pursue the computation}
\label{sec:appOrdresSuivants}
The solution given in the main part \ref{sec:contAsymptChara} and reproduced above are for infinite regularization $r$, keeping only the first constant order. We briefly show how to pursue the computation at any order.

The self-consistent equations for $V_{qh}$, $V_h$ and $\hat V_h$ at any order can be phrased as, rewritting eqs.~(\ref{eqApp:pointFixeC_Vqh}-\ref{eqApp:pointFixeC_VC}) and extending the matrices by continuity:
\begin{align}
& \frac{1}{t}V_{qh} = \mathcal P\left(D_{qh}^{-1,T}\sum_{a\ge 0}\left(-G_hD_{qh}^{-1}G_0D_{qh}^{-1,T}\right)^a\right)\ , \label{eqApp:edpTotVqh} \\
& V_h = D_{qh}^{-1}\mathcal P\left(G_0\sum_{a\ge 0}\left(-D_{qh}^{-1,T}G_hD_{qh}^{-1}G_0\right)^a\right)D_{qh}^{-1,T}\ , \label{eqApp:edpTotVh} \\
& \hat V_h = t^2D_{qh}^{-1,T}\mathcal P\left(G_h\sum_{a\ge 0}\left(-D_{qh}^{-1}G_0D_{qh}^{-1,T}G_h\right)^a\right)D_{qh}^{-1} \label{eqApp:edpTotVhC}
\end{align}
where we remind that $G_0=t^2V_h+V_w\delta(x,z)=\mathcal O(1/r)$, $G_h=\hat V_h+\bar s\delta(1-x,1-z)$ and $D_{qh}=D-t\delta_\mathrm{e}V_{qh}^T$.
These equations form a system of non-linear integral equations. A perturbative approach with expansion in powers of $1/r$ should allow to solve it. At each order one has to solve linear integral equations whose resolvant is $\Phi$ for $V_h$ and $\hat V_h$, the previously determined resolvant to the constant order. The perturbations have to summed and the resulting $V_{qh}$, $V_h$ and $\hat V_h$ can be used to express $h^*$, $h^{'*}$ and the other order parameters.

\subsection{Interpretation of terms of DMFT: computation}
\label{sec:appLienDMFT}
We prove the relations given in the main part \ref{sec:lienDMFT}, that state an equivalence between the order parameters $V_h$, $V_{qh}$ and $\hat V_h$ stemming from the replica computation and the correlation and response functions of the dynamical process that $h$ follows. We assume that the regularization $r$ is large and we derive the equalities to the constant order.

We introduce the tilting field $\eta(x)\in\mathbb R^N$ and the tilted Hamiltonian as
\begin{align}
& \frac{\mathrm dh}{\mathrm dx}(x) = \frac{t}{\sqrt N}\tilde A^\mathrm{e}h(x)+\eta(x)\ ,\\
& h(x) = \int_0^x\mathrm dx'e^{(x-x')\frac{t}{\sqrt N}\tilde A^\mathrm{e}}\left(\eta(x')+\delta(x')\frac{1}{\sqrt N}Xw\right)\ ,\\
& H(\eta) = \frac{1}{2}(y-h(1))^TR(y-h(1))+\frac{r}{2}w^Tw\ ,
\end{align}
where $R\in\mathbb R^{N\times N}$ diagonal accounts for the train and test nodes. We write $\langle \cdot\rangle_\beta$ the expectation under the density $e^{-\beta H(\eta)}/Z$ (normalized only at $\eta=0$, $Z$ is not a function of $\eta$).

For $V_h$ we have:
\begin{align}
& \frac{\beta}{N}\tr\left[\langle h(x)h(z)^T\rangle_\beta-\langle h(x)\rangle_\beta\langle h(z)^T\rangle_\beta\right]|_{\eta=0} \\
&\quad =\frac{1}{N}\tr\left(e^{\frac{tx}{\sqrt N}\tilde A^\mathrm{e}}\frac{1}{N}X(\langle ww^T\rangle_\beta-\langle w\rangle_\beta\langle w^T\rangle_\beta)X^Te^{\frac{tz}{\sqrt N}\tilde A^\mathrm{e}}\right)\\
&\quad = \frac{V_w}{N}\left \{\begin{array}{lc}
\tr\left(e^{\frac{tx}{\sqrt N}\tilde A}e^{\frac{tz}{\sqrt N}\tilde A^T}\right) & \mathrm{if}\,\delta_\mathrm{e}=0 \\
\tr\left(e^{\frac{tx+tz}{\sqrt N}\tilde A^\mathrm{s}}\right) & \mathrm{if}\,\delta_\mathrm{e}=1
\end{array}\right.\ .
\end{align}
We used that in the large regularization limit the covariance of $w$ is $I_M/r$ and $V_w=r\alpha$. We distinguish the two cases symmetrized or not. For the symmetrized case we have
\begin{align}
\frac{V_w}{N}\tr\left(e^{\frac{tx+tz}{\sqrt N}\tilde A^\mathrm{s}}\right) &= \int_{-2}^{+2}\frac{\mathrm d\hat\lambda}{2\pi}\sqrt{4-\hat\lambda^2}e^{\hat\lambda t(x+z)} \\
&= V_w\frac{I_1(2t(x+z))}{t(x+z)}\ ,
\end{align}
where we used that the spectrum of $\tilde A^\mathrm{s}/\sqrt N$ follows the semi-circle law up to negligible corrections. For the asymmetric case we expand the two exponentials. $\tilde A\approx \Xi$ has independent Gaussian entries.
\begin{align}
& \frac{V_w}{N}\tr\left(e^{\frac{tx}{\sqrt N}\tilde A}e^{\frac{tz}{\sqrt N}\tilde A^T}\right) \\
&\quad =\sum_{n,m\ge 0}\frac{V_w}{N^{1+\frac{n+m}{2}}}\frac{(tx)^n(tz)^m}{n!m!} \\
& \qquad \sum_{i_1,\ldots,i_n}\sum_{j_1,\ldots,j_m}\Xi_{i_1i_2}\ldots\Xi_{i_{n-1}i_n}\Xi_{i_nj_1}\Xi_{j_2j_1}\Xi_{j_3j_2}\ldots\Xi_{i_1j_m} \nonumber \\
&\quad = V_w\sum_{n}\frac{(t^2xz)^n}{(n!)^2} = V_wI_0(2t\sqrt{xz})\ .
\end{align}
In the sum only contribute the terms where $j_2=i_n, \ldots, j_m=i_2$ for $m=n$. Consequently in both cases we obtain that
\begin{align}
V_h(x,z) = \frac{\beta}{N}\tr\left[\langle h(x)h(z)^T\rangle_\beta-\langle h(x)\rangle_\beta\langle h(z)^T\rangle_\beta\right]|_{\eta=0}
\end{align}
$V_h$ is the correlation function between the states $h(x)\in\mathbb R^N$ of the network, under the dynamic defined by the Hamiltonian \eqref{eq:hamiltonien}. 

This derivation can be used to compute the resolvant $\Phi=V_h/V_w$ in the symmetrized case, instead of solving the integral equation that defines it eq.~\eqref{eq:resPhi}, that is $\Phi(x,z) = D_{qh}^{-1}(t^2\Phi(x,z)+\delta(x,z))D_{qh}^{-1,T}$. As a consequence of the two equivalent definitions we obtain the following mathematical identity, for all $x$ and $z$:
\begin{align}
&\int_{0,0}^{x,z}\mathrm dx'\mathrm dz'\,\frac{I_1(2(x-x'))}{x-x'}\frac{I_1(2(x'+z'))}{x'+z'}\frac{I_1(2(z-z'))}{z-z'} \nonumber \\
& \quad = \frac{I_1(2(x+z))}{x+z}-\frac{I_1(2x)I_1(2z)}{xz}\ .
\end{align}

For $V_{qh}$ we have:
\begin{align}
& \frac{t}{N}\tr\frac{\partial}{\partial\eta(z)}\langle h(x)\rangle_\beta|_{\eta=0} \\
&\quad = \frac{t}{N}\tr e^{(x-z)\frac{t}{\sqrt N}\tilde A^\mathrm{e}}\theta(x-z) \\
&\quad = \left \{\begin{array}{lc}
\theta(x-z) & \mathrm{if}\,\delta_\mathrm{e}=0 \\
\theta(x-z)(x-z)^{-1}I_1(2t(x-z)) & \mathrm{if}\,\delta_\mathrm{e}=1
\end{array}\right. \\
&\quad = V_{qh}(x,z)\ .
\end{align}
We neglected the terms of order $1/r$ stemming from $w$. We integrated over the spectrum of $\tilde A^\mathrm{e}$, which follows the semi-circle law (symmetric case) or the circular law (asymmetric) up to negligeable corrections. We obtain that $V_{qh}$ is the response function oh $h$.

Last for $\hat V_h$ we have:
\begin{align}
& \frac{t^2}{\beta^2N}\tr\frac{\partial^2}{\partial\eta(x)\partial\eta(z)}\langle 1\rangle_\beta|_{\eta=0} \\
&\quad = \frac{t^2}{N}\tr \left[R\langle (y-h(1))^{\otimes 2}\rangle_\beta|_{\eta=0} \right. \nonumber\\
&\qquad\qquad \left. Re^{(1-z)\frac{t}{\sqrt N}\tilde A^\mathrm{e}}e^{(1-x)\frac{t}{\sqrt N}(\tilde A^\mathrm{e})^T}\right] \\
&\quad = \frac{\rho t^2}{N}\tr e^{(1-z)\frac{t}{\sqrt N}\tilde A^\mathrm{e}}e^{(1-x)\frac{t}{\sqrt N}(\tilde A^\mathrm{e})^T}\\
&\quad = \hat V_h(x,z)\ .
\end{align}
We neglected the terms of order $1/\beta$ obtained by differenciating only once $e^{-\beta H}$ and these of order $1/r$, i.e. $y-h(1)\approx y$. We obtain that $\hat V_h$ is the correlation function between the responses.

\subsection{Limiting cases}
\label{sec:contCasLimites}
To obtain insights on the behaviour of the test accuracy and to make connections with already studied models we expand \eqref{eqApp:précisionGCNcontinu} around the limiting cases $t\to 0$ and $t\to\infty$.

At $t\to 0$ we use that $\varphi(x)=1+\lambda^\mathrm{e}tx+O(t^2)$ and $\Phi(x,z)=1+O(t^2)$; this simplifies several terms. We obtain the following expansions at the first order in $t$:
\begin{align}
& V_w=\frac{1}{r\alpha}\ ,\quad V(x,z)=\frac{1}{r\alpha}\ ,\\
& \hat m_w=\rho\sqrt\mu\ ,\quad \hat m(x)=\rho\lambda^\mathrm{e}t\ ,\\
& m_w=\frac{\rho}{r\alpha}\sqrt\mu\ ,\quad m(x)=\frac{\rho}{r\alpha}(1+\mu)(1+\lambda^\mathrm{e}t(x+1))\ ,\\
& \hat Q_w=\rho\ ,\quad \hat Q_h(x,z)=0\ , \\
& Q_w=\frac{\rho+\rho^2\mu}{r\alpha}\ ,\quad Q_{qh}=O(t)\ ,\\
& Q_h(1,1)=Q_w+m(0)^2+\rho(1-\rho)V_w^2+2\frac{\rho^2}{r^2\alpha^2}(1+\mu)^2\lambda^\mathrm{e}t\ .
\end{align}
Pluging them in eq.~\eqref{eqApp:précisionGCNcontinu} we obtain the expression given in the main part \ref{sec:continuConsequences}:
\begin{equation}
\mathrm{Acc}_\mathrm{test} = \frac{1}{2}\left(1+\mathrm{erf}\left(\frac{1}{\sqrt 2}\sqrt\frac{\rho}{\alpha}\frac{\mu+\lambda^\mathrm{e}t(2+\mu)}{\sqrt{1+\rho\mu}}\right)\right)\ .
\end{equation}

At $t\to\infty$ we assume that $\lambda^\mathrm{e}>1$. We distinguish the two cases asymmetric or symmetrized. For asymmetric we have $\varphi(x)=\exp(\lambda^\mathrm{e}tx)$ and $\log\Phi(x,z)=\Theta(2t\sqrt{xz})$. For the symmetrized we have
\begin{align}
\varphi(x) &= \frac{1}{tx}\frac{\partial}{\partial \lambda^\mathrm{e}}\sum_{\nu\ge 0}^\infty(\lambda^\mathrm{e})^\nu I_\nu(2tx) \\
&\approx \frac{1}{tx}\frac{\partial}{\partial \lambda^\mathrm{e}}\sum_{\nu=-\infty}^{+\infty}(\lambda^\mathrm{e})^\nu I_\nu(2tx) \\
&= \frac{1}{tx}\frac{\partial}{\partial \lambda^\mathrm{e}}e^{tx(\lambda^\mathrm{e}+1/\lambda^\mathrm{e})} \\
&= (1-(\lambda^\mathrm{e})^{-2})e^{tx(\lambda^\mathrm{e}+1/\lambda^\mathrm{e})}
\end{align}
and 
$\log\Phi(x,z)=\Theta(2t(x+z))$. In the two cases, only the few dominant terms scaling like $e^{2\lambda^\mathrm{e}t}$ or $e^{2(\lambda^\mathrm{e}+1/\lambda^\mathrm{e})t}$ dominate in \eqref{eqApp:précisionGCNcontinu}. We obtain
\begin{align}
& \mathrm{Acc}_\mathrm{test} \approx \frac{1}{2}\left(1+\mathrm{erf}\left(\frac{m(1)}{\sqrt 2\sqrt{Q(1,1)-m(1)^2}}\right)\right) \\
& m(x) = \frac{\rho}{r\alpha}\varphi(1)\varphi(x)(1+\mu+C(\lambda^\mathrm{e})) \\
& C(\lambda^\mathrm{e}) \mkern-5mu = \mkern-5mu \int_0^\infty\mkern-9mu\mathrm dx'\mathrm dz'
\left \{\begin{array}{lc}
I_0(2\sqrt{x'z'})e^{-(x'+z')\lambda^\mathrm{e}} & \mathrm{if}\,\delta_\mathrm{e}=0 \\
\mkern-5mu\frac{I_1(2(x'+z'))}{x'+z'}e^{-(x'+z')(\lambda^\mathrm{e}+1/\lambda^\mathrm{e})} & \mathrm{if}\,\delta_\mathrm{e}=1
\end{array}\right . \\
& Q(1,1)\approx\int_0^1\mathrm dx'\mathrm dz'\Phi(1-x',1-z')(\lambda^\mathrm{e})^2t^2m(x')m(z')
\end{align}
where in $m$ we performed the changes of variables $x'\to x'/t$ and $z'\to z'/t$ and took the limit $t\to\infty$ in the integration bounds to remove the dependency in $t$ and $x$. Performing a change of variables $1-x'\to x'/t$ and $1-z'\to z'/t$ in $Q(1,1)$ we can express $\mathrm{Acc}_\mathrm{test}$ solely in terms of $C(\lambda^\mathrm{e})$. Last we use the identity
\begin{equation}
C(\lambda^\mathrm{e})=\frac{1}{(\lambda^\mathrm{e})^2-1}\ ,
\end{equation}
valid in the two cases asymmetric or not, to obtain the expression given in the main part \ref{sec:continuConsequences}:
\begin{align}
\mathrm{Acc}_\mathrm{test} &\underset{t\to\infty}{\longrightarrow} \frac{1}{2}\left(1+\mathrm{erf}\left(\frac{\lambda^\mathrm{e} q_\mathrm{PCA}}{\sqrt 2}\right)\right)\ ,\\
q_\mathrm{PCA} &= \sqrt{1-(\lambda^\mathrm{e})^{-2}}
\end{align}

%% file: main.bbl
\providecommand{\noopsort}[1]{}\providecommand{\singleletter}[1]{#1}%
\begin{thebibliography}{56}%
\makeatletter
\providecommand \@ifxundefined [1]{%
 \@ifx{#1\undefined}
}%
\providecommand \@ifnum [1]{%
 \ifnum #1\expandafter \@firstoftwo
 \else \expandafter \@secondoftwo
 \fi
}%
\providecommand \@ifx [1]{%
 \ifx #1\expandafter \@firstoftwo
 \else \expandafter \@secondoftwo
 \fi
}%
\providecommand \natexlab [1]{#1}%
\providecommand \enquote  [1]{``#1''}%
\providecommand \bibnamefont  [1]{#1}%
\providecommand \bibfnamefont [1]{#1}%
\providecommand \citenamefont [1]{#1}%
\providecommand \href@noop [0]{\@secondoftwo}%
\providecommand \href [0]{\begingroup \@sanitize@url \@href}%
\providecommand \@href[1]{\@@startlink{#1}\@@href}%
\providecommand \@@href[1]{\endgroup#1\@@endlink}%
\providecommand \@sanitize@url [0]{\catcode `\\12\catcode `\$12\catcode
  `\&12\catcode `\#12\catcode `\^12\catcode `\_12\catcode `\%12\relax}%
\providecommand \@@startlink[1]{}%
\providecommand \@@endlink[0]{}%
\providecommand \url  [0]{\begingroup\@sanitize@url \@url }%
\providecommand \@url [1]{\endgroup\@href {#1}{\urlprefix }}%
\providecommand \urlprefix  [0]{URL }%
\providecommand \Eprint [0]{\href }%
\providecommand \doibase [0]{https://doi.org/}%
\providecommand \selectlanguage [0]{\@gobble}%
\providecommand \bibinfo  [0]{\@secondoftwo}%
\providecommand \bibfield  [0]{\@secondoftwo}%
\providecommand \translation [1]{[#1]}%
\providecommand \BibitemOpen [0]{}%
\providecommand \bibitemStop [0]{}%
\providecommand \bibitemNoStop [0]{.\EOS\space}%
\providecommand \EOS [0]{\spacefactor3000\relax}%
\providecommand \BibitemShut  [1]{\csname bibitem#1\endcsname}%
\let\auto@bib@innerbib\@empty
\bibitem [{\citenamefont {Wang}\ \emph {et~al.}(2023)\citenamefont {Wang},
  \citenamefont {Li},\ and\ \citenamefont
  {Barati~Farimani}}]{wang23gnnMolecules}%
  \BibitemOpen
  \bibfield  {author} {\bibinfo {author} {\bibfnamefont {Y.}~\bibnamefont
  {Wang}}, \bibinfo {author} {\bibfnamefont {Z.}~\bibnamefont {Li}},\ and\
  \bibinfo {author} {\bibfnamefont {A.}~\bibnamefont {Barati~Farimani}},\
  }\bibinfo {title} {Graph neural networks for molecules},\ in\ \href@noop {}
  {\emph {\bibinfo {booktitle} {Machine Learning in Molecular Sciences}}}\
  (\bibinfo  {publisher} {Springer International Publishing},\ \bibinfo {year}
  {2023})\ p.\ \bibinfo {pages} {21–66},\ \bibinfo {note}
  {arXiv:2209.05582}\BibitemShut {NoStop}%
\bibitem [{\citenamefont {Li}\ \emph {et~al.}(2022)\citenamefont {Li},
  \citenamefont {Huang},\ and\ \citenamefont {Zitnik}}]{li22gnnBiomedecine}%
  \BibitemOpen
  \bibfield  {author} {\bibinfo {author} {\bibfnamefont {M.~M.}\ \bibnamefont
  {Li}}, \bibinfo {author} {\bibfnamefont {K.}~\bibnamefont {Huang}},\ and\
  \bibinfo {author} {\bibfnamefont {M.}~\bibnamefont {Zitnik}},\ }\bibfield
  {title} {\bibinfo {title} {Graph representation learning in biomedicine and
  healthcare},\ }\href@noop {} {\bibfield  {journal} {\bibinfo  {journal}
  {Nature Biomedical Engineering}\ }\textbf {\bibinfo {volume} {6}},\ \bibinfo
  {pages} {1353–1369} (\bibinfo {year} {2022})},\ \bibinfo {note}
  {arXiv:2104.04883}\BibitemShut {NoStop}%
\bibitem [{\citenamefont {Bessadok}\ \emph {et~al.}(2021)\citenamefont
  {Bessadok}, \citenamefont {Mahjoub},\ and\ \citenamefont
  {Rekik}}]{bessadok21gnnNeuroscience}%
  \BibitemOpen
  \bibfield  {author} {\bibinfo {author} {\bibfnamefont {A.}~\bibnamefont
  {Bessadok}}, \bibinfo {author} {\bibfnamefont {M.~A.}\ \bibnamefont
  {Mahjoub}},\ and\ \bibinfo {author} {\bibfnamefont {I.}~\bibnamefont
  {Rekik}},\ }\href@noop {} {\bibinfo {title} {Graph neural networks in network
  neuroscience}} (\bibinfo {year} {2021}),\ \bibinfo {note}
  {arXiv:2106.03535}\BibitemShut {NoStop}%
\bibitem [{\citenamefont {Sanchez-Gonzalez}\ \emph {et~al.}(2020)\citenamefont
  {Sanchez-Gonzalez}, \citenamefont {Godwin}, \citenamefont {Pfaff},
  \citenamefont {Ying}, \citenamefont {Leskovec},\ and\ \citenamefont
  {Battaglia}}]{sanchez2020gnnPhysique}%
  \BibitemOpen
  \bibfield  {author} {\bibinfo {author} {\bibfnamefont {A.}~\bibnamefont
  {Sanchez-Gonzalez}}, \bibinfo {author} {\bibfnamefont {J.}~\bibnamefont
  {Godwin}}, \bibinfo {author} {\bibfnamefont {T.}~\bibnamefont {Pfaff}},
  \bibinfo {author} {\bibfnamefont {R.}~\bibnamefont {Ying}}, \bibinfo {author}
  {\bibfnamefont {J.}~\bibnamefont {Leskovec}},\ and\ \bibinfo {author}
  {\bibfnamefont {P.~W.}\ \bibnamefont {Battaglia}},\ }\bibfield  {title}
  {\bibinfo {title} {Learning to simulate complex physics with graph
  networks},\ }in\ \href@noop {} {\emph {\bibinfo {booktitle} {Proceedings of
  the 37th International Conference on Machine Learning}}}\ (\bibinfo {year}
  {2020})\ \bibinfo {note} {arXiv:2002.09405}\BibitemShut {NoStop}%
\bibitem [{\citenamefont {Shlomi}\ \emph {et~al.}(2020)\citenamefont {Shlomi},
  \citenamefont {Battaglia},\ and\ \citenamefont
  {Vlimant}}]{shlomi20gnnPhysiquePart}%
  \BibitemOpen
  \bibfield  {author} {\bibinfo {author} {\bibfnamefont {J.}~\bibnamefont
  {Shlomi}}, \bibinfo {author} {\bibfnamefont {P.}~\bibnamefont {Battaglia}},\
  and\ \bibinfo {author} {\bibfnamefont {J.-R.}\ \bibnamefont {Vlimant}},\
  }\bibfield  {title} {\bibinfo {title} {Graph neural networks in particle
  physics},\ }\href@noop {} {\bibfield  {journal} {\bibinfo  {journal} {Machine
  Learning: Science and Technology}\ }\textbf {\bibinfo {volume} {2}} (\bibinfo
  {year} {2020})},\ \bibinfo {note} {arXiv:2007.13681}\BibitemShut {NoStop}%
\bibitem [{\citenamefont {Peng}\ \emph {et~al.}(2021)\citenamefont {Peng},
  \citenamefont {Choi},\ and\ \citenamefont {Xu}}]{peng20gnnOpt}%
  \BibitemOpen
  \bibfield  {author} {\bibinfo {author} {\bibfnamefont {Y.}~\bibnamefont
  {Peng}}, \bibinfo {author} {\bibfnamefont {B.}~\bibnamefont {Choi}},\ and\
  \bibinfo {author} {\bibfnamefont {J.}~\bibnamefont {Xu}},\ }\bibfield
  {title} {\bibinfo {title} {Graph learning for combinatorial optimization: A
  survey of state-of-the-art},\ }\href@noop {} {\bibfield  {journal} {\bibinfo
  {journal} {Data Science and Engineering}\ }\textbf {\bibinfo {volume} {6}},\
  \bibinfo {pages} {119} (\bibinfo {year} {2021})},\ \bibinfo {note}
  {arXiv:2008.12646}\BibitemShut {NoStop}%
\bibitem [{\citenamefont {Cappart}\ \emph {et~al.}(2023)\citenamefont
  {Cappart}, \citenamefont {Chételat}, \citenamefont {Khalil}, \citenamefont
  {Lodi}, \citenamefont {Morris},\ and\ \citenamefont
  {Veličković}}]{cappart23gnnOpt}%
  \BibitemOpen
  \bibfield  {author} {\bibinfo {author} {\bibfnamefont {Q.}~\bibnamefont
  {Cappart}}, \bibinfo {author} {\bibfnamefont {D.}~\bibnamefont {Chételat}},
  \bibinfo {author} {\bibfnamefont {E.}~\bibnamefont {Khalil}}, \bibinfo
  {author} {\bibfnamefont {A.}~\bibnamefont {Lodi}}, \bibinfo {author}
  {\bibfnamefont {C.}~\bibnamefont {Morris}},\ and\ \bibinfo {author}
  {\bibfnamefont {P.}~\bibnamefont {Veličković}},\ }\bibfield  {title}
  {\bibinfo {title} {Combinatorial optimization and reasoning with graph neural
  networks},\ }\href@noop {} {\bibfield  {journal} {\bibinfo  {journal}
  {Journal of Machine Learning Research}\ }\textbf {\bibinfo {volume} {24}},\
  \bibinfo {pages} {1} (\bibinfo {year} {2023})},\ \bibinfo {note}
  {arXiv:2102.09544}\BibitemShut {NoStop}%
\bibitem [{\citenamefont {Morris}\ \emph {et~al.}(2024)\citenamefont {Morris},
  \citenamefont {Frasca}, \citenamefont {Dym}, \citenamefont {Maron},
  \citenamefont {Ceylan}, \citenamefont {Levie}, \citenamefont {Lim},
  \citenamefont {Bronstein}, \citenamefont {Grohe},\ and\ \citenamefont
  {Jegelka}}]{morris24positionGNN}%
  \BibitemOpen
  \bibfield  {author} {\bibinfo {author} {\bibfnamefont {C.}~\bibnamefont
  {Morris}}, \bibinfo {author} {\bibfnamefont {F.}~\bibnamefont {Frasca}},
  \bibinfo {author} {\bibfnamefont {N.}~\bibnamefont {Dym}}, \bibinfo {author}
  {\bibfnamefont {H.}~\bibnamefont {Maron}}, \bibinfo {author} {\bibfnamefont
  {I.~I.}\ \bibnamefont {Ceylan}}, \bibinfo {author} {\bibfnamefont
  {R.}~\bibnamefont {Levie}}, \bibinfo {author} {\bibfnamefont
  {D.}~\bibnamefont {Lim}}, \bibinfo {author} {\bibfnamefont {M.}~\bibnamefont
  {Bronstein}}, \bibinfo {author} {\bibfnamefont {M.}~\bibnamefont {Grohe}},\
  and\ \bibinfo {author} {\bibfnamefont {S.}~\bibnamefont {Jegelka}},\
  }\bibfield  {title} {\bibinfo {title} {Position: Future directions in the
  theory of graph machine learning},\ }in\ \href@noop {} {\emph {\bibinfo
  {booktitle} {Proceedings of the 41st International Conference on Machine
  Learning}}}\ (\bibinfo {year} {2024})\BibitemShut {NoStop}%
\bibitem [{\citenamefont {Li}\ \emph {et~al.}(2018)\citenamefont {Li},
  \citenamefont {Han},\ and\ \citenamefont {Wu}}]{li18oversmoothing}%
  \BibitemOpen
  \bibfield  {author} {\bibinfo {author} {\bibfnamefont {Q.}~\bibnamefont
  {Li}}, \bibinfo {author} {\bibfnamefont {Z.}~\bibnamefont {Han}},\ and\
  \bibinfo {author} {\bibfnamefont {X.-M.}\ \bibnamefont {Wu}},\ }\bibfield
  {title} {\bibinfo {title} {Deeper insights into graph convolutional networks
  for semi-supervised learning},\ }in\ \href@noop {} {\emph {\bibinfo
  {booktitle} {Thirty-Second AAAI conference on artificial intelligence}}}\
  (\bibinfo {year} {2018})\ \bibinfo {note} {arXiv:1801.07606}\BibitemShut
  {NoStop}%
\bibitem [{\citenamefont {Oono}\ and\ \citenamefont
  {Suzuki}(2020)}]{oono20oversmoothing}%
  \BibitemOpen
  \bibfield  {author} {\bibinfo {author} {\bibfnamefont {K.}~\bibnamefont
  {Oono}}\ and\ \bibinfo {author} {\bibfnamefont {T.}~\bibnamefont {Suzuki}},\
  }\bibfield  {title} {\bibinfo {title} {Graph neural networks exponentially
  lose expressive power for node classification},\ }in\ \href@noop {} {\emph
  {\bibinfo {booktitle} {International conference on learning
  representations}}}\ (\bibinfo {year} {2020})\ \bibinfo {note}
  {arXiv:1905.10947}\BibitemShut {NoStop}%
\bibitem [{\citenamefont {Li}\ \emph {et~al.}(2019)\citenamefont {Li},
  \citenamefont {Müller}, \citenamefont {Thabet},\ and\ \citenamefont
  {Ghanem}}]{li19GNNskipConn}%
  \BibitemOpen
  \bibfield  {author} {\bibinfo {author} {\bibfnamefont {G.}~\bibnamefont
  {Li}}, \bibinfo {author} {\bibfnamefont {M.}~\bibnamefont {Müller}},
  \bibinfo {author} {\bibfnamefont {A.}~\bibnamefont {Thabet}},\ and\ \bibinfo
  {author} {\bibfnamefont {B.}~\bibnamefont {Ghanem}},\ }\bibfield  {title}
  {\bibinfo {title} {Deep{GCN}s: Can {GCN}s go as deep as {CNN}s?},\ }in\
  \href@noop {} {\emph {\bibinfo {booktitle} {ICCV}}}\ (\bibinfo {year}
  {2019})\ \bibinfo {note} {arXiv:1904.03751}\BibitemShut {NoStop}%
\bibitem [{\citenamefont {Chen}\ \emph {et~al.}(2020)\citenamefont {Chen},
  \citenamefont {Wei}, \citenamefont {Huang}, \citenamefont {Ding},\ and\
  \citenamefont {Li}}]{chen20GNNskipConn}%
  \BibitemOpen
  \bibfield  {author} {\bibinfo {author} {\bibfnamefont {M.}~\bibnamefont
  {Chen}}, \bibinfo {author} {\bibfnamefont {Z.}~\bibnamefont {Wei}}, \bibinfo
  {author} {\bibfnamefont {Z.}~\bibnamefont {Huang}}, \bibinfo {author}
  {\bibfnamefont {B.}~\bibnamefont {Ding}},\ and\ \bibinfo {author}
  {\bibfnamefont {Y.}~\bibnamefont {Li}},\ }\bibfield  {title} {\bibinfo
  {title} {Simple and deep graph convolutional networks},\ }in\ \href@noop {}
  {\emph {\bibinfo {booktitle} {Proceedings of the 37th International
  Conference on Machine Learning}}}\ (\bibinfo {year} {2020})\ \bibinfo {note}
  {arXiv:2007.02133}\BibitemShut {NoStop}%
\bibitem [{\citenamefont {Ju}\ \emph {et~al.}(2023)\citenamefont {Ju},
  \citenamefont {Li}, \citenamefont {Sharma},\ and\ \citenamefont
  {Zhang}}]{ju23pacBayes}%
  \BibitemOpen
  \bibfield  {author} {\bibinfo {author} {\bibfnamefont {H.}~\bibnamefont
  {Ju}}, \bibinfo {author} {\bibfnamefont {D.}~\bibnamefont {Li}}, \bibinfo
  {author} {\bibfnamefont {A.}~\bibnamefont {Sharma}},\ and\ \bibinfo {author}
  {\bibfnamefont {H.~R.}\ \bibnamefont {Zhang}},\ }\bibfield  {title} {\bibinfo
  {title} {Generalization in graph neural networks: Improved {PAC}-{B}ayesian
  bounds on graph diffusion},\ }in\ \href@noop {} {\emph {\bibinfo {booktitle}
  {AISTATS}}}\ (\bibinfo {year} {2023})\ \bibinfo {note}
  {arXiv:2302.04451}\BibitemShut {NoStop}%
\bibitem [{\citenamefont {Tang}\ and\ \citenamefont
  {Liu}(2023)}]{tang23generalizationGNN}%
  \BibitemOpen
  \bibfield  {author} {\bibinfo {author} {\bibfnamefont {H.}~\bibnamefont
  {Tang}}\ and\ \bibinfo {author} {\bibfnamefont {Y.}~\bibnamefont {Liu}},\
  }\href@noop {} {\bibinfo {title} {Towards understanding the generalization of
  graph neural networks}} (\bibinfo {year} {2023}),\ \bibinfo {note}
  {arXiv:2305.08048}\BibitemShut {NoStop}%
\bibitem [{\citenamefont {Cong}\ \emph {et~al.}(2021)\citenamefont {Cong},
  \citenamefont {Ramezani},\ and\ \citenamefont {Mahdavi}}]{cong21depthGCN}%
  \BibitemOpen
  \bibfield  {author} {\bibinfo {author} {\bibfnamefont {W.}~\bibnamefont
  {Cong}}, \bibinfo {author} {\bibfnamefont {M.}~\bibnamefont {Ramezani}},\
  and\ \bibinfo {author} {\bibfnamefont {M.}~\bibnamefont {Mahdavi}},\
  }\bibfield  {title} {\bibinfo {title} {On provable benefits of depth in
  training graph convolutional networks},\ }in\ \href@noop {} {\emph {\bibinfo
  {booktitle} {35th Conference on Neural Information Processing Systems}}}\
  (\bibinfo {year} {2021})\ \bibinfo {note} {arxiv:2110.15174}\BibitemShut
  {NoStop}%
\bibitem [{\citenamefont {Esser}\ \emph {et~al.}(2021)\citenamefont {Esser},
  \citenamefont {Vankadara},\ and\ \citenamefont
  {Ghoshdastidar}}]{esser21generalizationCSBM}%
  \BibitemOpen
  \bibfield  {author} {\bibinfo {author} {\bibfnamefont {P.~M.}\ \bibnamefont
  {Esser}}, \bibinfo {author} {\bibfnamefont {L.~C.}\ \bibnamefont
  {Vankadara}},\ and\ \bibinfo {author} {\bibfnamefont {D.}~\bibnamefont
  {Ghoshdastidar}},\ }\bibfield  {title} {\bibinfo {title} {Learning theory can
  (sometimes) explain generalisation in graph neural networks},\ }in\
  \href@noop {} {\emph {\bibinfo {booktitle} {35th Conference on Neural
  Information Processing Systems}}}\ (\bibinfo {year} {2021})\ \bibinfo {note}
  {arXiv:2112.03968}\BibitemShut {NoStop}%
\bibitem [{\citenamefont {Seung}\ \emph {et~al.}(1992)\citenamefont {Seung},
  \citenamefont {Sompolinsky},\ and\ \citenamefont
  {Tishby}}]{seung1992statistical}%
  \BibitemOpen
  \bibfield  {author} {\bibinfo {author} {\bibfnamefont {H.~S.}\ \bibnamefont
  {Seung}}, \bibinfo {author} {\bibfnamefont {H.}~\bibnamefont {Sompolinsky}},\
  and\ \bibinfo {author} {\bibfnamefont {N.}~\bibnamefont {Tishby}},\
  }\bibfield  {title} {\bibinfo {title} {Statistical mechanics of learning from
  examples},\ }\href@noop {} {\bibfield  {journal} {\bibinfo  {journal}
  {Physical review A}\ }\textbf {\bibinfo {volume} {45}},\ \bibinfo {pages}
  {6056} (\bibinfo {year} {1992})}\BibitemShut {NoStop}%
\bibitem [{\citenamefont {Loureiro}\ \emph {et~al.}(2021)\citenamefont
  {Loureiro}, \citenamefont {Gerbelot}, \citenamefont {Cui}, \citenamefont
  {Goldt}, \citenamefont {Krzakala}, \citenamefont {Mezard},\ and\
  \citenamefont {Zdeborov{\'a}}}]{loureiro2021learning}%
  \BibitemOpen
  \bibfield  {author} {\bibinfo {author} {\bibfnamefont {B.}~\bibnamefont
  {Loureiro}}, \bibinfo {author} {\bibfnamefont {C.}~\bibnamefont {Gerbelot}},
  \bibinfo {author} {\bibfnamefont {H.}~\bibnamefont {Cui}}, \bibinfo {author}
  {\bibfnamefont {S.}~\bibnamefont {Goldt}}, \bibinfo {author} {\bibfnamefont
  {F.}~\bibnamefont {Krzakala}}, \bibinfo {author} {\bibfnamefont
  {M.}~\bibnamefont {Mezard}},\ and\ \bibinfo {author} {\bibfnamefont
  {L.}~\bibnamefont {Zdeborov{\'a}}},\ }\bibfield  {title} {\bibinfo {title}
  {Learning curves of generic features maps for realistic datasets with a
  teacher-student model},\ }\href@noop {} {\bibfield  {journal} {\bibinfo
  {journal} {Advances in Neural Information Processing Systems}\ }\textbf
  {\bibinfo {volume} {34}},\ \bibinfo {pages} {18137} (\bibinfo {year}
  {2021})}\BibitemShut {NoStop}%
\bibitem [{\citenamefont {Mei}\ and\ \citenamefont
  {Montanari}(2022)}]{mei2022generalization}%
  \BibitemOpen
  \bibfield  {author} {\bibinfo {author} {\bibfnamefont {S.}~\bibnamefont
  {Mei}}\ and\ \bibinfo {author} {\bibfnamefont {A.}~\bibnamefont
  {Montanari}},\ }\bibfield  {title} {\bibinfo {title} {The generalization
  error of random features regression: Precise asymptotics and the double
  descent curve},\ }\href@noop {} {\bibfield  {journal} {\bibinfo  {journal}
  {Communications on Pure and Applied Mathematics}\ }\textbf {\bibinfo {volume}
  {75}},\ \bibinfo {pages} {667} (\bibinfo {year} {2022})}\BibitemShut
  {NoStop}%
\bibitem [{\citenamefont {Shi}\ \emph {et~al.}(2023)\citenamefont {Shi},
  \citenamefont {Pan}, \citenamefont {Hu},\ and\ \citenamefont
  {Dokmani{\'c}}}]{shi2022statistical}%
  \BibitemOpen
  \bibfield  {author} {\bibinfo {author} {\bibfnamefont {C.}~\bibnamefont
  {Shi}}, \bibinfo {author} {\bibfnamefont {L.}~\bibnamefont {Pan}}, \bibinfo
  {author} {\bibfnamefont {H.}~\bibnamefont {Hu}},\ and\ \bibinfo {author}
  {\bibfnamefont {I.}~\bibnamefont {Dokmani{\'c}}},\ }\bibfield  {title}
  {\bibinfo {title} {Homophily modulates double descent generalization in graph
  convolution networks},\ }\href@noop {} {\bibfield  {journal} {\bibinfo
  {journal} {PNAS}\ }\textbf {\bibinfo {volume} {121}} (\bibinfo {year}
  {2023})},\ \bibinfo {note} {arXiv:2212.13069}\BibitemShut {NoStop}%
\bibitem [{\citenamefont {Yan}\ and\ \citenamefont
  {Sarkar}(2021)}]{yan2021covariate}%
  \BibitemOpen
  \bibfield  {author} {\bibinfo {author} {\bibfnamefont {B.}~\bibnamefont
  {Yan}}\ and\ \bibinfo {author} {\bibfnamefont {P.}~\bibnamefont {Sarkar}},\
  }\bibfield  {title} {\bibinfo {title} {Covariate regularized community
  detection in sparse graphs},\ }\href@noop {} {\bibfield  {journal} {\bibinfo
  {journal} {Journal of the American Statistical Association}\ }\textbf
  {\bibinfo {volume} {116}},\ \bibinfo {pages} {734} (\bibinfo {year}
  {2021})},\ \bibinfo {note} {arxiv:1607.02675}\BibitemShut {NoStop}%
\bibitem [{\citenamefont {Deshpande}\ \emph {et~al.}(2018)\citenamefont
  {Deshpande}, \citenamefont {Sen}, \citenamefont {Montanari},\ and\
  \citenamefont {Mossel}}]{cSBM18}%
  \BibitemOpen
  \bibfield  {author} {\bibinfo {author} {\bibfnamefont {Y.}~\bibnamefont
  {Deshpande}}, \bibinfo {author} {\bibfnamefont {S.}~\bibnamefont {Sen}},
  \bibinfo {author} {\bibfnamefont {A.}~\bibnamefont {Montanari}},\ and\
  \bibinfo {author} {\bibfnamefont {E.}~\bibnamefont {Mossel}},\ }\bibfield
  {title} {\bibinfo {title} {Contextual stochastic block models},\ }in\
  \href@noop {} {\emph {\bibinfo {booktitle} {Advances in Neural Information
  Processing Systems}}},\ Vol.~\bibinfo {volume} {31},\ \bibinfo {editor}
  {edited by\ \bibinfo {editor} {\bibfnamefont {S.}~\bibnamefont {Bengio}},
  \bibinfo {editor} {\bibfnamefont {H.}~\bibnamefont {Wallach}}, \bibinfo
  {editor} {\bibfnamefont {H.}~\bibnamefont {Larochelle}}, \bibinfo {editor}
  {\bibfnamefont {K.}~\bibnamefont {Grauman}}, \bibinfo {editor} {\bibfnamefont
  {N.}~\bibnamefont {Cesa-Bianchi}},\ and\ \bibinfo {editor} {\bibfnamefont
  {R.}~\bibnamefont {Garnett}}}\ (\bibinfo {year} {2018})\ \bibinfo {note}
  {arxiv:1807.09596}\BibitemShut {NoStop}%
\bibitem [{\citenamefont {Chien}\ \emph {et~al.}(2021)\citenamefont {Chien},
  \citenamefont {Peng}, \citenamefont {Li},\ and\ \citenamefont
  {Milenkovic}}]{chien20pageRank}%
  \BibitemOpen
  \bibfield  {author} {\bibinfo {author} {\bibfnamefont {E.}~\bibnamefont
  {Chien}}, \bibinfo {author} {\bibfnamefont {J.}~\bibnamefont {Peng}},
  \bibinfo {author} {\bibfnamefont {P.}~\bibnamefont {Li}},\ and\ \bibinfo
  {author} {\bibfnamefont {O.}~\bibnamefont {Milenkovic}},\ }\bibfield  {title}
  {\bibinfo {title} {Adaptative universal generalized pagerank graph neural
  network},\ }in\ \href@noop {} {\emph {\bibinfo {booktitle} {Proceedings of
  the 39th International Conference on Learning Representations}}}\ (\bibinfo
  {year} {2021})\ \bibinfo {note} {arxiv:2006.07988}\BibitemShut {NoStop}%
\bibitem [{\citenamefont {Fu}\ \emph {et~al.}(2021)\citenamefont {Fu},
  \citenamefont {Zhao},\ and\ \citenamefont {Bian}}]{fu21pLaplacianGNN}%
  \BibitemOpen
  \bibfield  {author} {\bibinfo {author} {\bibfnamefont {G.}~\bibnamefont
  {Fu}}, \bibinfo {author} {\bibfnamefont {P.}~\bibnamefont {Zhao}},\ and\
  \bibinfo {author} {\bibfnamefont {Y.}~\bibnamefont {Bian}},\ }\bibfield
  {title} {\bibinfo {title} {p-{L}aplacian based graph neural networks},\ }in\
  \href@noop {} {\emph {\bibinfo {booktitle} {Proceedings of the 39th
  International Conference on Machine Learning}}}\ (\bibinfo {year} {2021})\
  \bibinfo {note} {arxiv:2111.07337}\BibitemShut {NoStop}%
\bibitem [{\citenamefont {Lei}\ \emph {et~al.}(2022)\citenamefont {Lei},
  \citenamefont {Wang}, \citenamefont {Li}, \citenamefont {Ding},\ and\
  \citenamefont {Wei}}]{lei22evenNet}%
  \BibitemOpen
  \bibfield  {author} {\bibinfo {author} {\bibfnamefont {R.}~\bibnamefont
  {Lei}}, \bibinfo {author} {\bibfnamefont {Z.}~\bibnamefont {Wang}}, \bibinfo
  {author} {\bibfnamefont {Y.}~\bibnamefont {Li}}, \bibinfo {author}
  {\bibfnamefont {B.}~\bibnamefont {Ding}},\ and\ \bibinfo {author}
  {\bibfnamefont {Z.}~\bibnamefont {Wei}},\ }\bibfield  {title} {\bibinfo
  {title} {Even{N}et: Ignoring odd-hop neighbors improves robustness of graph
  neural networks},\ }in\ \href@noop {} {\emph {\bibinfo {booktitle} {36th
  Conference on Neural Information Processing Systems}}}\ (\bibinfo {year}
  {2022})\ \bibinfo {note} {arxiv:2205.13892}\BibitemShut {NoStop}%
\bibitem [{\citenamefont {Duranthon}\ and\ \citenamefont
  {Zdeborová}(2024{\natexlab{a}})}]{dz24gcn}%
  \BibitemOpen
  \bibfield  {author} {\bibinfo {author} {\bibfnamefont {O.}~\bibnamefont
  {Duranthon}}\ and\ \bibinfo {author} {\bibfnamefont {L.}~\bibnamefont
  {Zdeborová}},\ }\bibfield  {title} {\bibinfo {title} {Asymptotic
  generalization error of a single-layer graph convolutional network},\ }in\
  \href@noop {} {\emph {\bibinfo {booktitle} {The Learning on Graphs
  Conference}}}\ (\bibinfo {year} {2024})\ \bibinfo {note}
  {arxiv:2402.03818}\BibitemShut {NoStop}%
\bibitem [{\citenamefont {Chen}\ \emph {et~al.}(2018)\citenamefont {Chen},
  \citenamefont {Rubanova}, \citenamefont {Bettencourt},\ and\ \citenamefont
  {Duvenaud}}]{chen18NODE}%
  \BibitemOpen
  \bibfield  {author} {\bibinfo {author} {\bibfnamefont {R.~T.~Q.}\
  \bibnamefont {Chen}}, \bibinfo {author} {\bibfnamefont {Y.}~\bibnamefont
  {Rubanova}}, \bibinfo {author} {\bibfnamefont {J.}~\bibnamefont
  {Bettencourt}},\ and\ \bibinfo {author} {\bibfnamefont {D.}~\bibnamefont
  {Duvenaud}},\ }\bibfield  {title} {\bibinfo {title} {Neural ordinary
  differential equations},\ }in\ \href@noop {} {\emph {\bibinfo {booktitle}
  {32nd Conference on Neural Information Processing Systems}}}\ (\bibinfo
  {year} {2018})\ \bibinfo {note} {arXiv:1806.07366}\BibitemShut {NoStop}%
\bibitem [{\citenamefont {Kipf}\ and\ \citenamefont
  {Welling}(2017)}]{kipf17GCN}%
  \BibitemOpen
  \bibfield  {author} {\bibinfo {author} {\bibfnamefont {T.~N.}\ \bibnamefont
  {Kipf}}\ and\ \bibinfo {author} {\bibfnamefont {M.}~\bibnamefont {Welling}},\
  }\bibfield  {title} {\bibinfo {title} {Semi-supervised classification with
  graph convolutional networks},\ }in\ \href@noop {} {\emph {\bibinfo
  {booktitle} {International Conference on Learning Representations}}}\
  (\bibinfo {year} {2017})\ \bibinfo {note} {arxiv:1609.02907}\BibitemShut
  {NoStop}%
\bibitem [{\citenamefont {Cui}\ \emph {et~al.}(2023)\citenamefont {Cui},
  \citenamefont {Krzakala},\ and\ \citenamefont
  {Zdeborová}}]{hugo23reseauProfondProp}%
  \BibitemOpen
  \bibfield  {author} {\bibinfo {author} {\bibfnamefont {H.}~\bibnamefont
  {Cui}}, \bibinfo {author} {\bibfnamefont {F.}~\bibnamefont {Krzakala}},\ and\
  \bibinfo {author} {\bibfnamefont {L.}~\bibnamefont {Zdeborová}},\ }\bibfield
   {title} {\bibinfo {title} {Bayes-optimal learning of deep random networks of
  extensive-width},\ }in\ \href@noop {} {\emph {\bibinfo {booktitle}
  {Proceedings of the 40th International Conference on Machine Learning}}}\
  (\bibinfo {year} {2023})\ \bibinfo {note} {arxiv:2302.00375}\BibitemShut
  {NoStop}%
\bibitem [{\citenamefont {McCallum}\ \emph {et~al.}(2000)\citenamefont
  {McCallum}, \citenamefont {Nigam}, \citenamefont {Rennie},\ and\
  \citenamefont {Seymore}}]{mccallum00cora}%
  \BibitemOpen
  \bibfield  {author} {\bibinfo {author} {\bibfnamefont {A.~K.}\ \bibnamefont
  {McCallum}}, \bibinfo {author} {\bibfnamefont {K.}~\bibnamefont {Nigam}},
  \bibinfo {author} {\bibfnamefont {J.}~\bibnamefont {Rennie}},\ and\ \bibinfo
  {author} {\bibfnamefont {K.}~\bibnamefont {Seymore}},\ }\bibfield  {title}
  {\bibinfo {title} {Automating the construction of internet portals with
  machine learning},\ }\href@noop {} {\bibfield  {journal} {\bibinfo  {journal}
  {Information Retrieval}\ }\textbf {\bibinfo {volume} {3}},\ \bibinfo {pages}
  {127–163} (\bibinfo {year} {2000})}\BibitemShut {NoStop}%
\bibitem [{\citenamefont {Shchur}\ \emph {et~al.}(2018)\citenamefont {Shchur},
  \citenamefont {Mumme}, \citenamefont {Bojchevski},\ and\ \citenamefont
  {Günnemann}}]{shchur18coauthor}%
  \BibitemOpen
  \bibfield  {author} {\bibinfo {author} {\bibfnamefont {O.}~\bibnamefont
  {Shchur}}, \bibinfo {author} {\bibfnamefont {M.}~\bibnamefont {Mumme}},
  \bibinfo {author} {\bibfnamefont {A.}~\bibnamefont {Bojchevski}},\ and\
  \bibinfo {author} {\bibfnamefont {S.}~\bibnamefont {Günnemann}},\ }\bibfield
   {title} {\bibinfo {title} {Pitfalls of graph neural network evaluation},\
  }\href@noop {} {\  (\bibinfo {year} {2018})},\ \bibinfo {note}
  {arXiv:1811.05868}\BibitemShut {NoStop}%
\bibitem [{\citenamefont {Giles}\ \emph {et~al.}(1998)\citenamefont {Giles},
  \citenamefont {Bollacker},\ and\ \citenamefont {Lawrenc}}]{giles98citeseer}%
  \BibitemOpen
  \bibfield  {author} {\bibinfo {author} {\bibfnamefont {C.~L.}\ \bibnamefont
  {Giles}}, \bibinfo {author} {\bibfnamefont {K.~D.}\ \bibnamefont
  {Bollacker}},\ and\ \bibinfo {author} {\bibfnamefont {S.}~\bibnamefont
  {Lawrenc}},\ }\bibfield  {title} {\bibinfo {title} {Citeseer: An automatic
  citation indexing system},\ }in\ \href@noop {} {\emph {\bibinfo {booktitle}
  {Proceedings of the third ACM conference on Digital libraries}}}\ (\bibinfo
  {year} {1998})\ p.\ \bibinfo {pages} {89–98}\BibitemShut {NoStop}%
\bibitem [{\citenamefont {Sen}\ \emph {et~al.}(2008)\citenamefont {Sen},
  \citenamefont {Namata}, \citenamefont {Bilgic}, \citenamefont {Getoor},
  \citenamefont {Galligher},\ and\ \citenamefont {Eliassi-Rad}}]{sen08pubmed}%
  \BibitemOpen
  \bibfield  {author} {\bibinfo {author} {\bibfnamefont {P.}~\bibnamefont
  {Sen}}, \bibinfo {author} {\bibfnamefont {G.}~\bibnamefont {Namata}},
  \bibinfo {author} {\bibfnamefont {M.}~\bibnamefont {Bilgic}}, \bibinfo
  {author} {\bibfnamefont {L.}~\bibnamefont {Getoor}}, \bibinfo {author}
  {\bibfnamefont {B.}~\bibnamefont {Galligher}},\ and\ \bibinfo {author}
  {\bibfnamefont {T.}~\bibnamefont {Eliassi-Rad}},\ }\bibfield  {title}
  {\bibinfo {title} {Collective classification in network data},\ }\href@noop
  {} {\bibfield  {journal} {\bibinfo  {journal} {AI magazine}\ }\textbf
  {\bibinfo {volume} {29}} (\bibinfo {year} {2008})}\BibitemShut {NoStop}%
\bibitem [{\citenamefont {Baranwal}\ \emph {et~al.}(2021)\citenamefont
  {Baranwal}, \citenamefont {Fountoulakis},\ and\ \citenamefont
  {Jagannath}}]{fountoulakis21GC}%
  \BibitemOpen
  \bibfield  {author} {\bibinfo {author} {\bibfnamefont {A.}~\bibnamefont
  {Baranwal}}, \bibinfo {author} {\bibfnamefont {K.}~\bibnamefont
  {Fountoulakis}},\ and\ \bibinfo {author} {\bibfnamefont {A.}~\bibnamefont
  {Jagannath}},\ }\bibfield  {title} {\bibinfo {title} {Graph convolution for
  semi-supervised classification: Improved linear separability and
  out-of-distribution generalization},\ }in\ \href@noop {} {\emph {\bibinfo
  {booktitle} {Proceedings of the 38th International Conference on Machine
  Learning}}}\ (\bibinfo {year} {2021})\ \bibinfo {note}
  {arxiv:2102.06966}\BibitemShut {NoStop}%
\bibitem [{\citenamefont {Baranwal}\ \emph {et~al.}(2023)\citenamefont
  {Baranwal}, \citenamefont {Fountoulakis},\ and\ \citenamefont
  {Jagannath}}]{baranwal23clipGNN}%
  \BibitemOpen
  \bibfield  {author} {\bibinfo {author} {\bibfnamefont {A.}~\bibnamefont
  {Baranwal}}, \bibinfo {author} {\bibfnamefont {K.}~\bibnamefont
  {Fountoulakis}},\ and\ \bibinfo {author} {\bibfnamefont {A.}~\bibnamefont
  {Jagannath}},\ }\bibfield  {title} {\bibinfo {title} {Optimality of
  message-passing architectures for sparse graphs},\ }in\ \href@noop {} {\emph
  {\bibinfo {booktitle} {37th Conference on Neural Information Processing
  Systems}}}\ (\bibinfo {year} {2023})\ \bibinfo {note}
  {arxiv:2305.10391}\BibitemShut {NoStop}%
\bibitem [{\citenamefont {Wang}\ \emph {et~al.}(2024)\citenamefont {Wang},
  \citenamefont {Baranwal},\ and\ \citenamefont
  {Fountoulakis}}]{wang24GCNmultiConv}%
  \BibitemOpen
  \bibfield  {author} {\bibinfo {author} {\bibfnamefont {R.}~\bibnamefont
  {Wang}}, \bibinfo {author} {\bibfnamefont {A.}~\bibnamefont {Baranwal}},\
  and\ \bibinfo {author} {\bibfnamefont {K.}~\bibnamefont {Fountoulakis}},\
  }\href@noop {} {\bibinfo {title} {Analysis of corrected graph convolutions}}
  (\bibinfo {year} {2024}),\ \bibinfo {note} {arXiv:2405.13987}\BibitemShut
  {NoStop}%
\bibitem [{\citenamefont {Mignacco}\ \emph {et~al.}(2020)\citenamefont
  {Mignacco}, \citenamefont {Krzakala}, \citenamefont {Lu},\ and\ \citenamefont
  {Zdeborov{\'a}}}]{mignacco20gaussMixt}%
  \BibitemOpen
  \bibfield  {author} {\bibinfo {author} {\bibfnamefont {F.}~\bibnamefont
  {Mignacco}}, \bibinfo {author} {\bibfnamefont {F.}~\bibnamefont {Krzakala}},
  \bibinfo {author} {\bibfnamefont {Y.~M.}\ \bibnamefont {Lu}},\ and\ \bibinfo
  {author} {\bibfnamefont {L.}~\bibnamefont {Zdeborov{\'a}}},\ }\bibfield
  {title} {\bibinfo {title} {The role of regularization in classification of
  high-dimensional noisy {G}aussian mixture},\ }in\ \href@noop {} {\emph
  {\bibinfo {booktitle} {International conference on learning
  representations}}}\ (\bibinfo {year} {2020})\ \bibinfo {note}
  {arxiv:2002.11544}\BibitemShut {NoStop}%
\bibitem [{\citenamefont {Aubin}\ \emph {et~al.}(2020)\citenamefont {Aubin},
  \citenamefont {Krzakala}, \citenamefont {Lu},\ and\ \citenamefont
  {Zdeborov{\'a}}}]{aubin20glmReg}%
  \BibitemOpen
  \bibfield  {author} {\bibinfo {author} {\bibfnamefont {B.}~\bibnamefont
  {Aubin}}, \bibinfo {author} {\bibfnamefont {F.}~\bibnamefont {Krzakala}},
  \bibinfo {author} {\bibfnamefont {Y.~M.}\ \bibnamefont {Lu}},\ and\ \bibinfo
  {author} {\bibfnamefont {L.}~\bibnamefont {Zdeborov{\'a}}},\ }\bibfield
  {title} {\bibinfo {title} {Generalization error in high-dimensional
  perceptrons: Approaching {B}ayes error with convex optimization},\ }in\
  \href@noop {} {\emph {\bibinfo {booktitle} {Advances in Neural Information
  Processing Systems}}}\ (\bibinfo {year} {2020})\ \bibinfo {note}
  {arxiv:2006.06560}\BibitemShut {NoStop}%
\bibitem [{\citenamefont {Duranthon}\ and\ \citenamefont
  {Zdeborová}(2024{\natexlab{b}})}]{dz23csbm}%
  \BibitemOpen
  \bibfield  {author} {\bibinfo {author} {\bibfnamefont {O.}~\bibnamefont
  {Duranthon}}\ and\ \bibinfo {author} {\bibfnamefont {L.}~\bibnamefont
  {Zdeborová}},\ }\bibfield  {title} {\bibinfo {title} {Optimal inference in
  contextual stochastic block models},\ }\href@noop {} {\bibfield  {journal}
  {\bibinfo  {journal} {Transactions on Machine Learning Research}\ } (\bibinfo
  {year} {2024}{\natexlab{b}})},\ \bibinfo {note}
  {arxiv:2306.07948}\BibitemShut {NoStop}%
\bibitem [{\citenamefont {Keriven}(2022)}]{keriven22oversm}%
  \BibitemOpen
  \bibfield  {author} {\bibinfo {author} {\bibfnamefont {N.}~\bibnamefont
  {Keriven}},\ }\bibfield  {title} {\bibinfo {title} {Not too little, not too
  much: a theoretical analysis of graph (over)smoothing},\ }in\ \href@noop {}
  {\emph {\bibinfo {booktitle} {36th Conference on Neural Information
  Processing Systems}}}\ (\bibinfo {year} {2022})\ \bibinfo {note}
  {arXiv:2205.12156}\BibitemShut {NoStop}%
\bibitem [{\citenamefont {He}\ \emph {et~al.}(2016)\citenamefont {He},
  \citenamefont {Zhang}, \citenamefont {Ren},\ and\ \citenamefont
  {Sun}}]{he15resnet}%
  \BibitemOpen
  \bibfield  {author} {\bibinfo {author} {\bibfnamefont {K.}~\bibnamefont
  {He}}, \bibinfo {author} {\bibfnamefont {X.}~\bibnamefont {Zhang}}, \bibinfo
  {author} {\bibfnamefont {S.}~\bibnamefont {Ren}},\ and\ \bibinfo {author}
  {\bibfnamefont {J.}~\bibnamefont {Sun}},\ }\bibfield  {title} {\bibinfo
  {title} {Deep residual learning for image recognition},\ }in\ \href@noop {}
  {\emph {\bibinfo {booktitle} {IEEE Conference on Computer Vision and Pattern
  Recognition}}}\ (\bibinfo {year} {2016})\ \bibinfo {note}
  {arXiv:1512.03385}\BibitemShut {NoStop}%
\bibitem [{\citenamefont {Pham}\ \emph {et~al.}(2017)\citenamefont {Pham},
  \citenamefont {Tran}, \citenamefont {Phung},\ and\ \citenamefont
  {Venkatesh}}]{pham17columnNet}%
  \BibitemOpen
  \bibfield  {author} {\bibinfo {author} {\bibfnamefont {T.}~\bibnamefont
  {Pham}}, \bibinfo {author} {\bibfnamefont {T.}~\bibnamefont {Tran}}, \bibinfo
  {author} {\bibfnamefont {D.}~\bibnamefont {Phung}},\ and\ \bibinfo {author}
  {\bibfnamefont {S.}~\bibnamefont {Venkatesh}},\ }\bibfield  {title} {\bibinfo
  {title} {Column networks for collective classification},\ }in\ \href@noop {}
  {\emph {\bibinfo {booktitle} {AAAI}}}\ (\bibinfo {year} {2017})\ \bibinfo
  {note} {arXiv:1609.04508}\BibitemShut {NoStop}%
\bibitem [{\citenamefont {Xu}\ \emph {et~al.}(2021)\citenamefont {Xu},
  \citenamefont {Zhang}, \citenamefont {Jegelka},\ and\ \citenamefont
  {Kawaguchi}}]{xu21GNNskipConnGradient}%
  \BibitemOpen
  \bibfield  {author} {\bibinfo {author} {\bibfnamefont {K.}~\bibnamefont
  {Xu}}, \bibinfo {author} {\bibfnamefont {M.}~\bibnamefont {Zhang}}, \bibinfo
  {author} {\bibfnamefont {S.}~\bibnamefont {Jegelka}},\ and\ \bibinfo {author}
  {\bibfnamefont {K.}~\bibnamefont {Kawaguchi}},\ }\bibfield  {title} {\bibinfo
  {title} {Optimization of graph neural networks: Implicit acceleration by skip
  connections and more depth},\ }in\ \href@noop {} {\emph {\bibinfo {booktitle}
  {Proceedings of the 38th International Conference on Machine Learning}}}\
  (\bibinfo {year} {2021})\ \bibinfo {note} {arXiv:2105.04550}\BibitemShut
  {NoStop}%
\bibitem [{\citenamefont {Sander}\ \emph {et~al.}(2022)\citenamefont {Sander},
  \citenamefont {Ablin},\ and\ \citenamefont
  {Peyré}}]{sander22convergenceNODE}%
  \BibitemOpen
  \bibfield  {author} {\bibinfo {author} {\bibfnamefont {M.~E.}\ \bibnamefont
  {Sander}}, \bibinfo {author} {\bibfnamefont {P.}~\bibnamefont {Ablin}},\ and\
  \bibinfo {author} {\bibfnamefont {G.}~\bibnamefont {Peyré}},\ }\bibfield
  {title} {\bibinfo {title} {Do residual neural networks discretize neural
  ordinary differential equations?},\ }in\ \href@noop {} {\emph {\bibinfo
  {booktitle} {36th Conference on Neural Information Processing Systems}}}\
  (\bibinfo {year} {2022})\ \bibinfo {note} {arXiv:2205.14612}\BibitemShut
  {NoStop}%
\bibitem [{\citenamefont {Ling}\ \emph {et~al.}(2016)\citenamefont {Ling},
  \citenamefont {Kurzawski},\ and\ \citenamefont
  {Templeton}}]{ling16turbulenceNN}%
  \BibitemOpen
  \bibfield  {author} {\bibinfo {author} {\bibfnamefont {J.}~\bibnamefont
  {Ling}}, \bibinfo {author} {\bibfnamefont {A.}~\bibnamefont {Kurzawski}},\
  and\ \bibinfo {author} {\bibfnamefont {J.}~\bibnamefont {Templeton}},\
  }\bibfield  {title} {\bibinfo {title} {Reynolds averaged turbulence modelling
  using deep neural networks with embedded invariance},\ }\href@noop {}
  {\bibfield  {journal} {\bibinfo  {journal} {Journal of Fluid Mechanics}\
  }\textbf {\bibinfo {volume} {807}},\ \bibinfo {pages} {155–166} (\bibinfo
  {year} {2016})}\BibitemShut {NoStop}%
\bibitem [{\citenamefont {Rackauckas}\ \emph {et~al.}(2020)\citenamefont
  {Rackauckas}, \citenamefont {Ma}, \citenamefont {Martensen}, \citenamefont
  {Warner}, \citenamefont {Zubov}, \citenamefont {Supekar}, \citenamefont
  {Skinner}, \citenamefont {Ramadhan},\ and\ \citenamefont
  {Edelman}}]{rackauckas20nodeScience}%
  \BibitemOpen
  \bibfield  {author} {\bibinfo {author} {\bibfnamefont {C.}~\bibnamefont
  {Rackauckas}}, \bibinfo {author} {\bibfnamefont {Y.}~\bibnamefont {Ma}},
  \bibinfo {author} {\bibfnamefont {J.}~\bibnamefont {Martensen}}, \bibinfo
  {author} {\bibfnamefont {C.}~\bibnamefont {Warner}}, \bibinfo {author}
  {\bibfnamefont {K.}~\bibnamefont {Zubov}}, \bibinfo {author} {\bibfnamefont
  {R.}~\bibnamefont {Supekar}}, \bibinfo {author} {\bibfnamefont
  {D.}~\bibnamefont {Skinner}}, \bibinfo {author} {\bibfnamefont
  {A.}~\bibnamefont {Ramadhan}},\ and\ \bibinfo {author} {\bibfnamefont
  {A.}~\bibnamefont {Edelman}},\ }\href@noop {} {\bibinfo {title} {Universal
  differential equations for scientific machine learning}} (\bibinfo {year}
  {2020}),\ \bibinfo {note} {arXiv:2001.04385}\BibitemShut {NoStop}%
\bibitem [{\citenamefont {Marion}(2023)}]{marion23genNODE}%
  \BibitemOpen
  \bibfield  {author} {\bibinfo {author} {\bibfnamefont {P.}~\bibnamefont
  {Marion}},\ }\href@noop {} {\bibinfo {title} {Generalization bounds for
  neural ordinary differential equations and deep residual networks}} (\bibinfo
  {year} {2023}),\ \bibinfo {note} {arXiv:2305.06648}\BibitemShut {NoStop}%
\bibitem [{\citenamefont {Poli}\ \emph {et~al.}(2019)\citenamefont {Poli},
  \citenamefont {Massaroli}, \citenamefont {Park}, \citenamefont {Yamashita},
  \citenamefont {Asama},\ and\ \citenamefont {Park}}]{poli19cGNN}%
  \BibitemOpen
  \bibfield  {author} {\bibinfo {author} {\bibfnamefont {M.}~\bibnamefont
  {Poli}}, \bibinfo {author} {\bibfnamefont {S.}~\bibnamefont {Massaroli}},
  \bibinfo {author} {\bibfnamefont {J.}~\bibnamefont {Park}}, \bibinfo {author}
  {\bibfnamefont {A.}~\bibnamefont {Yamashita}}, \bibinfo {author}
  {\bibfnamefont {H.}~\bibnamefont {Asama}},\ and\ \bibinfo {author}
  {\bibfnamefont {J.}~\bibnamefont {Park}},\ }\href@noop {} {\bibinfo {title}
  {Graph neural ordinary differential equations}} (\bibinfo {year} {2019}),\
  \bibinfo {note} {arXiv:1911.07532}\BibitemShut {NoStop}%
\bibitem [{\citenamefont {Xhonneux}\ \emph {et~al.}(2020)\citenamefont
  {Xhonneux}, \citenamefont {Qu},\ and\ \citenamefont {Tang}}]{xhonneux20cGNN}%
  \BibitemOpen
  \bibfield  {author} {\bibinfo {author} {\bibfnamefont {L.-P. A.~C.}\
  \bibnamefont {Xhonneux}}, \bibinfo {author} {\bibfnamefont {M.}~\bibnamefont
  {Qu}},\ and\ \bibinfo {author} {\bibfnamefont {J.}~\bibnamefont {Tang}},\
  }\bibfield  {title} {\bibinfo {title} {Continuous graph neural networks},\
  }in\ \href@noop {} {\emph {\bibinfo {booktitle} {Proceedings of the 37th
  International Conference on Machine Learning}}}\ (\bibinfo {year} {2020})\
  \bibinfo {note} {arXiv:1912.00967}\BibitemShut {NoStop}%
\bibitem [{\citenamefont {Han}\ \emph {et~al.}(2023)\citenamefont {Han},
  \citenamefont {Shi}, \citenamefont {Lin},\ and\ \citenamefont
  {Gao}}]{han23revueDynGNN}%
  \BibitemOpen
  \bibfield  {author} {\bibinfo {author} {\bibfnamefont {A.}~\bibnamefont
  {Han}}, \bibinfo {author} {\bibfnamefont {D.}~\bibnamefont {Shi}}, \bibinfo
  {author} {\bibfnamefont {L.}~\bibnamefont {Lin}},\ and\ \bibinfo {author}
  {\bibfnamefont {J.}~\bibnamefont {Gao}},\ }\href@noop {} {\bibinfo {title}
  {From continuous dynamics to graph neural networks: Neural diffusion and
  beyond}} (\bibinfo {year} {2023}),\ \bibinfo {note}
  {arXiv:2310.10121}\BibitemShut {NoStop}%
\bibitem [{\citenamefont {Lu}\ and\ \citenamefont {Sen}(2020)}]{cSBM20}%
  \BibitemOpen
  \bibfield  {author} {\bibinfo {author} {\bibfnamefont {C.}~\bibnamefont
  {Lu}}\ and\ \bibinfo {author} {\bibfnamefont {S.}~\bibnamefont {Sen}},\
  }\href@noop {} {\bibinfo {title} {Contextual stochastic block model: Sharp
  thresholds and contiguity}} (\bibinfo {year} {2020}),\ \bibinfo {note}
  {arXiv:2011.09841}\BibitemShut {NoStop}%
\bibitem [{\citenamefont {Wu}\ \emph {et~al.}(2019)\citenamefont {Wu},
  \citenamefont {Zhang}, \citenamefont {de~Souza~Jr.}, \citenamefont {Fifty},
  \citenamefont {Yu},\ and\ \citenamefont {Weinberger}}]{wu19simpleGCN}%
  \BibitemOpen
  \bibfield  {author} {\bibinfo {author} {\bibfnamefont {F.}~\bibnamefont
  {Wu}}, \bibinfo {author} {\bibfnamefont {T.}~\bibnamefont {Zhang}}, \bibinfo
  {author} {\bibfnamefont {A.~H.}\ \bibnamefont {de~Souza~Jr.}}, \bibinfo
  {author} {\bibfnamefont {C.}~\bibnamefont {Fifty}}, \bibinfo {author}
  {\bibfnamefont {T.}~\bibnamefont {Yu}},\ and\ \bibinfo {author}
  {\bibfnamefont {K.~Q.}\ \bibnamefont {Weinberger}},\ }\bibfield  {title}
  {\bibinfo {title} {Simplifying graph convolutional networks},\ }in\
  \href@noop {} {\emph {\bibinfo {booktitle} {Proceedings of the 36th
  International Conference on Machine Learning}}}\ (\bibinfo {year} {2019})\
  \bibinfo {note} {arxiv:1902.07153}\BibitemShut {NoStop}%
\bibitem [{\citenamefont {Zhu}\ and\ \citenamefont
  {Koniusz}(2021)}]{zhu21simpleGCN}%
  \BibitemOpen
  \bibfield  {author} {\bibinfo {author} {\bibfnamefont {H.}~\bibnamefont
  {Zhu}}\ and\ \bibinfo {author} {\bibfnamefont {P.}~\bibnamefont {Koniusz}},\
  }\bibfield  {title} {\bibinfo {title} {Simple spectral graph convolution},\
  }in\ \href@noop {} {\emph {\bibinfo {booktitle} {International Conference on
  Learning Representations}}}\ (\bibinfo {year} {2021})\BibitemShut {NoStop}%
\bibitem [{\citenamefont {Lesieur}\ \emph {et~al.}(2017)\citenamefont
  {Lesieur}, \citenamefont {Krzakala},\ and\ \citenamefont
  {Zdeborov{\'a}}}]{lesieur2017constrained}%
  \BibitemOpen
  \bibfield  {author} {\bibinfo {author} {\bibfnamefont {T.}~\bibnamefont
  {Lesieur}}, \bibinfo {author} {\bibfnamefont {F.}~\bibnamefont {Krzakala}},\
  and\ \bibinfo {author} {\bibfnamefont {L.}~\bibnamefont {Zdeborov{\'a}}},\
  }\bibfield  {title} {\bibinfo {title} {Constrained low-rank matrix
  estimation: Phase transitions, approximate message passing and
  applications},\ }\href@noop {} {\bibfield  {journal} {\bibinfo  {journal}
  {Journal of Statistical Mechanics: Theory and Experiment}\ }\textbf {\bibinfo
  {volume} {2017}},\ \bibinfo {pages} {073403} (\bibinfo {year} {2017})},\
  \bibinfo {note} {arxiv:1701.00858}\BibitemShut {NoStop}%
\bibitem [{\citenamefont {Duranthon}\ and\ \citenamefont
  {Zdeborová}(2023)}]{dz23glmSbm}%
  \BibitemOpen
  \bibfield  {author} {\bibinfo {author} {\bibfnamefont {O.}~\bibnamefont
  {Duranthon}}\ and\ \bibinfo {author} {\bibfnamefont {L.}~\bibnamefont
  {Zdeborová}},\ }\bibfield  {title} {\bibinfo {title} {Neural-prior
  stochastic block model},\ }\href@noop {} {\bibfield  {journal} {\bibinfo
  {journal} {Mach. Learn.: Sci. Technol.}\ } (\bibinfo {year} {2023})},\
  \bibinfo {note} {arxiv:2303.09995}\BibitemShut {NoStop}%
\bibitem [{\citenamefont {Baik}\ \emph {et~al.}(2005)\citenamefont {Baik},
  \citenamefont {Arous},\ and\ \citenamefont {Péché}}]{baik05transition}%
  \BibitemOpen
  \bibfield  {author} {\bibinfo {author} {\bibfnamefont {J.}~\bibnamefont
  {Baik}}, \bibinfo {author} {\bibfnamefont {G.~B.}\ \bibnamefont {Arous}},\
  and\ \bibinfo {author} {\bibfnamefont {S.}~\bibnamefont {Péché}},\
  }\bibfield  {title} {\bibinfo {title} {Phase transition of the largest
  eigenvalue for nonnull complex sample covariance matrices},\ }\href@noop {}
  {\bibfield  {journal} {\bibinfo  {journal} {Annals of Probability}\ ,\
  \bibinfo {pages} {1643}} (\bibinfo {year} {2005})}\BibitemShut {NoStop}%
\end{thebibliography}%
